\documentclass[preprint,a4paper,10pt,3p]{elsarticle}

\usepackage{amssymb}
\newdefinition{rmk}{Remark}

\usepackage{physics}

\usepackage{mathtools}

\usepackage{xcolor}

\usepackage{psfrag}

\usepackage{subfigure}

\usepackage{pgfplots}
\usepackage{pgfplotstable}
\pgfplotsset{
tick label style={font=\footnotesize},
label style={font=\footnotesize},
legend style={font=\footnotesize},
axis lines=box,
unbounded coords=jump,
scale only axis,
every axis plot/.append style = {line width = 0.5pt},
xlabel near ticks,
ylabel near ticks,
compat = 1.9
}

\journal{arXiv}

\begin{document}

\begin{frontmatter}


\title{An SPH framework for fluid-solid and contact interaction problems including thermo-mechanical coupling and reversible phase transitions}

\author[1,2]{Sebastian L. Fuchs}
\ead{fuchs@lnm.mw.tum.de}

\author[1]{Christoph Meier}
\ead{meier@lnm.mw.tum.de}

\author[1]{Wolfgang A. Wall}
\ead{wall@lnm.mw.tum.de}

\author[2,3]{Christian J. Cyron\corref{cor}}
\ead{christian.cyron@tuhh.de}

\cortext[cor]{corresponding author}

\address[1]{Institute for Computational Mechanics, Technical University of Munich, Boltzmannstrasse 15, 85748, Garching, Germany}

\address[2]{Institute of Continuum and Materials Mechanics, Hamburg University of Technology, Eissendorfer Str. 42, 21073, Hamburg, Germany}

\address[3]{Institute of Material Systems Modeling, Helmholtz-Zentrum Geesthacht, Max-Planck-Stra{\ss}e 1, 21502 Geesthacht, Germany}

\begin{abstract}
The present work proposes an approach for fluid-solid and contact interaction problems including thermo-mechanical coupling and reversible phase transitions. The solid field is assumed to consist of several arbitrarily-shaped, undeformable but mobile rigid bodies, that are evolved in time individually and allowed to get into mechanical contact with each other. The fluid field generally consists of multiple liquid or gas phases. All fields are spatially discretized using the method of smoothed particle hydrodynamics (SPH). This approach is especially suitable in the context of continually changing interface topologies and dynamic phase transitions without the need for additional methodological and computational effort for interface tracking as compared to mesh- or grid-based methods. Proposing a concept for the parallelization of the computational framework, in particular concerning a computationally efficient evaluation of rigid body motion, is an essential part of this work. Finally, the accuracy and robustness of the proposed framework is demonstrated by several numerical examples in two and three dimensions, involving multiple rigid bodies, two-phase flow, and reversible phase transitions, with a focus on two potential application scenarios in the fields of engineering and biomechanics: powder bed fusion additive manufacturing (PBFAM) and disintegration of food boluses in the human stomach. The efficiency of the parallel computational framework is demonstrated by a strong scaling analysis.
\end{abstract}



\begin{keyword}
rigid body motion \sep two-phase flow \sep reversible phase transitions \sep smoothed particle hydrodynamics \sep metal additive manufacturing \sep gastric fluid mechanics
\end{keyword}

\end{frontmatter}

\section{Introduction} \label{sec:intro}

In many applications in science and engineering, like for example in some areas of biomechanics, fluid-solid and contact interaction problems characterized by a large number of solid bodies immersed in a fluid flow and undergoing reversible phase transitions, are of great interest. Often, explicitly considering the deformation of solid bodies can be neglected, which reduces the complexity of the problem to the treatment of undeformable but mobile rigid bodies, in favor of simplified modeling. Most current mesh- or grid-based methods, e.g., the finite element method (FEM), the finite difference method (FDM), or the finite volume method (FVM), require substantial methodological and computational efforts to model the motion of rigid bodies in fluid flow. To overcome those issues, several approaches, e.g., based on the particle finite element method (PFEM)~\cite{Idelsohn2003b,Idelsohn2006,Onate2008}, or on smoothed particle hydrodynamics (SPH)~\cite{Hashemi2012,Bouscasse2013,Bian2014,Polfer2016,Dong2019,Dietemann2020,Kijanski2020}, have been proposed. SPH as a mesh-free discretization scheme is, due to its Lagrangian nature, very well suited for flow problems involving multiple phases, dynamic and reversible phase transitions, and complex interface topologies. This makes SPH very appropriate for a wide range of applications in engineering, e.g., in metal additive manufacturing melt pool modeling~\cite{Meier2017b,Furstenau2020}, or in biomechanics, e.g., for modeling the digestion of food in the human stomach~\cite{Brandstaeter2019}. All aforementioned SPH formulations for modeling rigid body motion in fluid flow have in common, that rigid bodies are spatially discretized as (clusters of) particles. It is generally accepted that advanced boundary particle methods, e.g., based on the extrapolation of field quantities from fluid to boundary particles~\cite{Morris1997,Basa2009,Adami2012}, are beneficial, because one can model the fluid field close to the boundary with high accuracy. In many of the aforementioned applications, an exact representation of the fluid-solid interface plays an important role. Therefore, herein a formulation of this kind proposed in~\cite{Adami2012} is utilized. To the best of the authors’ knowledge non of the aforementioned SPH formulations modeling rigid body motion in fluid flow simultaneously consider thermal conduction, reversible phase transitions, and multiple (liquid and gas) phases.

To help close this gap, this contribution proposes a smoothed particle hydrodynamics framework for fluid-solid and contact interaction problems including thermo-mechanical coupling and reversible phase transitions. The solid field is assumed to consist of several arbitrarily-shaped, undeformable but mobile rigid bodies, that are evolved in time individually. Based on a temperature field, provided by solving the heat equation, reversible phase transitions, i.e., melting and solidification, are evaluated between the fluid and the solid field. As a result, the shape and the total number of rigid bodies may vary over time. In addition, contact between rigid bodies is considered by employing a spring-dashpot model.

While parallel implementation aspects along with detailed scalability studies are not in the focus of the aforementioned references, in this work, a concept for the parallelization of the computational framework is proposed, setting the focus in particular on an efficient evaluation of rigid body motion. The parallel behavior is demonstrated, confirming that detailed studies at a large scale become possible. It shall be noted, that the parallel implementation of such a computational framework is far from trivial but indispensable when examining numerical examples that are of practical relevance. Note that the introduced concept for the parallelization of the computational framework is applicable not only when using SPH as a discretization scheme, but also for other particle-based methods, e.g., discrete element method (DEM), or molecular dynamics (MD).

The remainder of this work is organized as follows: Section~\ref{sec:goveq} outlines the governing equations for a fluid-solid and contact interaction problem including thermal conduction and phase transitions. In Section~\ref{sec:nummeth} details of the computational implementation are discussed. Finally, in Section~\ref{sec:numex} the accuracy and robustness of the proposed formulation is demonstrated by several numerical examples.

\section{Governing equations} \label{sec:goveq}

\begin{figure}[htbp]
\centering
\newcommand*{\scaletext}{1.0}
\newcommand*{\scalefig}{0.75}
\psfrag{os1}{\scalebox{\scaletext}{$\Omega^{s}_{k}$}}
\psfrag{os2}{\scalebox{\scaletext}{$\Omega^{s}_{\tilde{k}}$}}
\psfrag{os3}{\scalebox{\scaletext}{$\Omega^{s}_{\hat{k}}$}}
\psfrag{of}{\scalebox{\scaletext}{$\Omega^{f}$}}
\psfrag{gss}{\scalebox{\scaletext}{$\Gamma^{ss}_{k,\hat{k}}$}}
\psfrag{gfd}{\scalebox{\scaletext}{$\Gamma^{f}_{D}$}}
\psfrag{gfn}{\scalebox{\scaletext}{$\Gamma^{f}_{N}$}}
\psfrag{gfs1}{\scalebox{\scaletext}{$\Gamma^{fs}_{k}$}}
\psfrag{gfs2}{\scalebox{\scaletext}{$\Gamma^{fs}_{\tilde{k}}$}}
\psfrag{gfs3}{\scalebox{\scaletext}{$\Gamma^{fs}_{\hat{k}}$}}
\includegraphics[scale=\scalefig]{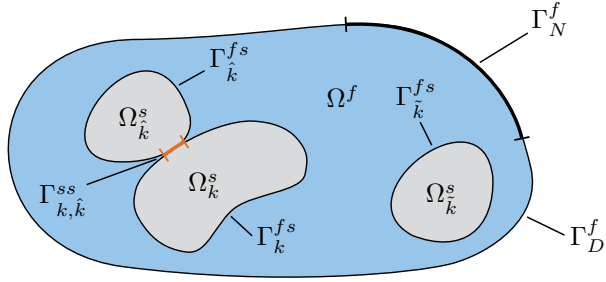}
\caption{Domain~$\Omega$ consisting of several disjunct domains, the fluid domain~$\Omega^{f}$ and the solid sub-domains $\Omega^{s}_{k}$ representing rigid bodies~$k$, with fluid-solid interface~$\Gamma^{fs}_{k}$ and solid-solid interface~$\Gamma^{ss}_{k,\hat{k}}$ in the event of contact between the rigid bodies $k$ and $\hat{k}$.}
\label{fig:domain_continuous}
\end{figure}

Consider a domain~$\Omega$ of a fluid-solid interaction problem that consists at each time~$t \in \qty[0,T]$ of the non-overlapping fluid domain~$\Omega^{f}$ and solid domain~$\Omega^{s}$ that share a common interface~$\Gamma^{fs}$, with $\Omega = \Omega^{f} \cup \Omega^{s}$ and $\Omega^{f} \cap \Omega^{s} = \Gamma^{fs}$. In general, the fluid domain~$\Omega^{f}$ may consist of multiple (liquid and gas) phases. For ease of notation, in the following it will not be distinguished between the different fluid phases. The solid domain~$\Omega^{s}$ is composed of several non-overlapping sub-domains~$\Omega^{s}_{k}$, which represent rigid bodies~$k$, such that $\Omega^{s} = \bigcup_{k} \Omega^{s}_{k}$. In the event of contact between two rigid bodies $k$ and $\hat{k}$, a common interface~$\Gamma^{ss}_{k,\hat{k}} = \Omega^{s}_{k} \cap \Omega^{s}_{\hat{k}}$ exists, separating the two solid sub-domains $\Omega^{s}_{k}$ and $\Omega^{s}_{\hat{k}}$. A detailed illustration of the problem is given in Figure~\ref{fig:domain_continuous}. In the following the (standard) governing equations of the fluid and the solid field as well as the respective coupling conditions are briefly given. In addition, reversible phase transitions between the fluid and the solid field, e.g., temperature-induced melting and solidification, may occur. For this reason, the temperature field is modeled solving the heat equation.

% begin figure
% end figure

%%
\subsection{Fluid field} \label{subsec:goveq_fluid}

The fluid field is governed by the instationary Navier-Stokes equations in the domain~$\Omega^{f}$, which consist in convective form of the mass continuity equation and the momentum equation
\begin{equation} \label{eq:goveq_fluid_conti}
\dv{\rho^{f}}{t} = -\rho^{f} \div \vectorbold{u}^{f} \qin \Omega^{f} \, ,
\end{equation}
\begin{equation} \label{eq:goveq_fluid_momentum}
\dv{\vectorbold{u}^{f}}{t} = -\frac{1}{\rho^{f}} \grad{p^{f}} + \vectorbold{f}_{\nu} + \vectorbold{b}^{f} \qin \Omega^{f} \, ,
\end{equation}
with viscous force~$\vectorbold{f}_{\nu}$ and body force~$\vectorbold{b}^{f}$ each per unit mass. For a Newtonian fluid the viscous force is $\vectorbold{f}_{\nu} = \nu^{f} \laplacian{\vectorbold{u}^{f}}$ with kinematic viscosity~$\nu^{f}$. The mass continuity equation~\eqref{eq:goveq_fluid_conti} and the momentum equation~\eqref{eq:goveq_fluid_momentum} represent a system of $d+1$ equations with the $d+2$ unknowns, velocity~$\vectorbold{u}^{f}$, density~$\rho^{f}$, and pressure~$p^{f}$, in $d$-dimensional space. The system of equations is closed with an equation of state $p^{f} = p^{f}\qty(\rho^{f})$ relating fluid density~$\rho^{f}$ and pressure~$p^{f}$, cf. Section~\ref{subsec:nummeth_fluid_eos}. The Navier-Stokes equations~\eqref{eq:goveq_fluid_conti} and~\eqref{eq:goveq_fluid_momentum} are subject to the following initial conditions
\begin{equation}
\rho^{f} = \rho^{f}_{0} \qand \vectorbold{u}^{f} = \vectorbold{u}^{f}_{0} \qin \Omega^{f} \qq{at} t = 0
\end{equation}
with initial density~$\rho^{f}_{0}$ and initial velocity~$\vectorbold{u}^{f}_{0}$. In addition, Dirichlet and Neumann boundary conditions are applied on the fluid boundary~$\Gamma^{f} = \partial\Omega^{f} \setminus \Gamma^{fs}$
\begin{equation}
\vectorbold{u}^{f} = \vectorbold{\hat{u}}^{f} \qq{on} \Gamma^{f}_{D} \qand \vectorbold{t}^{f} = \vectorbold{\hat{t}}^{f} \qq{on} \Gamma^{f}_{N} \, ,
\end{equation}
with prescribed boundary velocity~$\vectorbold{\hat{u}}^{f}$ and boundary traction~$\vectorbold{\hat{t}}^{f}$, where $\Gamma^{f} = \Gamma^{f}_{D} \cup \Gamma^{f}_{N}$ and $\Gamma^{f}_{D} \cap \Gamma^{f}_{N} = \emptyset$. Furthermore, on the fluid-solid interface $\Gamma^{fs} = \bigcup_{k} \Gamma^{fs}_{k}$ the so-called kinematic and dynamic coupling conditions are
\begin{equation}
\vectorbold{u}^{f} = \vectorbold{u}^{fs}_{k} \qand
\vectorbold{t}^{f} = \vectorbold{t}^{fs}_{k} \qq{on} \Gamma^{fs}_{k} \quad \forall{k} \, ,
\end{equation}
resembling a no-slip boundary condition and ensuring equilibrium of fluid and solid traction across the interface~$\Gamma^{fs}$. Herein, $\vectorbold{u}^{fs}_{k}$ and $\vectorbold{t}^{fs}_{k}$ denote the velocity respectively traction of a rigid body~$k$ on the fluid-solid interface~$\Gamma^{fs}_{k}$.

% begin remark
\begin{rmk}
In the equations~\eqref{eq:goveq_fluid_conti}-\eqref{eq:goveq_fluid_momentum} governing the fluid field, all time derivatives follow the motion of material points, i.e., are material derivatives $\dv{\qty(\cdot)}{t} = \pdv{\qty(\cdot)}{t} + \vectorbold{u} \vdot \grad{\qty(\cdot)}$. Besides, $\grad{\qty(\cdot)}$ denotes derivatives with respect to spatial coordinates.
\end{rmk}
% end remark

%%
\subsection{Solid field} \label{subsec:goveq_solid}

The solid field is assumed to consist of several mobile rigid bodies~$k$ each represented by a sub-domain~$\Omega^{s}_{k}$ embedded in the fluid domain~$\Omega^{f}$. Thus, the interface of a rigid body~$k$ is $\Gamma^{s}_{k} = \Gamma^{fs}_{k} \cup \qty( \bigcup_{\hat{k}} \Gamma^{ss}_{k, \hat{k}} )$ with contacting rigid bodies~$\hat{k}$, cf. Figure~\ref{fig:domain_continuous}. The kinematics of each rigid body~$k$ are uniquely defined by three respectively six degrees of freedom in two- and three-dimensional space, i.e., the position of the center of mass~$\vectorbold{r}^{s}_{k}$ and the orientation~$\vectorbold*{\psi}^{s}_{k}$. As a result, the equations of motion of an individual rigid body~$k$ are described by the balance of linear and angular momentum
\begin{equation} \label{eq:goveq_solid_linearmomentum}
m^{s}_{k} \dv[2]{\vectorbold{r}^{s}_{k}}{t} = \vectorbold{f}^{fs}_{k} + \sum_{\hat{k}} \vectorbold{f}^{ss}_{k,\hat{k}} + m^{s}_{k} \vectorbold{b}^{s}_{k} \qin \Omega^{s}_{k} \, ,
\end{equation}
\begin{equation} \label{eq:goveq_solid_angularmomentum}
\vectorbold{I}^{s}_{k} \dv{\vectorbold*{\omega}^{s}_{k}}{t} = \vectorbold{m}^{fs}_{k} + \sum_{\hat{k}} \vectorbold{m}^{ss}_{k,\hat{k}} \qin \Omega^{s}_{k} \, ,
\end{equation}
with mass~$m^{s}_{k}$ and mass moment of inertia~$\vectorbold{I}^{s}_{k}$ with respect to the center of mass position~$\vectorbold{r}^{s}_{k}$. Herein, $\vectorbold*{\omega}^{s}_{k}$ denotes the angular velocity of a rigid body~$k$, cf. Remark \ref{rmk:goveq_solid_angular_vel}. Furthermore, $\vectorbold{f}^{fs}_{k}$ and $\vectorbold{m}^{fs}_{k}$ describe the resultant coupling force respectively torque acting on the fluid-solid interface~$\Gamma^{fs}_{k}$ of rigid body~$k$. Contacting rigid bodies $k$ and $\hat{k}$ exchange the resultant contact force~$\vectorbold{f}^{ss}_{k,\hat{k}}$ respectively torque~$\vectorbold{m}^{ss}_{k,\hat{k}}$ at the solid-solid interface $\Gamma^{ss}_{k, \hat{k}}$. Finally, the body force~$\vectorbold{b}^{s}_{k}$ given per unit mass is contributing to the balance of linear momentum.

% begin remark
\begin{rmk} \label{rmk:goveq_solid_angular_vel}
The orientation~$\vectorbold*{\psi}^{s}_{k}$ is expressed by a (pseudo-)vector whose direction and magnitude represent the axis and angle of rotation. Note that in general, the angular velocity~$\vectorbold*{\omega}^{s}_{k}$ of a rigid body~$k$ is different from the time derivative of the orientation~$\vectorbold*{\psi}^{s}_{k}$, i.e., $\vectorbold*{\omega}^{s}_{k} \neq \dv*{\vectorbold*{\psi}^{s}_{k}}{t}$, due to the non-additivity of large rotations~\cite{Meier2019a,Meier2019c,Cardona1988,Simo1986}. Direct evolution of the orientation~$\vectorbold*{\psi}^{s}_{k}$ of a rigid body~$k$ requires a special class of time integration schemes, so-called Lie group time integrators~\cite{Bruls2010,Romero2008}.
\end{rmk}
% end remark

%%
\subsection{Thermal conduction} \label{subsec:goveq_thermo}

Thermal conduction in the combined fluid and solid domain~$\Omega = \Omega^{f} \cup \Omega^{s}$ in the absence of heat sources or heat sinks (which are neglected herein for simplicity) is governed by the heat equation
\begin{equation} \label{eq:goveq_thermo_heat_equation}
c_{p}^{\phi} \dv{T}{t} = \frac{1}{\rho^{\phi}} \grad \qty( \kappa^{\phi} \grad{T} ) \qin \Omega \, ,
\end{equation}
with temperature~$T$ and heat flux~$\vectorbold{q}=-\kappa^{\phi} \grad{T}$. The material parameters heat capacity~$c_{p}^{\phi}$ and thermal conductivity~$\kappa^{\phi}$ are in general different for fluid and solid field, and hence for clarity are denoted by the index~$\qty(\cdot)^{\phi}$ with $\phi \in \qty{f,s}$. The heat equation~\eqref{eq:goveq_thermo_heat_equation} is subject to the following initial condition
\begin{equation}
T = T_{0} \qin \Omega \qq{at} t = 0
\end{equation}
with initial temperature~$T_{0}$. In addition, Dirichlet and Neumann boundary conditions are required on the domain boundary~$\Gamma = \partial\Omega$
\begin{equation}
T = \hat{T} \qq{on} \Gamma_{D} \qand \vectorbold{q} = \vectorbold{\hat{q}} \qq{on} \Gamma_{N} \, ,
\end{equation}
with prescribed boundary temperature~$\hat{T}$ and boundary heat flux~$\vectorbold{\hat{q}}$, where $\Gamma = \Gamma_{D} \cup \Gamma_{N}$ and $\Gamma_{D} \cap \Gamma_{N} = \emptyset$.

\subsection{Reversible phase transition} \label{subsec:goveq_phasetransition}

Reversible phase transitions, i.e., melting and solidification, are considered between the solid and the fluid field. Within this publication, solid material points that exceed a transition temperature~$T_{t}$ undergo a phase transition to a fluid material point and vice versa, cf. Remark~\ref{rmk:goveq_phasetransition_latentheat}. Consequently, the shape of a rigid body~$k$, i.e., its sub-domain~$\Omega^{s}_{k}$, is changing due to a loss or gain of material points resulting in a varying mass~$m_{k}$, center of mass position~$\vectorbold{r}_{k}$, and mass moment of inertia~$\vectorbold{I}_{k}$.

% begin remark
\begin{rmk} \label{rmk:goveq_phasetransition_latentheat}
For the sake of simplicity, only temperature-independent parameters are considered herein, and latent heat is neglected. Latent heat could be included by an apparent capacity scheme relying on an increased heat capacity~$c_{p}$ within a finite temperature interval~\cite{Proell2020} in a straightforward manner.
\end{rmk}
% end remark

% begin remark
\begin{rmk} \label{rmk:goveq_diffusion}
The proposed framework is general enough to model also chemically-induced phase transitions based on a concentration field. For this purpose, the diffusion equation $\dv*{C}{t} = \flatfrac{1}{\rho^{\phi}} \grad \qty( D^{\phi} \grad{C} )$ with diffusivity~$D^{\phi}$ modeling the transport of a concentration~$C$ is solved. Considering the similarity between the heat equation~\eqref{eq:goveq_thermo_heat_equation} and the diffusion equation, the latter can be discretized following a similar SPH discretization~\cite{Monaghan2005a,Monaghan2005b} as applied for the heat equation, cf. Section~\ref{subsec:nummeth_thermo}. Similarly, modeling phase transitions a transition concentration~$C_{t}$ is defined.
\end{rmk}
% end remark

%%
\section{Numerical methods and parallel computational framework} \label{sec:nummeth}

This section presents the methods applied for discretization and numerical solution of a fluid-solid and contact interaction problem with phase transitions as described in Section~\ref{sec:goveq}. The presented parallel computational framework is implemented in the in-house parallel multiphysics research code BACI (Bavarian Advanced Computational Initiative) \cite{Baci} using the Message Passing Interface (MPI) for distributed-memory parallel programming.

\subsection{Spatial discretization via smoothed particle hydrodynamics} \label{subsec:nummeth_sph}

For the spatial discretization smoothed particle hydrodynamics (SPH) is used, allowing for a straightforward particle-based evaluation of fluid-solid coupling conditions. In the following, the basics of this method are recapitulated briefly.

\subsubsection{Approximation of field quantities applying a smoothing kernel} \label{subsec:nummeth_sph_kernel}

The fundamental concept of SPH is based on the approximation of a field quantity~$f$ via a smoothing operation and on the discretization of the domain~$\Omega$ with discretization points, so-called particles $j$, each occupying a volume $V_{j}$. Introducing a smoothing kernel~$W\qty(r,h)$ that fulfills certain consistency properties~\cite{Monaghan2005a,Liu2010}, cf. Remark~\ref{rmk:nummeth_sph_kernel_properties}, leads to an approximation of the field quantity~$f$ based on summation of contributions from all particles~$j$ in the domain~$\Omega$
\begin{equation} \label{eq:nummeth_sph_approximation}
f\qty(\vectorbold{r})
\approx \int_{\Omega} f\qty(\vectorbold{r}') W\qty(\qty| \vectorbold{r} - \vectorbold{r}' |, h) \dd{\vectorbold{r}'}
\approx \sum_{j} V_{j} f\qty(\vectorbold{r}_{j}) W\qty(\qty| \vectorbold{r} - \vectorbold{r}_{j} |, h)
\end{equation}
which includes a smoothing error and a discretization error~\cite{Quinlan2006}.

% begin remark
\begin{rmk} \label{rmk:nummeth_sph_kernel_properties}
The smoothing kernel~$W\qty(r,h)$ is a monotonically decreasing, smooth function that depends on a distance~$r$ and a smoothing length~$h$. The smoothing length~$h$ together with a scaling factor~$\kappa$ define the support radius~$r_{c} = \kappa h$ of the smoothing kernel. Compact support, i.e., $W\qty(r, h) = 0$ for $r > r_{c}$, as well as positivity, i.e., $W\qty(r, h) \geq 0$ for $r \leq r_{c}$, are typical properties of standard smoothing kernels~$W\qty(r,h)$. In addition, the normalization property requires that $\int_{\Omega} W\qty(\qty| \vectorbold{r} - \vectorbold{r}' |, h) \dd{\vectorbold{r}'} = 1$. The Dirac delta function property $\lim_{h \rightarrow 0}{ W\qty(r, h) } = \delta\qty(r)$ ensures an exact representation of a field quantity~$f$ in the limit $h \rightarrow 0$.
\end{rmk}
% end remark

A straightforward approach in SPH to determine the gradient of a quantity~$f$ follows directly by differentiation of \eqref{eq:nummeth_sph_approximation} resulting in
\begin{equation}
\grad{f\qty(\vectorbold{r})}
\approx \int_{\Omega} f\qty(\vectorbold{r}') \grad{W\qty(\qty| \vectorbold{r} - \vectorbold{r}_{j} |, h)} \dd{\vectorbold{r}'}
\approx \sum_{j} V_{j} f\qty(\vectorbold{r}_{j}) \grad{W\qty(\qty| \vectorbold{r} - \vectorbold{r}_{j} |, h)} \, .
\end{equation}
Note that this (simple) gradient approximation shows some particular disadvantages. Hence, more advanced approximations for gradients are given in the literature~\cite{Monaghan2005a} and will also be applied in the following. In sum, the concept of SPH allows to reduce partial differential equations to a system of coupled ordinary differential equations (with as many equations as particles) that is solved in the domain~$\Omega$. Thereby, all field quantities are evaluated at and associated with particle positions, meaning each particle carries its corresponding field quantities. Finally, in a post-processing step a continuous field quantity $f$ is recovered from the discrete quantities~$f\qty(\vectorbold{r}_{j})$ of particles~$j$ in the domain~$\Omega$ using the approximation \eqref{eq:nummeth_sph_approximation} and the commonly known Shepard filter
\begin{equation} \label{eq:nummeth_sph_postprocessing}
\hat{f}\qty(\vectorbold{r}) \approx \frac{ \sum_{j} V_{j} f\qty(\vectorbold{r}_{j}) W\qty(\qty| \vectorbold{r} - \vectorbold{r}_{j} |, h) }{ \sum_{j} V_{j} W\qty(\qty| \vectorbold{r} - \vectorbold{r}_{j} |, h) } \, .
\end{equation}
Note that the denominator typically takes on values close to one inside the fluid domain and is mainly relevant for boundary regions with reduced support due to a lack of neighboring particles.

% begin remark
\begin{rmk}
In the following, a quantity~$f$ evaluated for particle~$i$ at position~$\vectorbold{r}_{i}$ is written as~$f_{i} = f\qty(\vectorbold{r}_{i})$. The short notation $W_{ij} = W\qty(r_{ij}, h)$ denotes the smoothing kernel~$W$ evaluated for particle~$i$ at position~$\vectorbold{r}_{i}$ with neighboring particle~$j$ at position~$\vectorbold{r}_{j}$, where $r_{ij} = \qty|\vectorbold{r}_{ij}| = \qty|\vectorbold{r}_{i} - \vectorbold{r}_{j}|$ is the absolute distance between particles~$i$ and~$j$. The derivative of the smoothing kernel~$W$ with respect to the absolute distance~$r_{ij}$ is denoted by $\pdv*{W}{r_{ij}} = \pdv*{W\qty(r_{ij}, h)}{r_{ij}}$.
\end{rmk}
% end remark

% begin remark
\begin{rmk} \label{rmk:nummeth_sph_applied_kernel}
Herein, a quintic spline smoothing kernel~$W\qty(r, h)$ as defined in~\cite{Morris1997} with smoothing length~$h$ and compact support of the smoothing kernel with support radius~$r_{c} = \kappa h$ and scaling factor~$\kappa = 3$ is used.
\end{rmk}
% end remark

%%
\subsubsection{Initial particle spacing} \label{subsec:nummeth_sph_spacing}

Within this contribution, the domain~$\Omega$ is initially filled with particles located on a regular grid with particle spacing~$\Delta{}x$, thus in the $d$-dimensional space each particle initially occupies an effective volume~$V_{eff}=\qty(\Delta{}x)^{d}$. A particle in the fluid domain~$\Omega^{f}$ is called a fluid particle~$i$, whereas a particle in the solid domain~$\Omega^{s}_{k}$ of a rigid body~$k$ is called a rigid particle~$r$. Naturally, the choice of the particle spacing~$\Delta{}x$ influences the accuracy of the interface representation between fluid and solid domain. The mass of a particle is initially assigned using the reference density of the respective phase, i.e., $\rho^{f}_{0}$ for the fluid phase and $\rho^{s}_{0}$ for the solid phase, and the effective volume~$V_{eff}$.

% begin remark
\begin{rmk}
Within this work, the smoothing length~$h$ of the smoothing kernel~$W\qty(r, h)$, cf. Remark~\ref{rmk:nummeth_sph_applied_kernel}, is set equal to the initial particle spacing~$\Delta{}x$. Consequently, in a convergence analysis with decreasing particle spacing~$\Delta{}x$ the ratio $\flatfrac{\Delta{}x}{h}$ remains constant \cite{Quinlan2006}.
\end{rmk}
% end remark

%%
\subsection{Parallelization via spatial decomposition of the domain} \label{subsec:nummeth_parallelization}

For the problems studied herein, an efficient parallel computational framework capable of handling systems constituted of a large number of particles is required. This requires addressing in particular two aspects, namely, an efficient evaluation of particle interactions, and a parallel load distribution strategy while keeping communication overhead at an acceptable level. In the literature, several approaches for parallel particle frameworks have been proposed~\cite{Clark1994,Plimpton1995,Verlet1967,Allen2017,Oger2016,Dominguez2011}. In the present work, a spatial decomposition approach with neighbor pair detection and a combination of Verlet-lists and cell-linked lists based on~\cite{Plimpton1995} is applied. The general idea of the spatial decomposition approach is briefly explained in the following, however, for detailed information, the interested reader is referred to the original publication~\cite{Plimpton1995}.

% begin figure
\begin{figure}[htbp]
\centering
\newcommand*{\scaletext}{1.0}
\newcommand*{\scalefig}{0.75}
\psfrag{rkp}{\scalebox{\scaletext}{processor $p$}}
\psfrag{rbk}{\scalebox{\scaletext}{rigid body ${k}$}}
\psfrag{fp}{\scalebox{\scaletext}{fluid particle $i$}}
\psfrag{rp}{\scalebox{\scaletext}{rigid particle $r$}}
\psfrag{pd}{\scalebox{\scaletext}{processor domain boundary}}
\psfrag{bc}{\scalebox{\scaletext}{cell boundary}}
\psfrag{gc}{\scalebox{\scaletext}{ghosted cell on processor~$p$}}
\includegraphics[scale=\scalefig]{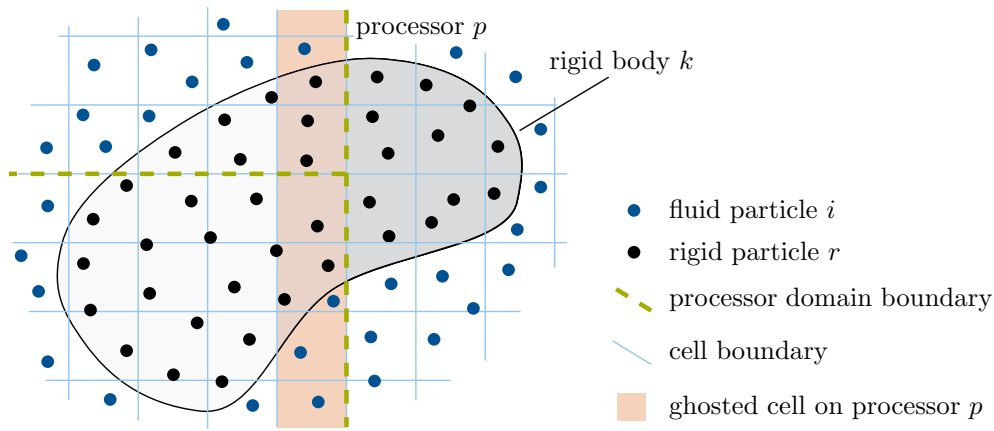}
\caption{A fluid-solid interaction problem consisting of a rigid body~$k$ with affiliated rigid particles~$r$ and surrounding fluid particles~$i$ distributed over several processors~$p$ according to a spatial decomposition approach.}
\label{fig:rigidbody_spatialdecomposition}
\end{figure}
% end figure

The evaluation of particle interactions in SPH requires knowledge of neighboring particles within a geometrically limited interaction distance, i.e., within the support radius~$r_{c}$ of the smoothing kernel, cf. Section~\ref{subsec:nummeth_sph}. Thus, the computational domain is divided into several cubic cells forming a uniform lattice, while each particle is uniquely assigned to one of those cells according to its current spatial position, cf. Figure~\ref{fig:rigidbody_spatialdecomposition}. The size of the cells is chosen such that neighboring particles are either located in the same cell or in adjacent cells, i.e., the size of the cells is at least equal to the support radius~$r_{c}$ of the smoothing kernel.

Following a spatial decomposition approach, the cells together with assigned particles are distributed over all involved processors, i.e., forming so-called processor domains. To keep the computational load balanced between all processors and to minimize the communication overhead, cubic processor domains are defined such that each contains (nearly) the same number of particles. The cells occupied by each processor are called owned cells. On each processor the position of particles located in its processor domain, i.e., the position of so-called owned particles, is evolved. This requires the evaluation of interactions of owned particles with their neighboring particles. However, the correct evaluation of particle interactions close to processor domain boundaries requires that each processor has information not only about its owned particles but also about particles in cells adjacent to its processor domain. To this end, each processor is provided full information not only about its own domain but additionally about a layer of ghosted cells (with ghosted particles) around its own domain. Keeping the information about ghosted cells and particles continuously updated requires communication between processors.

% begin remark
\begin{rmk} \label{rmk:nummeth_parallelization_overhead}
To exemplify the cost of communication overhead, consider a perfectly cubic processor domain occupying $n_{o}$ owned cells. Consequently, assuming one layer of ghosted cells surrounding the processor domain, a total of $n_{g} = \qty( \sqrt[3]{n_{o}} + 2 )^{3} - n_{o}$ cells are ghosted. That is, the communication overhead scales with the ratio $\flatfrac{n_{g}}{n_{o}}$ of ghosted cells $n_{g}$ to owned cells $n_{o}$. Furthermore, the (average) number of particles per cell, and, consequently, also the communication overhead, scale with the ratio $\flatfrac{r_{c}}{\Delta{}x}$ of the support radius~$r_{c}$ and the initial particle spacing~$\Delta{}x$.
\end{rmk}
% end remark

As a consequence of the spatial decomposition approach, the affiliated rigid particles~$r$ of a rigid body~$k$ might be distributed over several processors, cf. Figure~\ref{fig:rigidbody_spatialdecomposition}. However, note that the balance of linear and angular momentum, cf. equations~\eqref{eq:goveq_solid_linearmomentum}-\eqref{eq:goveq_solid_angularmomentum}, describing the motion of a rigid body~$k$ are given with respect to the center of mass position~$\vectorbold{r}^{s}_{k}$. Thus, the evaluation of mass quantities, i.e., mass~$m_k$, center of mass position~$\vectorbold{r}^{s}_{k}$, and mass moment of inertia~$\vectorbold{I}_{k}$, as well as the evaluation of resultant force~$\vectorbold{f}_{k}$ and torque~$\vectorbold{m}_{k}$ acting on a rigid body~$k$, requires special communication between all processors hosting rigid particles~$r$ belonging to rigid body~$k$ and the single processor owning rigid body~$k$, cf. Section~\ref{subsec:nummeth_rigid}.

\subsection{Modeling fluid flow using weakly compressible SPH} \label{subsec:nummeth_fluid}

For modeling fluid flow using SPH, several different formulations each with its own characteristics and benefits can be derived. Here, the instationary Navier-Stokes equations~\eqref{eq:goveq_fluid_conti} and~\eqref{eq:goveq_fluid_momentum} are discretized by a weakly compressible approach~\cite{Monaghan2005a,Liu2010,Price2012}. This section gives a brief overview of this formulation applied already in~\cite{Fuchs2020,Meier2020}. For ease of notation, in the following the index~$\qty(\cdot)^{f}$ denoting fluid quantities, as used in Section~\ref{sec:goveq}, is dropped.

\subsubsection{Density summation} \label{subsec:nummeth_fluid_denssum}

The density of a particle~$i$ is determined via summation of the respective smoothing kernel contributions of all neighboring particles~$j$ within the support radius~$r_{c}$
\begin{equation} \label{eq:nummeth_fluid_densum}
\rho_{i} = m_{i} \sum_{j} W_{ij}
\end{equation}
with mass~$m_{i}$ of particle~$i$. This approach is typically denoted as density summation and results in an exact conservation of mass in the fluid domain, which can be shown in a straightforward manner considering the commonly applied normalization of the smoothing kernel to unity. It shall be noted that the density field may alternatively be obtained by discretization and integration of the mass continuity equation~\eqref{eq:goveq_fluid_conti}~\cite{Liu2010}.

\subsubsection{Momentum equation} \label{subsec:nummeth_fluid_momentum}

The momentum equation~\eqref{eq:goveq_fluid_momentum} is discretized following~\cite{Adami2012, Adami2013} including a transport velocity formulation to suppress the problem of tensile instability. It will be briefly recapitulated in the following. The transport velocity formulation relies on a constant background pressure~$p_{b}$ that is applied to all particles and results in a contribution to the particle accelerations for in general disordered particle distributions. However, these additional acceleration contributions vanish for particle distributions fulfilling the partition of unity of the smoothing kernel, thus fostering these desirable configurations. For the sake of brevity, the definition of the modified advection velocity and the additional terms in the momentum equation from the aforementioned transport velocity formulation are not discussed in the following and the reader is referred to the original publication~\cite{Adami2013}. Altogether, the acceleration $\vectorbold{a}_{i} = \dv*{\vectorbold{u}_{i}}{t}$ of a particle~$i$ results from summation of all acceleration contributions due to interaction with neighboring particles~$j$ and a body force as
\begin{equation} \label{eq:nummeth_fluid_momentum}
\vectorbold{a}_{i} = \frac{1}{m_{i}} \sum_{j} \qty(V_{i}^{2}+V_{j}^{2}) \qty[ - \tilde{p}_{ij} \pdv{W}{r_{ij}} \vectorbold{e}_{ij} + \tilde{\eta}_{ij} \frac{\vectorbold{u}_{ij}}{r_{ij}} \pdv{W}{r_{ij}} ] + \vectorbold{b}_{i} \, ,
\end{equation}
with volume~$V_{i} = m_{i}/\rho_{i}$ of particle~$i$, unit vector $\vectorbold{e}_{ij} = \flatfrac{\vectorbold{r}_{i} - \vectorbold{r}_{j}}{\qty|\vectorbold{r}_{i} - \vectorbold{r}_{j}|} = \flatfrac{\vectorbold{r}_{ij}}{r_{ij}}$, relative velocity $\vectorbold{u}_{ij} = \vectorbold{u}_{i}-\vectorbold{u}_{j}$, and density-weighted inter-particle averaged pressure and inter-particle averaged viscosity
\begin{equation}
\tilde{p}_{ij} = \frac{\rho_{j}p_{i}+\rho_{i}p_{j}}{\rho_{i} + \rho_{j}} \qand
\tilde{\eta}_{ij} = \frac{2\eta_{i}\eta_{j}}{\eta_{i}+\eta_{j}} \, .
\end{equation}
In the following, the acceleration contribution of a neighboring particle~$j$ to particle~$i$ is, for ease of notation, denoted as~$\vectorbold{a}_{ij}$, where $\vectorbold{a}_{i} = \sum_{j} \vectorbold{a}_{ij} + \vectorbold{b}_{i}$. The above given momentum equation~\eqref{eq:nummeth_fluid_momentum} exactly conserves linear momentum due to pairwise anti-symmetric particle forces
\begin{equation} \label{eq:nummeth_fluid_conservlinmom}
m_{i} \vectorbold{a}_{ij} = - m_{j} \vectorbold{a}_{ji} \, ,
\end{equation}
which follows from the property $\pdv*{W}{r_{ij}} = \pdv*{W}{r_{ji}}$ of the smoothing kernel.
\subsubsection{Equation of state} \label{subsec:nummeth_fluid_eos}
Following a weakly compressible approach, the density~$\rho_{i}$ and pressure~$p_{i}$ of a particle~$i$ are linked via the equation of state
\begin{equation} \label{eq:nummeth_fluid_eos}
p_{i}\qty(\rho_{i}) = c^{2} \qty(\rho_{i} - \rho_{0}) = p_{0} \qty(\frac{\rho_{i}}{\rho_{0}} - 1)
\end{equation}
with reference density~$\rho_{0}$, reference pressure~$p_{0} = \rho_{0} c^{2}$ and artificial speed of sound~$c$. Note that this commonly applied approach can only capture deviations from the reference pressure, i.e., $p_{i}\qty(\rho_{0}) = 0$, and not the total pressure. To limit density fluctuations to an acceptable level, while still avoiding too severe time step restrictions, strategies are discussed in~\cite{Morris1997} how to choose the artificial speed of sound~$c$.

\subsubsection{Boundary and coupling conditions} \label{subsec:nummeth_fluid_bdrycoupcond}

Herein, both rigid wall boundary conditions as well as rigid body coupling conditions, are modeled following~\cite{Adami2012}. In the former case, at least $q = \mathrm{floor}\qty(\flatfrac{r_{c}}{\Delta{}x})$ layers of boundary particles~$b$ are placed parallel to the fluid boundary~$\Gamma^{f}_{D}$ with a distance of~$\flatfrac{\Delta{}x}{2}$ outside of the fluid domain~$\Omega^{f}$ in order to maintain full support of the smoothing kernel. In the latter case, rigid particles~$r$ of rigid bodies~$k$ are considered, while naturally describing the fluid-solid interface~$\Gamma^{fs}$. In both cases, a boundary particle~$b$ or a rigid particle~$r$ contribute to the density summation~\eqref{eq:nummeth_fluid_densum} and to the momentum equation~\eqref{eq:nummeth_fluid_momentum} evaluated for a fluid particle~$i$ considered as neighboring particle~$j$. The respective quantities of boundary particles~$b$ respectively rigid particles~$r$ are extrapolated from the fluid field based on a local force balance as described in~\cite{Adami2012}. Consequently, striving for conservation of linear momentum, cf. equation~\eqref{eq:nummeth_fluid_conservlinmom}, the force acting on a rigid particle~$r$ stemming from interaction with fluid particle~$i$, cf. equation~\eqref{eq:nummeth_fluid_momentum}, is given as
\begin{equation} \label{eq:nummeth_fluid_couplingforce}
\vectorbold{f}_{ri} = - m_{i} \vectorbold{a}_{ir} \, .
\end{equation}

% begin remark
\begin{rmk}
The $\mathrm{floor}$ operator used herein is defined by $\mathrm{floor}\qty(x) := \max\qty{ k \in \mathbb{Z} \mid k \leq x}$ and returns the largest integer that is less than or equal to its argument~$x$.
\end{rmk}
% end remark

%%
\subsection{Modeling the motion of rigid bodies discretized by particles} \label{subsec:nummeth_rigid}

Within this formulation, each rigid body~$k$ is composed of several rigid particles~$r$ that are fixed relative to a rigid body frame, i.e., there is no relative motion among rigid particles of a rigid body. Thus, the rigid particles of a rigid body are not evolved in time individually, but follow the motion of the rigid body, cf. Section~\ref{subsec:nummeth_timint}, described by the balance of linear and angular momentum, cf. equations~\eqref{eq:goveq_solid_linearmomentum}-\eqref{eq:goveq_solid_angularmomentum}. 

As a consequence of the spatial decomposition approach, cf. Section~\ref{subsec:nummeth_parallelization}, special communication between all processors hosting rigid particles~$r$ of a rigid body~$k$ in terms of the evaluation of mass quantities, cf. Section~\ref{subsec:nummeth_rigid_massquantities}, respectively the evaluation of resultant forces and torques, cf. Section~\ref{subsec:nummeth_rigid_forces}, is required. For ease of notation, in the following the index~$\qty(\cdot)^{s}$ denoting solid quantities, as used in Section~\ref{sec:goveq}, is dropped.

\subsubsection{Orientation of rigid bodies} \label{subsec:nummeth_rigid_orientation}

\begin{figure}[htbp]
\centering
\newcommand*{\scaletext}{1.0}
\newcommand*{\scalefig}{0.75}
\psfrag{rbk}{\scalebox{\scaletext}{rigid body ${k}$}}
\psfrag{mqk}{\scalebox{\scaletext}{$m_{k}$}, $\vectorbold{r}_{k}$, $\vectorbold{I}_{k}$}
\psfrag{cmk}{\scalebox{\scaletext}{center of mass}}
\psfrag{rp}{\scalebox{\scaletext}{rigid particle $r$}}
\psfrag{qk}{\scalebox{\scaletext}{$\vectorbold{q}_{k}$}}
\psfrag{rk}{\scalebox{\scaletext}{$\vectorbold{r}_{rk}$}}
\includegraphics[scale=\scalefig]{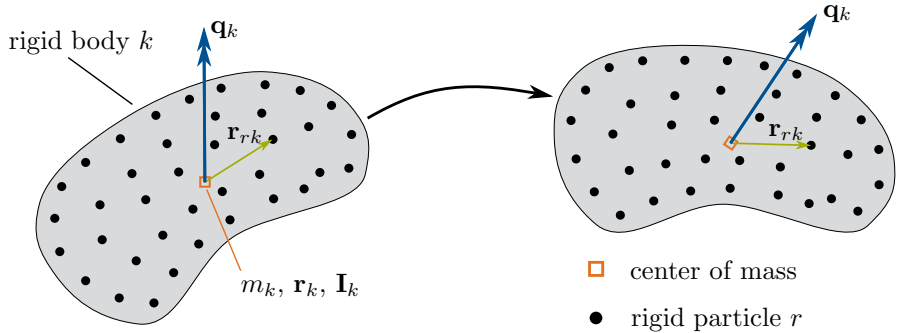}
\caption{Orientation of a rigid body~$k$ with rigid particles~$r$ and their relative position~$\vectorbold{r}_{rk}$ described via a unit quaternion~$\vectorbold{q}_{k}$ at different times.}
\label{fig:rigidbody_orientation}
\end{figure}

The orientation~$\vectorbold*{\psi}_{k}$ of a rigid body~$k$, described by one respectively three degrees of freedom in two- and three-dimensional space, is introduced in Section~\ref{subsec:goveq_solid} without explicitly defining a specific parameterization of the underlying rotation, e.g., via Euler angles or Rodriguez parameters. Moreover, as stated in Remark~\ref{rmk:goveq_solid_angular_vel}, explicit evolution of the orientation~$\vectorbold*{\psi}_{k}$ requires special Lie group time integrators. A straightforward approach to overcome aforementioned issues is to describe the orientation of a rigid body via quaternion algebra, cf. Remark~\ref{rmk:nummeth_rigid_quaternion_algebra}. Consequently, in the following it is assumed that at all times the orientation~$\vectorbold*{\psi}_{k}$ of a rigid body~$k$ can be uniquely described by a unit quaternion~$\vectorbold{q}_{k}$, cf. Figure~\ref{fig:rigidbody_orientation}. Once a local rigid body frame is defined, this allows to transform the relative position of rigid particles~$\vectorbold{r}_{rk}$, cf. Remark~\ref{rmk:nummeth_rigid_relative_position}, from that rigid body frame to the reference frame, e.g., a global cartesian system.

% begin figure
% end figure

% begin remark
\begin{rmk} \label{rmk:nummeth_rigid_relative_position}
Note that the relative position of rigid particles~$\vectorbold{r}_{rk}$ expressed in the rigid body frame is, in general, a known (and constant) quantity, that only needs to be updated in case the center of mass position~$\vectorbold{r}_{k}$ changes, i.e., due to phase transitions.
\end{rmk}
% end remark

% begin remark
\begin{rmk} \label{rmk:nummeth_rigid_quaternion_algebra}
For the sake of brevity, the principals of quaternion algebra are not delineated herein. It remains the definition of operator $\circ$ denoting quaternion multiplication as used in the following.
\end{rmk}
% end remark

%%
\subsubsection{Parallel evaluation of mass-related quantities} \label{subsec:nummeth_rigid_massquantities}

\begin{figure}[htbp]
\centering
\newcommand*{\scaletext}{1.0}
\newcommand*{\scalefig}{0.75}
\psfrag{rkp}{\scalebox{\scaletext}{processor $p$}}
\psfrag{rbk}{\scalebox{\scaletext}{rigid body ${k}$}}
\psfrag{mqk}{\scalebox{\scaletext}{$m_{k}$}, $\vectorbold{r}_{k}$, $\vectorbold{I}_{k}$}
\psfrag{mqp}{\scalebox{\scaletext}{$\prescript{}{p}{m}_{k}$}, $\prescript{}{p}{\vectorbold{r}}_{k}$, $\prescript{}{p}{\vectorbold{I}}_{k}$}
\psfrag{cmk}{\scalebox{\scaletext}{center of mass}}
\psfrag{cmp}{\scalebox{\scaletext}{processor-wise center of mass}}
\psfrag{rp}{\scalebox{\scaletext}{rigid particle $r$}}
\psfrag{pd}{\scalebox{\scaletext}{processor domain boundary}}
\includegraphics[scale=\scalefig]{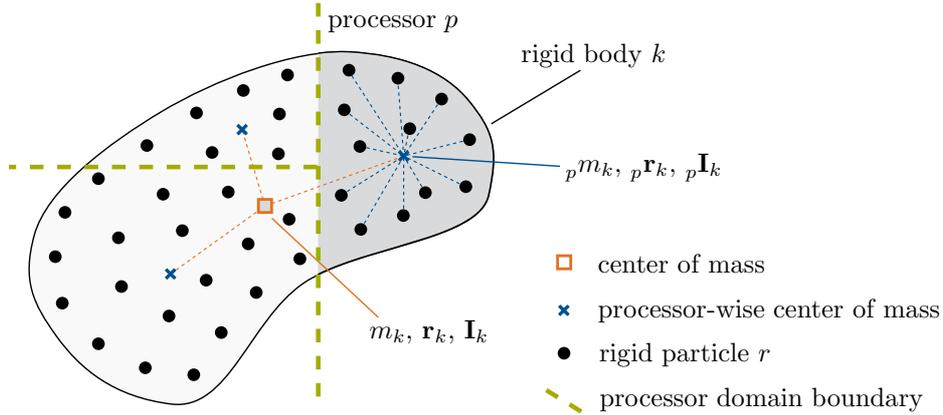}
\caption{Parallel distribution of a rigid body~$k$ with rigid particles~$r$ over several processors~$p$ illustrating the evaluation of mass~$m_{k}$, center of mass position~$\vectorbold{r}_{k}$, and mass moment of inertia~$\vectorbold{I}_{k}$ via processor-wise mass~$\prescript{}{p}{m}_{k}$, center of mass position~$\prescript{}{p}{\vectorbold{r}}_{k}$, and mass moment of inertia~$\prescript{}{p}{\vectorbold{I}}_{k}$.}
\label{fig:rigidbody_massquantities}
\end{figure}

In a first step, on each processor~$p$ the processor-wise mass~$\prescript{}{p}{m}_{k}$ and center of mass position~$\prescript{}{p}{\vectorbold{r}}_{k}$ of a rigid body~$k$ are computed as
\begin{equation} \label{eq:nummeth_rigid_procwise_centerofmass}
\prescript{}{p}{m}_{k} = \sum_{r} m_{r}
\qand
\prescript{}{p}{\vectorbold{r}}_{k} = \frac{\sum_{r} m_{r} \vectorbold{r}_{r}}{\sum_{r} m_{r}}
\end{equation}
considering the mass~$m_{r}$ and position~$\vectorbold{r}_{r}$ of all affiliated rigid particles~$r$ being located in the computational domain of processor~$p$, cf. Figure~\ref{fig:rigidbody_massquantities} for an illustration. Accordingly, the processor-wise mass moment of inertia~$\prescript{}{p}{\vectorbold{I}}_{k}$ of a rigid body~$k$ follows componentwise (in index notation) as
\begin{equation} \label{eq:nummeth_rigid_procwise_inertia}
\prescript{}{p}{I}_{k,ij} = \sum_{r} \qty[ I_{r} \, \delta_{ij} + \qty[ \sum_{q} \qty( \prescript{}{p}{r}_{k,q} - r_{r,q} )^{2} \delta_{ij} - \qty( \prescript{}{p}{r}_{k,i} - r_{r,i} ) \qty( \prescript{}{p}{r}_{k,j} - r_{r,j} ) ] m_{r} ]
\end{equation}
with mass~$m_{r}$ and mass moment of inertia~$I_{r}$ of a rigid particle $r$, cf. Remark~\ref{rmk:nummeth_rigid_particle_inertia}, and Kronecker delta $\delta_{ij}$, cf. Remark \ref{rmk:nummeth_rigid_kronecker_delta}. The computed processor-wise quantities, i.e., mass~$\prescript{}{p}{m}_{k}$, center of mass position~$\prescript{}{p}{\vectorbold{r}}_{k}$, and mass moment of inertia~$\prescript{}{p}{\vectorbold{I}}_{k}$, are communicated to the owning processor of rigid body~$k$. In a second step, on the owning processor the total mass~$m_{k}$ and center of mass position~$\vectorbold{r}_{k}$ of rigid body~$k$ are computed over all processors~$p$ as
\begin{equation} \label{eq:nummeth_rigid_global_centerofmass}
m_{k} = \sum_{p} \prescript{}{p}{m}_{k}
\qand
\vectorbold{r}_{k} = \frac{\sum_{p} \prescript{}{p}{m}_{k} \prescript{}{p}{\vectorbold{r}}_{k}}{\sum_{p} \prescript{}{p}{m}_{k}}
\end{equation}
making use of the received processor-wise quantities. Similar to \eqref{eq:nummeth_rigid_procwise_inertia} the mass moment of inertia~$\vectorbold{I}_{k}$ of rigid body~$k$ follows componentwise (in index notation) as
\begin{equation} \label{eq:nummeth_rigid_global_inertia}
I_{k,ij} = \sum_{p} \qty[ \prescript{}{p}{I}_{k,ij} + \qty[ \sum_{q} \qty( r_{k,q} - \prescript{}{p}{r}_{k,q} )^{2} \delta_{ij} - \qty( r_{k,i} - \prescript{}{p}{r}_{k,i} ) \qty( r_{k,j} - \prescript{}{p}{r}_{k,j} ) ] \prescript{}{p}{m}_{k} ]
\end{equation}
again considering the received processor-wise quantities. Finally, the determined global quantities, i.e., mass~$m_{k}$, center of mass position~$\vectorbold{r}_{k}$, and mass moment of inertia~$\vectorbold{I}_{k}$, are communicated from the owning processor to all hosting processors of rigid body~$k$.

% begin figure
% end figure

% begin remark
\begin{rmk} \label{rmk:nummeth_rigid_particle_inertia}
The mass moment of inertia~$I_{r}$ of a rigid particle $r$ with mass~$m_{r}$ is computed based on the effective volume~$V_{eff}=\qty(\Delta{}x)^{d}$ with initial particle spacing~$\Delta{}x$, cf. Section \ref{subsec:nummeth_sph_spacing}. In two-dimensional space ($d=2$) assuming circular disk-shaped particles results in $I_{r} = 0.5 m_{r} r_{eff}$ with effective radius~$r_{eff} = \flatfrac{\Delta{}x}{\sqrt{\pi}}$. Accordingly, in three-dimensional space ($d=3$) assuming spherical-shaped particles results in $I_{r} = 0.4 m_{r} r_{eff}$ with effective radius~$r_{eff} = \sqrt[3]{\flatfrac{0.75}{\pi}} \Delta{}x$.
\end{rmk}
% end remark

% begin remark
\begin{rmk} \label{rmk:nummeth_rigid_kronecker_delta}
The Kronecker delta $\delta_{ij}$ used in equations~\eqref{eq:nummeth_rigid_procwise_inertia} and~\eqref{eq:nummeth_rigid_global_inertia} to compute the mass moments of inertia is defined by $\delta _{ij} = {\begin{cases} 1 \qq{if} i=j \, , \\ 0 \qq{otherwise.} \end{cases}}$
\end{rmk}
% end remark

% begin remark
\begin{rmk} \label{rmk:nummeth_rigid_huygens_steiner}
The computation of the mass moment of inertia, cf. equations~\eqref{eq:nummeth_rigid_procwise_inertia} respectively~\eqref{eq:nummeth_rigid_global_inertia}, is based on the Huygens-Steiner theorem, also called parallel axis theorem.
\end{rmk}
% end remark

%%
\subsubsection{Parallel evaluation of resultant force and torque} \label{subsec:nummeth_rigid_forces}

To begin with, the resultant coupling and contact force acting on a rigid particle~$r$ of rigid body~$k$ is given as
\begin{equation} \label{eq:nummeth_rigid_rigidparticle_force}
\vectorbold{f}_{r} = \sum_{i} \vectorbold{f}_{ri} + \sum_{\hat{k}} \sum_{\hat{r}} \vectorbold{f}_{r\hat{r}}
\end{equation}
with coupling forces~$\vectorbold{f}_{ri}$ stemming from interaction with neighboring fluid particles~$i$, cf. equation~\eqref{eq:nummeth_fluid_couplingforce}, and contact forces~$\vectorbold{f}_{r\hat{r}}$ stemming from interaction with rigid particles~$\hat{r}$ of contacting rigid bodies~$\hat{k}$, cf. equation~\eqref{eq:nummeth_rigid_contactforce}. Similar to the computation of mass-related quantities as described in Section~\ref{subsec:nummeth_rigid_massquantities}, the resultant force~$\vectorbold{f}_{k}$ and torque~$\vectorbold{m}_{k}$ acting on a rigid body~$k$ are determined considering the parallel distribution of the affiliated rigid particles~$r$ on hosting processors~$p$, cf. Figure~\ref{fig:rigidbody_massquantities}. Thus, in a first step the processor-wise resultant force~$\prescript{}{p}{\vectorbold{f}}_{k}$ and torque~$\prescript{}{p}{\vectorbold{m}}_{k}$ acting on rigid body~$k$ are computed as
\begin{equation} \label{eq:nummeth_rigid_procwise_force}
\prescript{}{p}{\vectorbold{f}}_{k} = \sum_{r} \vectorbold{f}_{r}
\qand
\prescript{}{p}{\vectorbold{m}}_{k} = \sum_{r} \vectorbold{r}_{rk} \cross \vectorbold{f}_{r}
\end{equation}
with $\vectorbold{r}_{rk} = \vectorbold{r}_{r} - \vectorbold{r}_{k}$ while considering the resultant forces~$\vectorbold{f}_{r}$ acting on all rigid particles~$r$ being located in the computational domain of processor~$p$. For correct computation of the processor-wise resultant torque~$\prescript{}{p}{\vectorbold{m}}_{k}$ the knowledge of the global center of mass position~$\vectorbold{r}_{k}$ is required on all processors, cf. Section~\ref{subsec:nummeth_rigid_massquantities}. Finally, the computed processor-wise forces~$\prescript{}{p}{\vectorbold{f}}_{k}$ and torques~$\prescript{}{p}{\vectorbold{m}}_{k}$ are communicated to the owning processor of rigid body~$k$ and summed up to the global resultant force and torque acting on rigid body~$k$
\begin{equation} \label{eq:nummeth_rigid_global_force}
\vectorbold{f}_{k} = \sum_{p} \prescript{}{p}{\vectorbold{f}}_{k}
\qand
\vectorbold{m}_{k} = \sum_{p} \prescript{}{p}{\vectorbold{m}}_{k} \, .
\end{equation}

% begin remark
\begin{rmk} \label{rmk:nummeth_rigid_serial_massquantities}
In the case of a computation on a single processor, the evaluation of mass~$m_{k}$, center of mass position~$\vectorbold{r}_{k}$, and mass moment of inertia~$\vectorbold{I}_{k}$ of a rigid body~$k$ follow directly from equations~\eqref{eq:nummeth_rigid_procwise_centerofmass} and \eqref{eq:nummeth_rigid_procwise_inertia}, while the resultant force~$\vectorbold{f}_{k}$ and torque~$\vectorbold{m}_{k}$ directly follow from equation~\eqref{eq:nummeth_rigid_procwise_force}, in each case without the need for special communication.
\end{rmk}
% end remark

%%
\subsubsection{Contact evaluation between neighboring rigid bodies} \label{subsec:nummeth_rigid_contact}

For the modeling of frictionless contact between neighboring rigid bodies~$k$ and~$\hat{k}$, a contact normal force law based on a spring-dashpot model, similar to~\cite{Meier2019a}, is employed. The contact force is acting between pairs of neighboring rigid particles~$r$ and~$\hat{r}$ of contacting rigid bodies, i.e., for distances $r_{r\hat{r}} < \Delta{}x$ with $r_{r\hat{r}} = \qty|\vectorbold{r}_{r\hat{r}}| = \qty|\vectorbold{r}_{r} - \vectorbold{r}_{\hat{r}}|$. Accordingly, the contact force acting on a particle~$r$ of rigid body~$k$ due to contact with a particle~$\hat{r}$ of neighboring rigid body~$\hat{k}$ is given as
\begin{equation} \label{eq:nummeth_rigid_contactforce}
\vectorbold{f}_{r\hat{r}} =
\begin{cases}
- \qty\Big[ \min \qty\Big(0, k_{c} \qty( r_{r\hat{r}} - \Delta{}x ) + d_{c} \qty( \vectorbold{e}_{r\hat{r}} \vdot \vectorbold{v}_{r\hat{r}} ) ) ] \, \vectorbold{e}_{r\hat{r}} & \qq*{if} r_{r\hat{r}} < \Delta{}x \, , \\
0 & \qq*{otherwise,}
\end{cases}
\end{equation}
with unit vector $\vectorbold{e}_{r\hat{r}} = \flatfrac{\vectorbold{r}_{r\hat{r}}}{r_{r\hat{r}}}$, stiffness constant~$k_{c}$, and damping constant~$d_{c}$. The $\min$ operator in equation~\eqref{eq:nummeth_rigid_contactforce} ensures that only repulsive forces between the rigid particles are considered (tension cut-off).

% begin remark
\begin{rmk}
Contact between a rigid body~$k$ and a rigid wall is modeled similar to equation~\eqref{eq:nummeth_rigid_contactforce} considering rigid particles~$r$ of a rigid body~$k$ and boundary particles~$b$ of a discretized rigid wall, cf. Section~\ref{subsec:nummeth_fluid_bdrycoupcond}.
\end{rmk}
% end remark

% begin remark
\begin{rmk}
The applied contact evaluation between rigid bodies is for simplicity based on a contact normal force law evaluated between rigid particles while neglecting frictional effects. Generally, following a macroscopic approach of contact mechanics with non-penetration constraint, the normal distance between the contacting bodies, typically determined via closest point projections, is the contact-relevant kinematic quantity. Accordingly, the concept applied in this work, can be interpeted as a microscale approach based on a repulsive/steric interaction potential~\cite{Grill2020} defined between pairs of rigid particles of contacting rigid bodies. In the current work, this approach has been chosen for reasons of simplicity and numerical robustness. An extension to a macroscale approach, i.e., a normal distance-based contact interaction~\cite{Dong2019,Junior2019}, is possible in a straightforward manner.
\end{rmk}
% end remark

%%
\subsection{Discretization of the heat equation using SPH} \label{subsec:nummeth_thermo}

Thermal conduction in the combined fluid and solid domain governed by the heat equation~\eqref{eq:goveq_thermo_heat_equation} is discretized using smoothed particle hydrodynamics following a formulation proposed by Cleary and Monaghan~\cite{Cleary1999}
\begin{equation} \label{eq:nummeth_thermo_heat_equation}
c_{p,a} \dv{T_{a}}{t} = \frac{1}{\rho_{a}} \sum_{b} V_{b} \frac{4\kappa_{a} \kappa_{b}}{\kappa_{a} + \kappa_{b}} \frac{T_{ab}}{r_{ab}} \pdv{W}{r_{ab}}
\end{equation}
with volume~$V_{b} = \flatfrac{m_{b}}{\rho_{b}}$ of particle~$b$ and temperature difference~$T_{ab} = T_{a} - T_{b}$ between particle~$a$ and particle~$b$. The discretization of the conductive term is especially suited for problems involving a different thermal conductivity among the fields~\cite{Cleary1999}. In the equation above, the index~$\qty(\cdot)^{\phi}$ with $\phi \in \qty{f,s}$ for fluid and solid field, as introduced in Section~\ref{sec:goveq}, is dropped for ease of notation. Accordingly, the particles~$a$ and~$b$ may denote fluid particles~$i$ as well as rigid particles~$j$, respectively.

\subsection{Modeling thermally driven reversible phase transitions} \label{subsec:nummeth_phasetransition}

Due to the Lagrangian nature of SPH, each (material) particle carries its phase information. This allows for direct evaluation of the discretized heat equation~\eqref{eq:nummeth_thermo_heat_equation} for fluid and rigid particles with corresponding phase-specific parameters of the particle~$a$ itself and of neighboring particles~$b$, cf. Section~\ref{subsec:nummeth_thermo}. Phase transitions in the form of melting of a rigid body occurs, in case the temperature~$T_{r}$ of a rigid particle~$r$ exceeds the transition temperature~$T_{t}$. The former rigid particle~$r$ changes phase to become a fluid particle~$i$. Conversely, phase transitions in form of solidification occurs, in case the temperature~$T_{i}$ of a fluid particle~$i$ falls below the transition temperature~$T_{t}$ and the former fluid particle~$i$ becomes a rigid particle~$r$.

Consequently, each time a rigid body~$k$ is subject to phase transition, its mass~$m_{k}$, center of mass position~$\vectorbold{r}_{k}$, and mass moment of inertia~$\vectorbold{I}_{k}$ are updated, cf. Section~\ref{subsec:nummeth_rigid_massquantities}. In addition, the velocity~$\vectorbold{u}_{k}$ after phase transition is determined based on quantities prior to phase transition indicated by index~$\qty(\cdot)^{\prime}$ as
\begin{equation}
\vectorbold{u}_{k} = \vectorbold{u}_{k}^{\prime} + \vectorbold*{\omega}_{k} \cross \qty( \vectorbold{r}_{rk} - \vectorbold{r}_{rk}^{\prime})
\end{equation}
following rigid body motion with (unchanged) angular velocity~$\vectorbold*{\omega}_{k}$.

\subsection{Time integration following a velocity-Verlet scheme} \label{subsec:nummeth_timint}

The discretized fluid and solid field are both integrated in time applying an explicit velocity-Verlet time integration scheme in kick-drift-kick form, also denoted as leapfrog scheme, that is of second order accuracy and reversible in time when dissipative effects are absent~\cite{Monaghan2005a}. Again, for ease of notation, in the following the indices~$\qty(\cdot)^{f}$ and~$\qty(\cdot)^{s}$ denoting fluid respectively solid quantities, as introduced in Section~\ref{sec:goveq}, are dropped. Altogether, for the fluid field the positions~$\vectorbold{r}_{i}$ of fluid particles~$i$ are evolved in time, while for the solid field the center of mass positions~$\vectorbold{r}_{k}$ and the orientations~$\vectorbold*{\psi}_{k}$ of all rigid bodies~$k$ are evolved in time. However, the positions~$\vectorbold{r}_{r}$ of rigid particles~$r$ are not evolved in time but directly follow the motion of corresponding affiliated rigid bodies~$k$.

In a first kick-step, the accelerations~$\vectorbold{a}_{i}^{n} = \qty(\dv*{\vectorbold{u}_{i}}{t})^{n}$, as determined in the previous time step~$n$, are used to compute the intermediate velocities
\begin{equation}
\vectorbold{u}_{i}^{n+1/2} = \vectorbold{u}_{i}^{n} + \frac{\Delta{}t}{2} \, \vectorbold{a}_{i}^{n}
\end{equation}
of fluid particles~$i$, where~$\Delta{}t$ is the time step size. Similar, for rigid bodies~$k$ the linear and angular accelerations~$\vectorbold{a}_{k}^{n} = \qty(\dv*[2]{\vectorbold{r}_{k}}{t})^{n}$ respectively~$\vectorbold*{\alpha}_{k}^{n} = \qty(\dv*{\vectorbold*{\omega}_{k}}{t})^{n}$ are used to compute the intermediate linear and angular velocities
\begin{equation}
\vectorbold{u}_{k}^{n+1/2} = \vectorbold{u}_{k}^{n} + \frac{\Delta{}t}{2} \, \vectorbold{a}_{k}^{n}
\qand
\vectorbold*{\omega}_{k}^{n+1/2} = \vectorbold*{\omega}_{k}^{n} + \frac{\Delta{}t}{2} \, \vectorbold*{\alpha}_{k}^{n}
\, .
\end{equation}
In a drift-step, the positions (and orientations) of fluid particles~$i$ and rigid bodies~$k$ are updated to time step~$n+1$ using the intermediate velocities. Accordingly, the positions of fluid particles~$i$ follow as
\begin{equation}
\vectorbold{r}_{i}^{n+1} = \vectorbold{r}_{i}^{n} + \Delta{}t \, \vectorbold{u}_{i}^{n+1/2}
\end{equation}
and the center of mass positions of rigid bodies~$k$ as
\begin{equation}
\vectorbold{r}_{k}^{n+1} = \vectorbold{r}_{k}^{n} + \Delta{}t \, \vectorbold{u}_{k}^{n+1/2} \, .
\end{equation}
The orientations of rigid bodies~$k$ are updated making use of quaternion algebra, cf. Section~\ref{subsec:nummeth_rigid_orientation}. First, the angular orientation increments from time step~$n$ to time step~$n+1$ are determined using the intermediate angular velocities of rigid bodies~$k$ following
\begin{equation}
\vectorbold*{\phi}_{k}^{n,n+1} = \Delta{}t \, \vectorbold*{\omega}_{k}^{n+1/2} \, .
\end{equation}
Next, the angular orientation increments are described by so-called transition quaternions~$\vectorbold{q}_{k}^{n,n+1}$. Finally, quaternion multiplication, cf. Remark~\ref{rmk:nummeth_rigid_quaternion_algebra}, gives the updated orientations of rigid bodies~$k$ at time step~$n+1$
\begin{equation}
\vectorbold{q}_{k}^{n+1} = \vectorbold{q}_{k}^{n,n+1} \circ \vectorbold{q}_{k}^{n} \, .
\end{equation}
Once the updated orientations (and thus also the updated rigid body frames) are known, the relative positions of rigid particles~$\vectorbold{r}_{rk}^{n+1}$ can be transformed from the rigid body frame to the reference frame. The velocities and the positions of rigid particles~$r$ are updated, considering the underlying rigid body motion of the corresponding rigid bodies~$k$, in consistency with the applied time integration scheme following
\begin{equation}
\vectorbold{u}_{r}^{n+1/2} = \vectorbold{u}_{k}^{n+1/2} + \vectorbold*{\omega}_{k}^{n+1/2} \cross \vectorbold{r}_{rk}^{n+1}
\qand
\vectorbold{r}_{r}^{n+1} = \vectorbold{r}_{k}^{n+1} + \vectorbold{r}_{rk}^{n+1}
\, .
\end{equation}
Using the positions~$\vectorbold{r}_{a}^{n+1}$ and the intermediate velocities~$\vectorbold{v}_{a}^{n+1/2}$ of fluid and rigid particles~$a \in \qty{i,r}$, the densities~$\rho_{i}^{n+1}$ of fluid particles~$i$ are computed via equation~\eqref{eq:nummeth_fluid_densum}. The densities~$\rho_{r}$ of rigid particles~$r$ are not evolved and remain constant. The temperature rates~$\qty(\dv*{T_{a}}{t})^{n+1}$ of fluid and rigid particles~$a$ are then updated on the basis of equation~\eqref{eq:nummeth_thermo_heat_equation} with the temperatures~$T_{a}^{n}$ as well as the positions~$\vectorbold{r}_{a}^{n+1}$ and densities~$\rho_{a}^{n+1}$. Finally, the temperatures of fluid and rigid particles~$a$ are computed as
\begin{equation}
T_{a}^{n+1} = T_{a}^{n} + \Delta{}t \qty(\dv{T_{a}}{t})^{n+1} \, .
\end{equation}
The accelerations~$\vectorbold{a}_{i}^{n+1}$ of fluid particles~$i$, cf. equation~\eqref{eq:nummeth_fluid_momentum}, and the forces~$\vectorbold{f}_{r}^{n+1}$ acting on rigid particles~$r$, cf. equation~\eqref{eq:nummeth_rigid_rigidparticle_force}, are concurrently computed using the positions~$\vectorbold{r}_{a}^{n+1}$, the intermediate velocities~$\vectorbold{v}_{a}^{n+1/2}$, and the densities~$\rho_{a}^{n+1}$ of fluid and rigid particles~$a$. Consequently, the resultant forces~$\vectorbold{f}_{k}^{n+1}$ and torques~$\vectorbold{m}_{k}^{n+1}$ acting on rigid bodies~$k$, cf. Section~\ref{subsec:nummeth_rigid_forces}, together with mass-related quantities, cf. Section~\ref{subsec:nummeth_rigid_massquantities}, give the linear and angular accelerations~$\vectorbold{a}_{k}^{n+1}$ respectively~$\vectorbold*{\alpha}_{k}^{n+1}$, cf. equations~\eqref{eq:goveq_solid_linearmomentum} and~\eqref{eq:goveq_solid_angularmomentum}. In a final kick-step, the velocities of fluid particles~$i$ at time step~$n+1$ are computed as
\begin{equation}
\vectorbold{u}_{i}^{n+1} = \vectorbold{u}_{i}^{n+1/2} + \frac{\Delta{}t}{2} \, \vectorbold{a}_{i}^{n+1} \, ,
\end{equation}
while the linear and angular velocities of rigid bodies~$k$ are
\begin{equation}
\vectorbold{u}_{k}^{n+1} = \vectorbold{u}_{k}^{n+1/2} + \frac{\Delta{}t}{2} \, \vectorbold{a}_{k}^{n+1}
\qand
\vectorbold*{\omega}_{k}^{n+1} = \vectorbold*{\omega}_{k}^{n+1/2} + \frac{\Delta{}t}{2} \, \vectorbold*{\alpha}_{k}^{n+1}
\, .
\end{equation}
Accordingly, the velocities of rigid particles~$r$ are determined following the motion of the corresponding rigid bodies~$k$ to
\begin{equation}
\vectorbold{u}_{r}^{n+1} = \vectorbold{u}_{k}^{n+1} + \vectorbold*{\omega}_{k}^{n+1} \cross \vectorbold{r}_{rk}^{n+1} \, .
\end{equation}
To maintain stability of the time integration scheme, the time step size~$\Delta{}t$ is restricted by the Courant-Friedrichs-Lewy (CFL) condition, the viscous condition, the body force condition, the contact condition, and the conductivity condition, refer to~\cite{Morris1997,Adami2013,OSullivan2004,Cleary1998} for more details,
\begin{equation} \label{eq:nummeth_timint_timestepcond}
\Delta{}t \leq \min\qty{
0.25\frac{h}{c+\qty|\vectorbold{u}_{max}|}, \quad
0.125\frac{h^{2}}{\nu}, \quad
0.25\sqrt{\frac{h}{\qty|\vectorbold{b}_{max}|}}, \quad
0.22 \sqrt{\frac{m_{r}}{k_{c}}}, \quad
0.1\frac{\rho c_{p} h^{2}}{\kappa}
} \, ,
\end{equation}
with maximum fluid velocity~$\vectorbold{u}_{max}$ and maximum body force~$\vectorbold{b}_{max}$.

\section{Numerical examples} \label{sec:numex}

The purpose of this section is to investigate the proposed numerical formulation for solving fluid-solid and contact interaction problems examining several numerical examples in two and three dimensions involving multiple mobile rigid bodies, two-phase flow, and reversible phase transitions. To begin with, several numerical examples of a single rigid body in a fluid flow, considering different spatial discretizations, are studied and compared to reference solutions, cf. Sections~\ref{subsec:numex_spatialdiscretcyl}, \ref{subsec:numex_cylindershearflow}, and~ \ref{subsec:numex_fallingcylinder}. In a next step, two examples close to potential application scenarios of the proposed formulation in the fields of engineering and biomechanics are investigated, cf. Sections~\ref{subsec:numex_melting} and~\ref{subsec:numex_gastric}. Finally, the capabilities of the proposed parallel computational framework are demonstrated performing a strong scaling analysis, cf. Section~\ref{subsec:numex_strongscaling}.

\subsection{Spatial discretization of a rigid circular disk} \label{subsec:numex_spatialdiscretcyl}

\begin{figure}[htbp]%
\centering
% begin subfigure
\subfigure %[caption of subfigure]
{
\begin{tikzpicture}[trim axis left,trim axis right]
\begin{axis}
[
width=0.35\textwidth,
height=0.12\textwidth,
xmin=0.0e-4, xmax=5.0e-4,
ymin=4.75e-3, ymax=5.25e-3,
every x tick scale label/.style={at={(1,0)},above right,inner sep=0pt,xshift=0.2em},
every y tick scale label/.style={at={(0,1)},above right,inner sep=0pt,yshift=0.2em},
xtick={0.0e-4, 1.0e-4, 2.0e-4, 3.0e-4, 4.0e-4, 5.0e-4},
xticklabels={4.0, 2.0, 1.0, 0.5, 0.25, 0.125},
ytick={4.75e-3, 5.0e-3, 5.25e-3},
xlabel={$\Delta{}x$},
ylabel={$m^{s}$},
yticklabel style={/pgf/number format/precision=2, /pgf/number format/fixed, /pgf/number format/fixed zerofill},
]
% begin plot
\addplot [color=black,solid,line width=0.5pt,mark=square] table [x expr={(\thisrow{"i"})*1.0e-4}, y expr={(\thisrow{"mass"})*1.0e-3}, col sep=comma] {data/rotcyl_massinertia.csv};
\addplot [color=black,dashed,line width=0.5pt] coordinates{(0.0e-4,4.908738521e-3) (5.0e-4,4.908738521e-3)};
% end plot
\end{axis}
\end{tikzpicture}
}
% end subfigure
\hspace{0.12\textwidth}
% begin subfigure
\subfigure %[caption of subfigure]
{
\begin{tikzpicture}[trim axis left,trim axis right]
\begin{axis}
[
width=0.35\textwidth,
height=0.12\textwidth,
xmin=0.0e-4, xmax=5.0e-4,
ymin=3.6e-9, ymax=4.4e-9,
every x tick scale label/.style={at={(1,0)},above right,inner sep=0pt,xshift=0.2em},
every y tick scale label/.style={at={(0,1)},above right,inner sep=0pt,yshift=0.2em},
xtick={0.0e-4, 1.0e-4, 2.0e-4, 3.0e-4, 4.0e-4, 5.0e-4},
xticklabels={4.0, 2.0, 1.0, 0.5, 0.25, 0.125},
ytick={3.6e-9, 4.0e-9, 4.4e-9},
xlabel={$\Delta{}x$},
ylabel={$I^{s}$},
yticklabel style={/pgf/number format/precision=1, /pgf/number format/fixed, /pgf/number format/fixed zerofill},
]
% begin plot
\addplot [color=black,solid,line width=0.5pt,mark=square] table [x expr={(\thisrow{"i"})*1.0e-4}, y expr={(\thisrow{"inertia_2"})*1.0e-9}, col sep=comma] {data/rotcyl_massinertia.csv};
\addplot [color=black,dashed,line width=0.5pt] coordinates{(0.0e-4,3.83495197e-9) (5.0e-4,3.83495197e-9)};
% end plot
\end{axis}
\end{tikzpicture}
}
% end subfigure
\caption{Spatial discretization of a rigid circular disk: mass~$m^{s}$ and mass moment of inertia~$I^{s}$ of a rigid circular disk for different values of the initial particle spacing~$\Delta{}x$ (solid line) compared to the analytical solution (dashed line).}
\label{fig:example_spatialdiscretcyl_mass_inertia}
\end{figure}
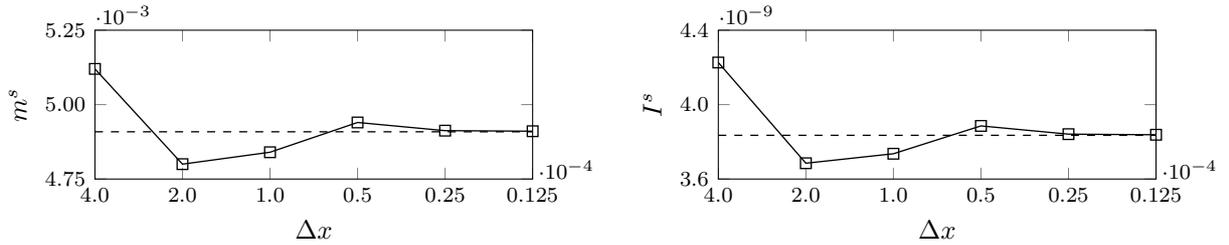

\begin{figure}[htbp]
\centering
% begin subfigure
\subfigure [$\Delta{}x = 4.0 \times 10^{-4}$]
{
\includegraphics[width=0.15\textwidth]{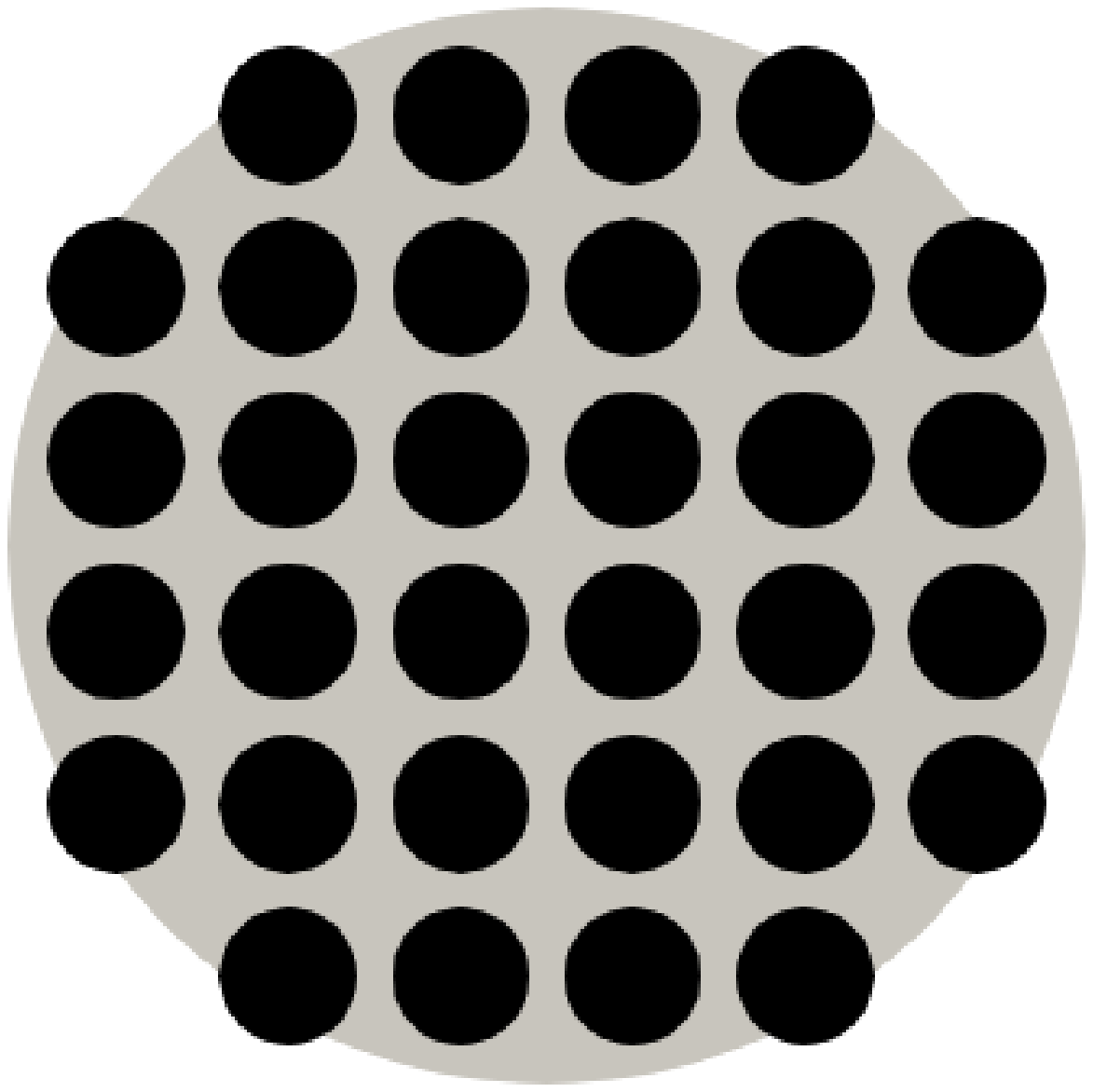}
}
% end subfigure
\hspace{0.05\textwidth}
% begin subfigure
\subfigure [$\Delta{}x = 2.0 \times 10^{-4}$]
{
\includegraphics[width=0.15\textwidth]{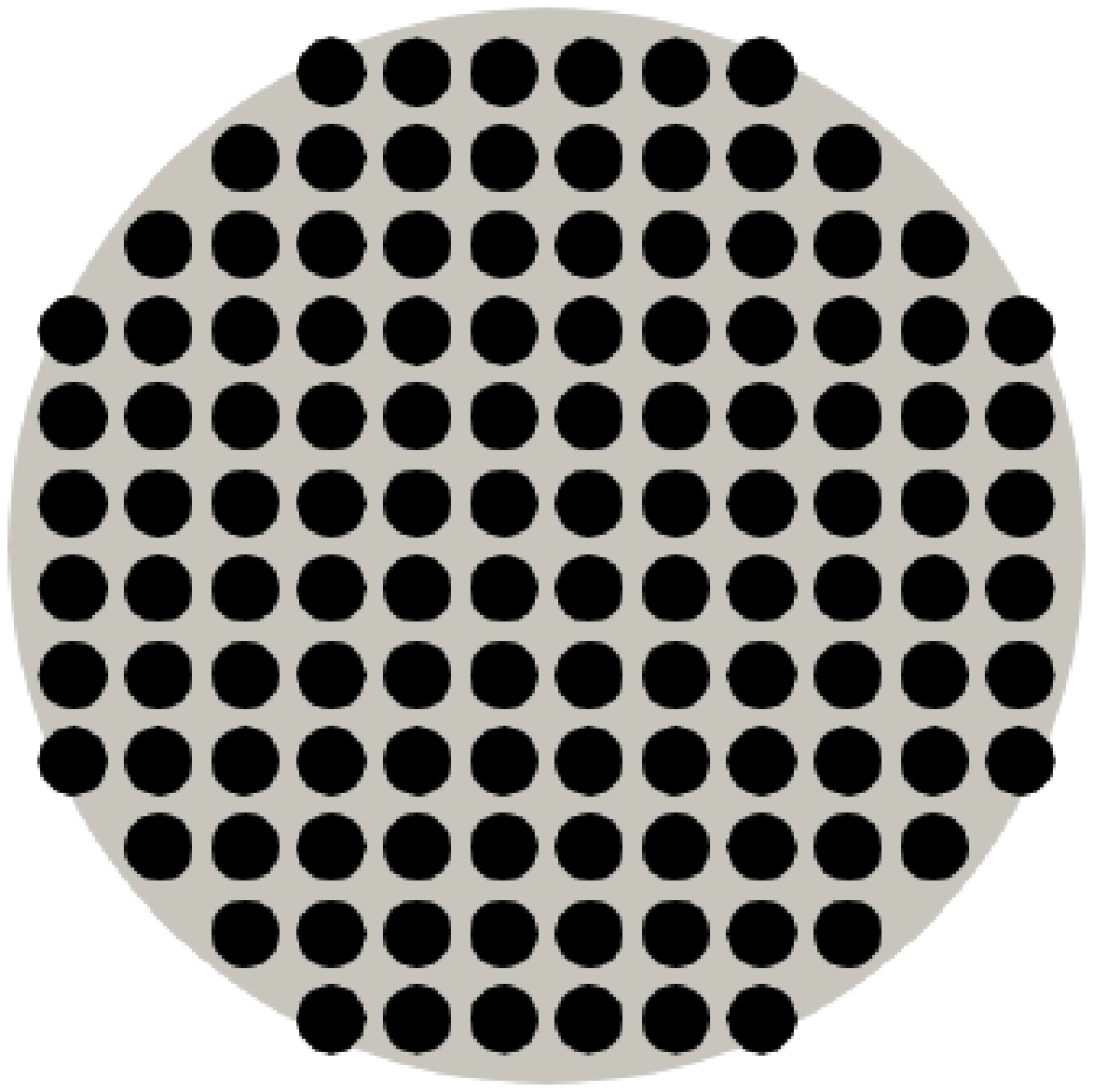}
}
% end subfigure
\hspace{0.05\textwidth}
% begin subfigure
\subfigure [$\Delta{}x = 1.0 \times 10^{-4}$]
{
\includegraphics[width=0.15\textwidth]{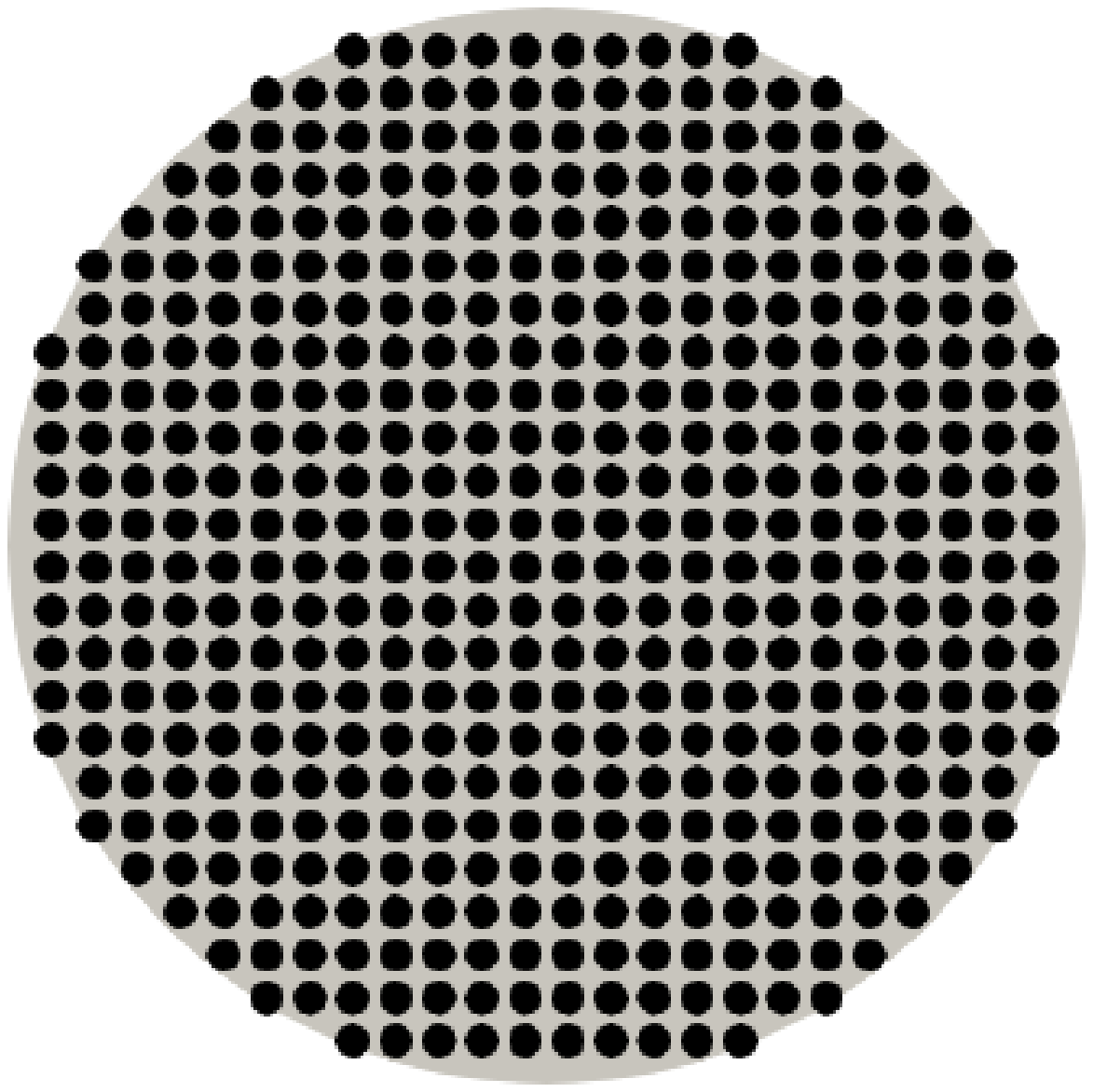}
}
% end subfigure
\hspace{0.05\textwidth}
% begin subfigure
\subfigure [$\Delta{}x = 0.5 \times 10^{-4}$]
{
\includegraphics[width=0.15\textwidth]{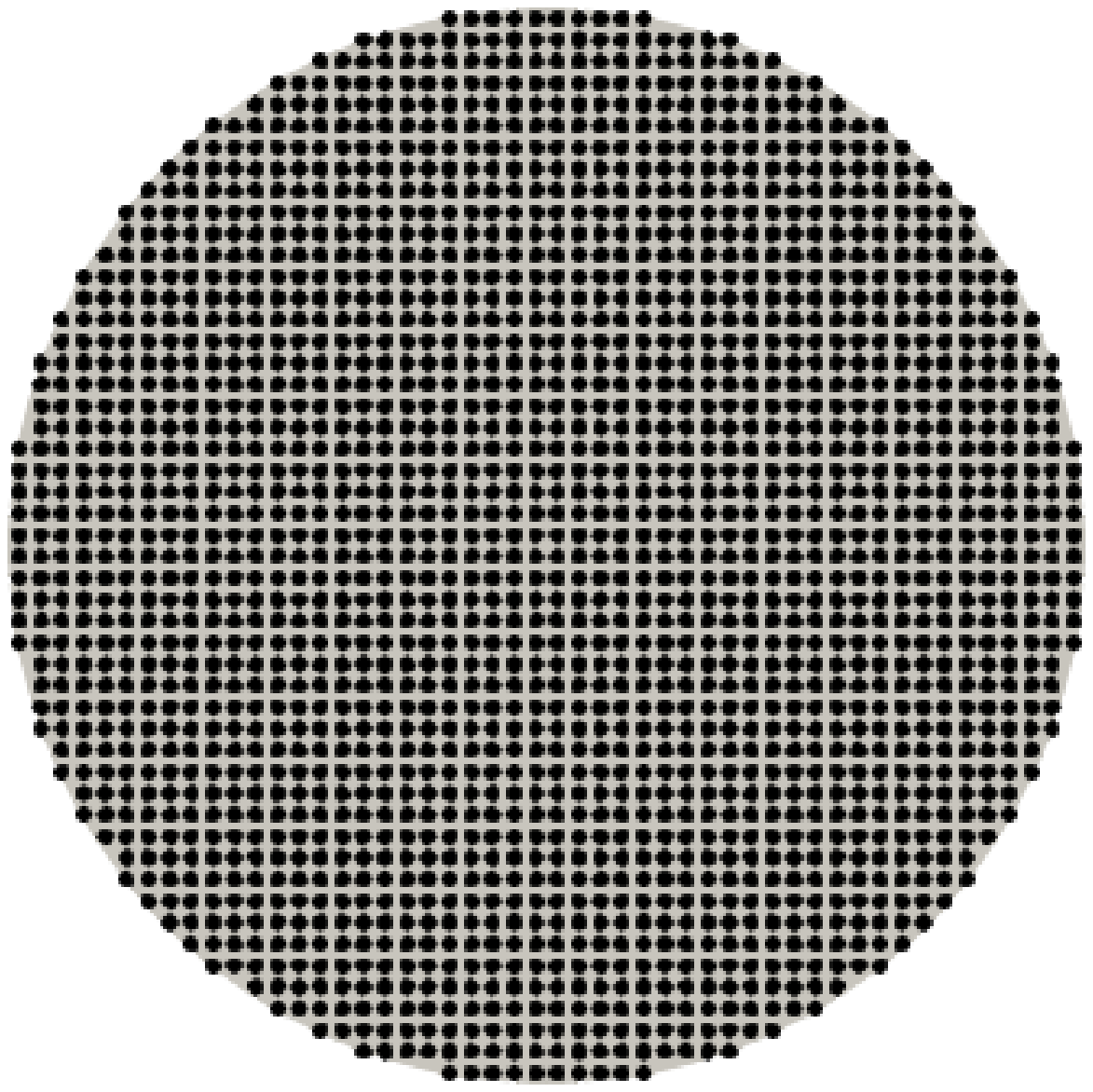}
}
% end subfigure
\caption{Spatial discretization of a rigid circular disk: resulting spatial discretization of a rigid circular disk with rigid particles for selected values of the initial particle spacing~$\Delta{}x$.}
\label{fig:example_spatialdiscretcyl_spatialdiscret}
\end{figure}

In the following, a rigid circular disk of diameter~$D = 2.5 \times 10^{-3}$ with density~$\rho^{s} = 1.0 \times 10^{3}$, motivated by the subsequent examples discussed in Sections~\ref{subsec:numex_cylindershearflow} and~\ref{subsec:numex_fallingcylinder}, is discretized with different values of the initial particle spacing~$\Delta{}x$. The mass~$m^{s}$ and the mass moment of inertia~$I^{s}$ (with respect to the axis of symmetry) of the circular disk are computed with the proposed formulation and shown in Figure~\ref{fig:example_spatialdiscretcyl_mass_inertia}. With decreasing initial particle spacing~$\Delta{}x$ the values for mass~$m^{s}$ and mass moment of inertia~$I^{s}$ converge to the analytical solution confirming the proposed formulation. To illustrate, the resulting spatial discretizations of the circular disk with rigid particles are shown in Figure~\ref{fig:example_spatialdiscretcyl_spatialdiscret}. Clearly, the approximation of the circular shape of the disk is of better accuracy for decreasing initial particle spacing~$\Delta{}x$. To keep the computational effort at a feasible level, in the examples discussed in Sections~\ref{subsec:numex_cylindershearflow} and~\ref{subsec:numex_fallingcylinder} the domain is discretized with an initial particle spacing of $\Delta{}x = 2.0 \times 10^{-4}$ and $\Delta{}x = 1.0 \times 10^{-4}$.

% begin figure
% end figure

% begin figure
% end figure

%%
\subsection{A rigid circular disk floating in a shear flow} \label{subsec:numex_cylindershearflow}

The following numerical examples are concerned with the motion of a rigid circular disk floating in a shear flow. First, the principal setup of the problem along with numerical parameters is described, thereafter, two distinct cases, cf. Sections~\ref{subsec:numex_cylindershearflow_case_1} and~\ref{subsec:numex_cylindershearflow_case_2}, are considered in detail. For validation, the results obtained with the proposed formulation are compared to~\cite{Hashemi2012} also applying SPH to discretize the fluid and the solid field.

\begin{figure}[htbp]
\centering
% begin subfigure
\subfigure [Case 1: Migration of a floating rigid circular disk to the center line of a channel.]
{
\newcommand*{\scaletext}{0.8}
\newcommand*{\scalefig}{0.36}
\psfrag{L}{\scalebox{\scaletext}{$L$}}
\psfrag{H}{\scalebox{\scaletext}{$H$}}
\psfrag{D}{\scalebox{\scaletext}{$D$}}
\psfrag{cx}{\scalebox{\scaletext}{$x$}}
\psfrag{cy}{\scalebox{\scaletext}{$y$}}
\psfrag{Uw}{\scalebox{\scaletext}{$\flatfrac{u_{w}}{2}$}}
\includegraphics[scale=\scalefig]{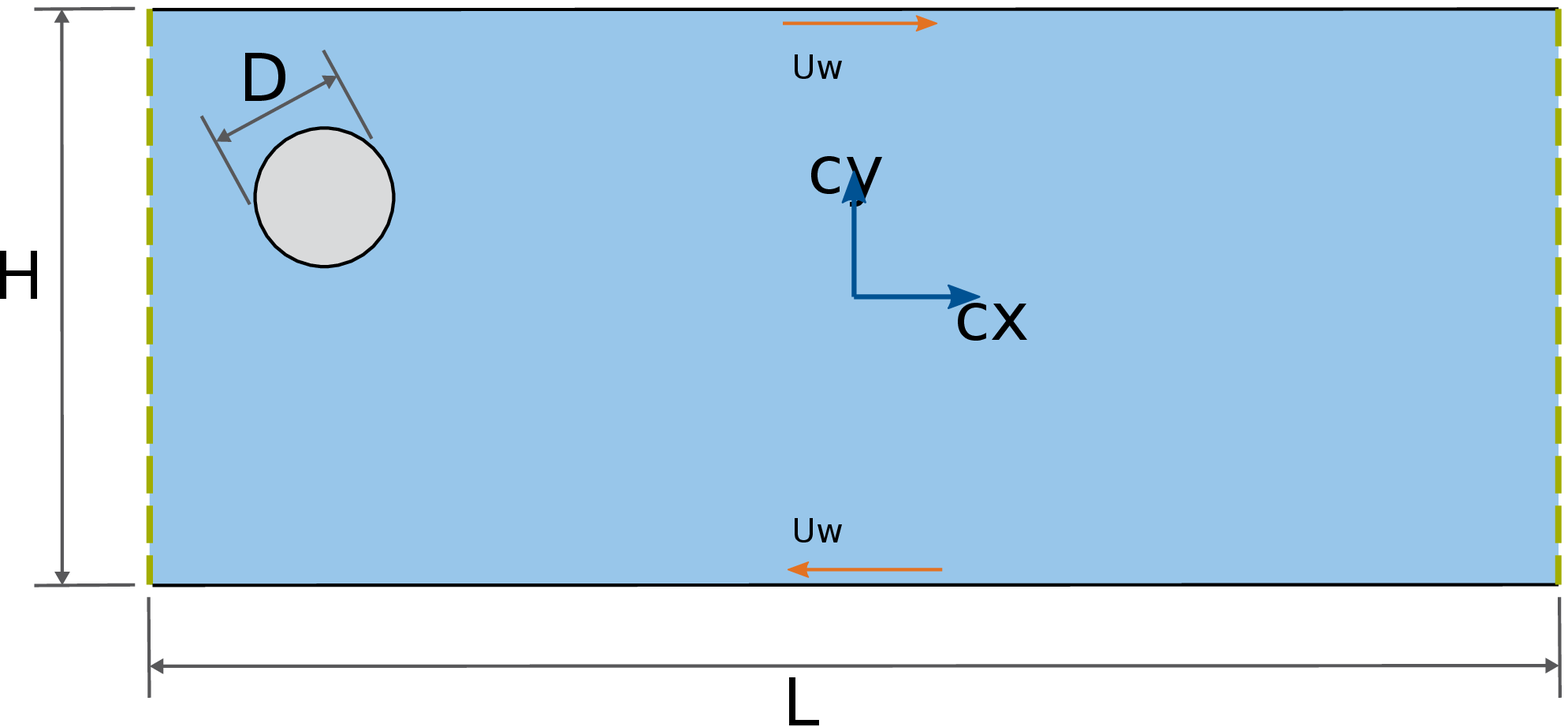}
\label{fig:example_cylindershearflow_case1}
}
% end subfigure
\hspace{0.02\textwidth}
% begin subfigure
\subfigure [Case 2: Interaction of a floating rigid circular disk with a fixed rigid circular disk.]
{
\newcommand*{\scaletext}{0.8}
\newcommand*{\scalefig}{0.36}
\psfrag{L}{\scalebox{\scaletext}{$L$}}
\psfrag{H}{\scalebox{\scaletext}{$H$}}
\psfrag{D}{\scalebox{\scaletext}{$D$}}
\psfrag{cx}{\scalebox{\scaletext}{$x$}}
\psfrag{cy}{\scalebox{\scaletext}{$y$}}
\psfrag{Uw}{\scalebox{\scaletext}{$\flatfrac{u_{w}}{2}$}}
\includegraphics[scale=\scalefig]{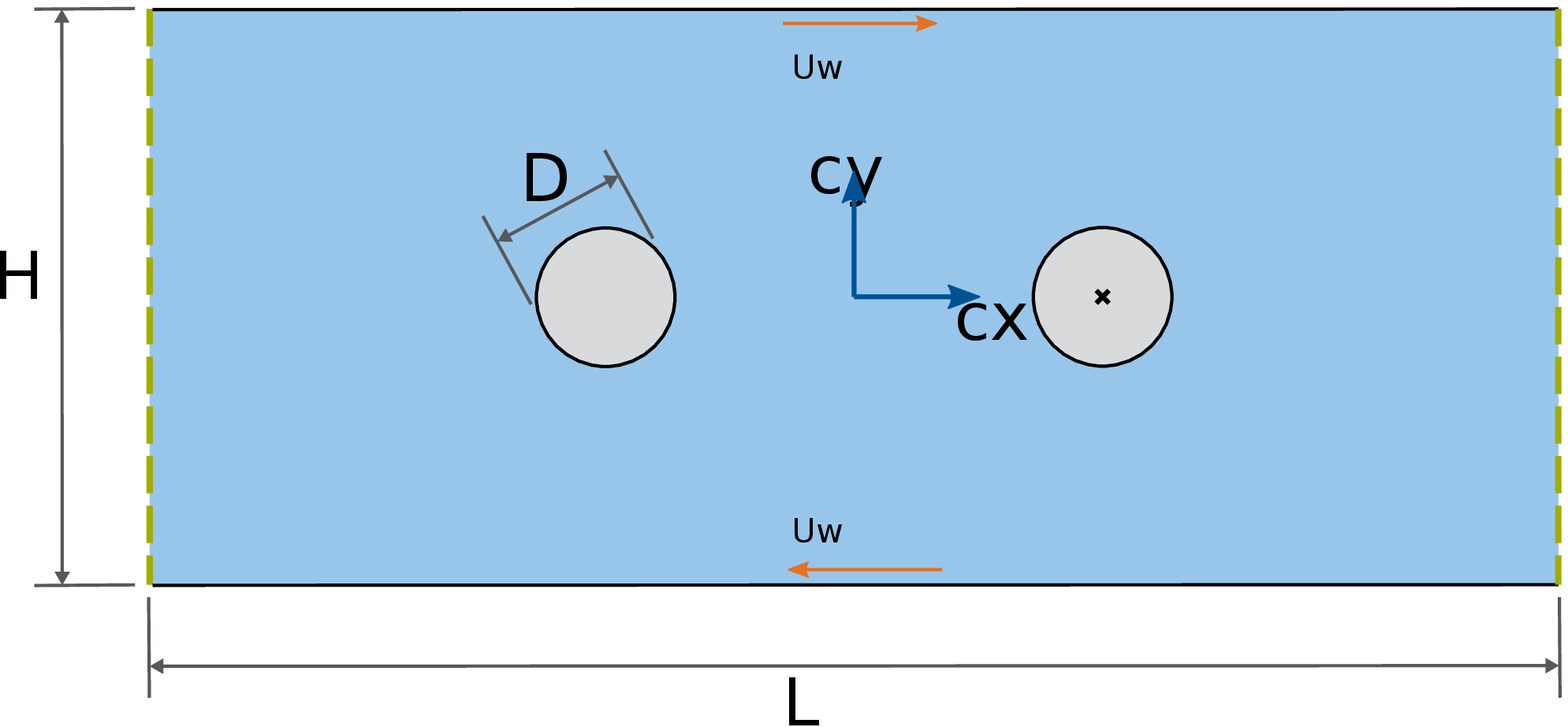}
\label{fig:example_cylindershearflow_case2}
}
% end subfigure
\caption{A rigid circular disk floating in a shear flow: geometry and boundary conditions of two different cases.}
\label{fig:example_cylindershearflow}
\end{figure}

A rigid circular disk of diameter~$D = 2.5 \times 10^{-3}$ with density~$\rho^{s} = 1.0 \times 10^{3}$ is allowed to move freely in a rectangular channel of length~$L = 5.0 \times 10^{-2}$ and height~$H = 1.0 \times 10^{-2}$, cf. Figure~\ref{fig:example_cylindershearflow}. The remainder of the channel is occupied by a Newtonian fluid with density~$\rho^{f} = 1.0 \times 10^{3}$ and kinematic viscosity~$\nu^{f} = 5.0 \times 10^{-6}$. The bottom and top channel walls move with velocity~$\flatfrac{u_{w}}{2}$ in opposite direction inducing a shear flow in the channel. The Reynolds number of the problem is given as $Re = \flatfrac{u_{w} D^{2}}{4 \nu^{f} H}$~\cite{Feng2002, Hashemi2012} taking into account the diameter of the circular disk~$D$ and the channel height~$H$. At the left and right end of the channel, periodic boundary conditions are applied, cf. Remark~\ref{rmk:numex_cylindershearflow_pbc}.

For the fluid phase, an artificial speed of sound~$c = 0.25$ is chosen, resulting in a reference pressure~$p_{0} = 62.5$ of the weakly compressible model. The background pressure~$p_{b}$ of the transport velocity formulation is set equal to the reference pressure~$p_{0}$. The motion of the bottom and top channel walls is modeled using moving boundary particles. The problem is solved for different values of the initial particle spacing~$\Delta{}x$, cf. Section~\ref{subsec:numex_spatialdiscretcyl}, for times~$t \in \qty[0, 60.0]$ with time step size~$\Delta{}t$ obeying respective conditions~\eqref{eq:nummeth_timint_timestepcond}.

% begin remark
\begin{rmk} \label{rmk:numex_cylindershearflow_pbc}
Imposing a periodic boundary condition in a specific spatial direction allows for particle interaction evaluation across opposite domain borders. Moreover, particles leaving the domain on one side are re-entering on the opposite side.
\end{rmk}
% end remark

% begin figure
% end figure

%%
\subsubsection{Case 1: Migration of a floating rigid circular disk to the center line of a channel} \label{subsec:numex_cylindershearflow_case_1}

This case is based on studies~\cite{Feng1994, Feng2002} stating that a rigid circular disk floating in a shear flow in a channel migrates to the center line of the channel independent of its initial position and initial velocity. Herein, the rigid circular disk is initially at rest placed at vertical position~$r_{y} = 2.5 \times 10^{-3}$ in the channel, cf. Figure~\ref{fig:example_cylindershearflow_case1}. The channel walls move in opposite direction with a velocity magnitude of $\flatfrac{u_{w}}{2} = 0.01$ resulting in the Reynolds number~$Re = 0.625$ of the problem.

\begin{figure}[htbp]%
\centering
% begin subfigure
\subfigure %[caption of subfigure]
{
\begin{tikzpicture}[trim axis left,trim axis right]
\begin{axis}
[
width=0.35\textwidth,
height=0.15\textwidth,
xmin=0.0, xmax=60.0,
ymin=0.0, ymax=0.0025,
%scaled ticks=false,
every x tick scale label/.style={at={(1,0)},above right,inner sep=0pt,xshift=0.2em},
every y tick scale label/.style={at={(0,1)},above right,inner sep=0pt,yshift=0.2em},
ytick={0.0, 0.0005, 0.001, 0.0015, 0.002, 0.0025},
yticklabel style={/pgf/number format/precision=1, /pgf/number format/fixed, /pgf/number format/fixed zerofill},
xlabel={$t$},
ylabel={$r_{y}$},
]
% begin plot
\addplot [color=black,solid,line width=0.5pt] table [x expr={(\thisrow{"Time"})}, y expr={(\thisrow{"position (1) (stats)"})}, col sep=comma] {data/cylinshearflow_data_h0_1.csv};
\addplot [color=red,densely dotted,line width=0.5pt] table [x expr={(\thisrow{"Time"})}, y expr={(\thisrow{"position (1) (stats)"})}, col sep=comma] {data/cylinshearflow_data_h0_2.csv};
\addplot [color=black,only marks,mark=+] table [x expr={(\thisrow{"time"})}, y expr={(\thisrow{"ypos"})}, col sep=comma] {data/cylinshearflow_hashemi2012_data_pos.csv};
% end plot
\end{axis}
\end{tikzpicture}
}
% end subfigure
\hspace{0.12\textwidth}
% begin subfigure
\subfigure %[caption of subfigure]
{
\begin{tikzpicture}[trim axis left,trim axis right]
\begin{axis}
[
width=0.35\textwidth,
height=0.15\textwidth,
xmin=0.0, xmax=60.0,
ymin=0.0, ymax=0.005,
%scaled ticks=false,
every x tick scale label/.style={at={(1,0)},above right,inner sep=0pt,xshift=0.2em},
every y tick scale label/.style={at={(0,1)},above right,inner sep=0pt,yshift=0.2em},
ytick={0.0, 0.001, 0.002, 0.003, 0.004, 0.005},
yticklabel style={/pgf/number format/precision=1, /pgf/number format/fixed, /pgf/number format/fixed zerofill},
xlabel={$t$},
ylabel={$u_{x}$},
]
% begin plot
\addplot [color=black,solid,line width=0.5pt] table [x expr={(\thisrow{"Time"})}, y expr={\thisrow{"velocity (0) (stats)"}}, col sep=comma] {data/cylinshearflow_data_h0_1.csv};
\addplot [color=red,densely dotted,line width=0.5pt] table [x expr={(\thisrow{"Time"})}, y expr={\thisrow{"velocity (0) (stats)"}}, col sep=comma] {data/cylinshearflow_data_h0_2.csv};
\addplot [color=black,only marks,mark=+] table [x expr={(\thisrow{"time"})}, y expr={\thisrow{"xvel"}}, col sep=comma] {data/cylinshearflow_hashemi2012_data_vel.csv};
% end plot
\end{axis}
\end{tikzpicture}
}
% end subfigure
\caption{Migration of a floating rigid circular disk to the center line of a channel: vertical position~$r_{y}$ and horizontal velocity~$u_{x}$ of the center of the floating circular disk in the channel computed with the proposed formulation and an initial particle spacing of $\Delta{}x = 2.0 \times 10^{-4}$ (red dashed line) and $\Delta{}x = 1.0 \times 10^{-4}$ (black solid line) compared to the reference solution~\cite{Hashemi2012} (crosses).}
\label{fig:example_cylindershearflow_case1_pos_vel}
\end{figure}
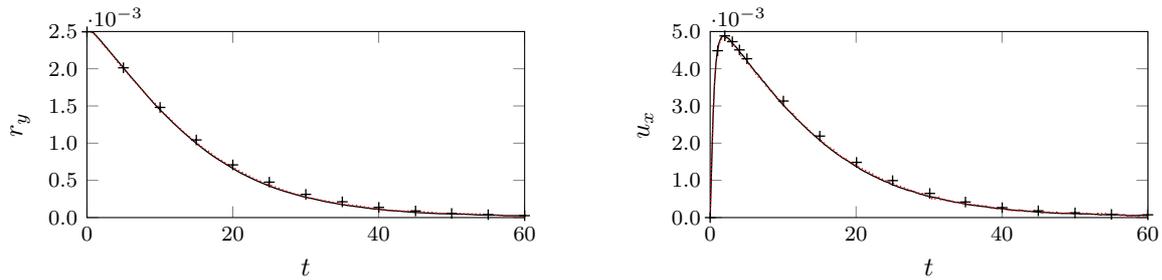

The obtained vertical position~$r_{y}$ and the horizontal velocity~$u_{x}$ of the center of the circular disk in the channel over time~$t$ are displayed in Figure~\ref{fig:example_cylindershearflow_case1_pos_vel} for two different  values of the initial particle spacing~$\Delta{}x$. The circular disk migrates to the center line of the channel as expected, showing no significant difference between the results obtained with different initial particle spacings~$\Delta{}x$. In addition, a comparison to the results of~\cite{Hashemi2012} shows very good agreement for the dynamics of the solution.

% begin figure
% end figure

%%
\subsubsection{Case 2: Interaction of a floating rigid circular disk with a fixed rigid circular disk} \label{subsec:numex_cylindershearflow_case_2}

In the presence of a rigid circular disk that is fixed at the center line of the channel, a rigid circular disk floating in a shear flow migrates to a specific position of equilibrium independent of its initial position and velocity as stated in~\cite{Yan2007}. Herein, the fixed and the floating rigid circular disks are initially placed on the center line of the channel at horizontal position~$r_{x} = \pm 3.75 \times 10^{-3}$, cf. Figure~\ref{fig:example_cylindershearflow_case2}. The channel walls move in opposite direction with a velocity magnitude of $\flatfrac{u_{w}}{2} = 0.012$ resulting in the Reynolds number~$Re = 0.75$ of the problem.

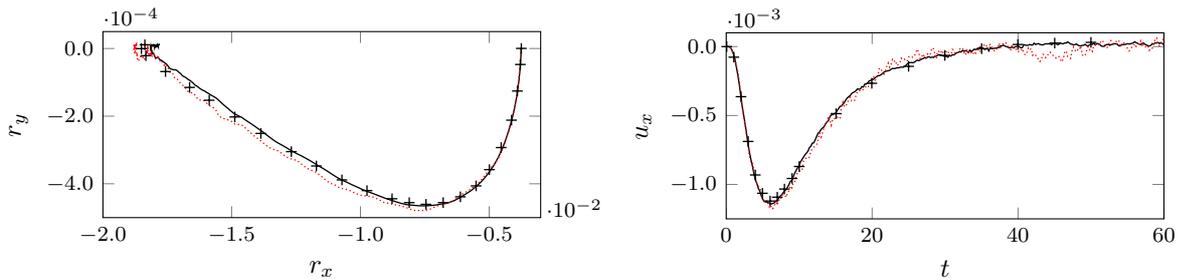
\begin{figure}[htbp]%
\centering
% begin subfigure
\subfigure %[caption of subfigure]
{
\begin{tikzpicture}[trim axis left,trim axis right]
\begin{axis}
[
width=0.35\textwidth,
height=0.15\textwidth,
xmin=-0.02, xmax=-0.003,
ymin=-0.0005, ymax=0.00005,
%scaled ticks=false,
every x tick scale label/.style={at={(1,0)},above right,inner sep=0pt,xshift=0.2em},
every y tick scale label/.style={at={(0,1)},above right,inner sep=0pt,yshift=0.2em},
xticklabel style={/pgf/number format/precision=1, /pgf/number format/fixed, /pgf/number format/fixed zerofill},
yticklabel style={/pgf/number format/precision=1, /pgf/number format/fixed, /pgf/number format/fixed zerofill},
xlabel={$r_{x}$},
ylabel={$r_{y}$},
]
% begin plot
\addplot [color=black,solid,line width=0.5pt] table [x expr={(\thisrow{"position (0) (stats)"})}, y expr={(\thisrow{"position (1) (stats)"})}, col sep=comma] {data/fixedfreecylinshearflow_data_h0_1.csv};
\addplot [color=red,densely dotted,line width=0.5pt] table [x expr={(\thisrow{"position (0) (stats)"})}, y expr={(\thisrow{"position (1) (stats)"})}, col sep=comma] {data/fixedfreecylinshearflow_data_h0_2.csv};
\addplot [color=black,only marks,mark=+] table [x expr={(\thisrow{"xpos"})}, y expr={-0.005+(\thisrow{"ypos"})}, col sep=comma] {data/fixedfreecylinshearflow_hashemi2012_data_pos.csv};
% end plot
\end{axis}
\end{tikzpicture}
}
% end subfigure
\hspace{0.12\textwidth}
% begin subfigure
\subfigure %[caption of subfigure]
{
\begin{tikzpicture}[trim axis left,trim axis right]
\begin{axis}
[
width=0.35\textwidth,
height=0.15\textwidth,
xmin=0.0, xmax=60.0,
ymin=-0.00125, ymax=0.0001,
%scaled ticks=false,
every x tick scale label/.style={at={(1,0)},above right,inner sep=0pt,xshift=0.2em},
every y tick scale label/.style={at={(0,1)},above right,inner sep=0pt,yshift=0.2em},
yticklabel style={/pgf/number format/precision=1, /pgf/number format/fixed, /pgf/number format/fixed zerofill},
xlabel={$t$},
ylabel={$u_{x}$},
]
% begin plot
\addplot [color=black,solid,line width=0.5pt] table [x expr={(\thisrow{"Time"})}, y expr={\thisrow{"velocity (0) (stats)"}}, col sep=comma] {data/fixedfreecylinshearflow_data_h0_1.csv};
\addplot [color=red,densely dotted,line width=0.5pt] table [x expr={(\thisrow{"Time"})}, y expr={\thisrow{"velocity (0) (stats)"}}, col sep=comma] {data/fixedfreecylinshearflow_data_h0_2.csv};
\addplot [color=black,only marks,mark=+] table [x expr={(\thisrow{"time"})}, y expr={(\thisrow{"xvel"})}, col sep=comma] {data/fixedfreecylinshearflow_hashemi2012_data_vel.csv};
% end plot
\end{axis}
\end{tikzpicture}
}
% end subfigure
\caption{Interaction of a floating rigid circular disk with a fixed rigid circular disk: trajectory and horizontal velocity~$u_{x}$ of the center of the floating circular disk in the channel computed with the proposed formulation and an initial particle spacing of $\Delta{}x = 2.0 \times 10^{-4}$ (red dashed line) and $\Delta{}x = 1.0 \times 10^{-4}$ (black solid line) compared to the reference solution~\cite{Hashemi2012} (crosses).}
\label{fig:example_cylindershearflow_case2_pos_vel}
\end{figure}

Figure~\ref{fig:example_cylindershearflow_case2_pos_vel} shows the obtained trajectory, i.e., vertical position~$r_{y}$ over horizontal position~$r_{x}$, and horizontal velocity~$u_{x}$ of the center of the floating circular disk in the channel for two different values of the initial particle spacing~$\Delta{}x$. The results obtained with initial particle spacing~$\Delta{}x = 1.0 \times 10^{-4}$ are in good agreement to the reference solution~\cite{Hashemi2012}. However, the results obtained with initial particle spacing~$\Delta{}x = 2.0 \times 10^{-4}$ show fluctuations of the horizontal velocity~$u_{x}$, which is why also the trajectory deviates from the reference solution~\cite{Hashemi2012}. This can be explained with disturbances of the density field due to relative particle movement~\cite{Morris1997}, that are more pronounced with a coarser spatial discretization, i.e., with larger initial particle spacing~$\Delta{}x$.

% begin figure
% end figure

%%
\subsection{A rigid circular disk falling in a fluid column} \label{subsec:numex_fallingcylinder}

\begin{figure}[htbp]
\centering
\newcommand*{\scaletext}{0.8}
\newcommand*{\scalefig}{0.36}
\psfrag{W}{\scalebox{\scaletext}{$W$}}
\psfrag{H}{\scalebox{\scaletext}{$H$}}
\psfrag{D}{\scalebox{\scaletext}{$D$}}
\psfrag{cx}{\scalebox{\scaletext}{$x$}}
\psfrag{cy}{\scalebox{\scaletext}{$y$}}
\psfrag{b}{\scalebox{\scaletext}{$\vectorbold{g}$}}
\includegraphics[scale=\scalefig]{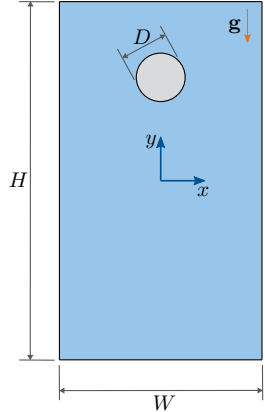}
\caption{A rigid circular disk falling in a fluid column: geometry and boundary conditions of the problem.}
\label{fig:example_fallingcylinder}
\end{figure}

A rigid circular disk of diameter~$D = 2.5 \times 10^{-3}$ with density~$\rho^{s} = 1.25 \times 10^{3}$ is initially at rest placed on the $y$-axis at vertical position~$r_{y} = 1.0 \times 10^{-2}$ in a closed rectangular box of height~$H = 6.0 \times 10^{-2}$ and width~$W = 2.0 \times 10^{-2}$, cf. Figure~\ref{fig:example_fallingcylinder}. The remainder of the box is occupied by a Newtonian fluid with density~$\rho^{f} = 1.0 \times 10^{3}$ and kinematic viscosity~$\nu^{f} = 1.0 \times 10^{-5}$. A gravitational acceleration of magnitude~$\qty|\vectorbold{g}| = 9.81$ shall act on both the fluid and solid field in negative $y$-direction. Following~\cite{Hashemi2012} this is modeled considering the buoyancy effect, i.e., body force~$\vectorbold{b}^{s} = \flatfrac{ \qty( \rho^{s} - \rho^{f} ) }{\rho^{s}} \vectorbold{g}$ is acting on the solid field while no body force~$\vectorbold{b}^{f} = 0.0$ is applied on the fluid field (each per unit mass). It is worth noting that, naturally, it would also be possible to directly set the gravitational acceleration for the fluid and solid field. For validation, the results obtained with the proposed formulation are compared to~\cite{Hashemi2012} also applying SPH to discretize the fluid and the solid field.

% begin figure
% end figure

For the fluid phase, an artificial speed of sound~$c = 0.5$ is chosen, resulting in a reference pressure~$p_{0} = 250.0$ of the weakly compressible model. The background pressure~$p_{b}$ of the transport velocity formulation is set equal to the reference pressure~$p_{0}$. The walls of the box are modeled using boundary particles. The problem is solved for different values of the initial particle spacing~$\Delta{}x$, cf. Section~\ref{subsec:numex_spatialdiscretcyl}, for times~$t \in \qty[0, 0.8]$ with time step size~$\Delta{}t$ obeying respective conditions~\eqref{eq:nummeth_timint_timestepcond}.

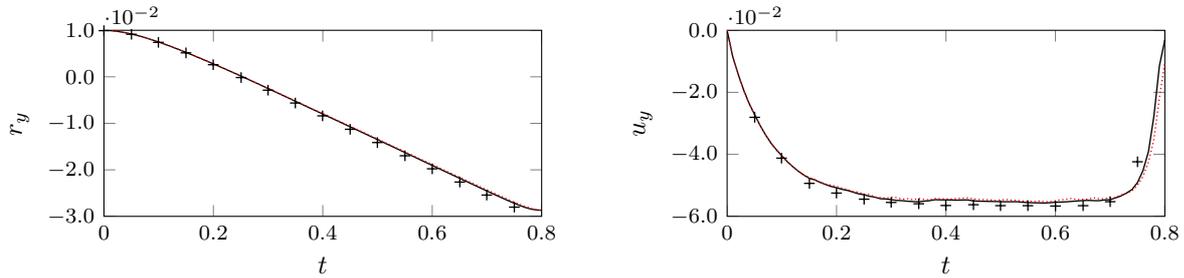
\begin{figure}[htbp]%
\centering
% begin subfigure
\subfigure %[caption of subfigure]
{
\begin{tikzpicture}[trim axis left,trim axis right]
\begin{axis}
[
width=0.35\textwidth,
height=0.15\textwidth,
xmin=0.0, xmax=0.8,
ymin=-0.03, ymax=0.01,
%scaled ticks=false,
every x tick scale label/.style={at={(1,0)},above right,inner sep=0pt,xshift=0.2em},
every y tick scale label/.style={at={(0,1)},above right,inner sep=0pt,yshift=0.2em},
yticklabel style={/pgf/number format/precision=1, /pgf/number format/fixed, /pgf/number format/fixed zerofill},
xlabel={$t$},
ylabel={$r_{y}$},
]
% begin plot
\addplot [color=black,solid,line width=0.5pt] table [x expr={(\thisrow{"Time"})}, y expr={(\thisrow{"position (1) (stats)"})}, col sep=comma] {data/fallingcyl_data_h0_1.csv};
\addplot [color=red,densely dotted,line width=0.5pt] table [x expr={(\thisrow{"Time"})}, y expr={(\thisrow{"position (1) (stats)"})}, col sep=comma] {data/fallingcyl_data_h0_2.csv};
\addplot [color=black,only marks,mark=+] table [x expr={(\thisrow{"time"})}, y expr={\thisrow{"ypos"}}, col sep=comma] {data/fallingcyl_hashemi2012_data_pos.csv};
% end plot
\end{axis}
\end{tikzpicture}
}
% end subfigure
\hspace{0.12\textwidth}
% begin subfigure
\subfigure %[caption of subfigure]
{
\begin{tikzpicture}[trim axis left,trim axis right]
\begin{axis}
[
width=0.35\textwidth,
height=0.15\textwidth,
xmin=0.0, xmax=0.8,
ymin=-0.06, ymax=0.0,
%scaled ticks=false,
every x tick scale label/.style={at={(1,0)},above right,inner sep=0pt,xshift=0.2em},
every y tick scale label/.style={at={(0,1)},above right,inner sep=0pt,yshift=0.2em},
yticklabel style={/pgf/number format/precision=1, /pgf/number format/fixed, /pgf/number format/fixed zerofill},
xlabel={$t$},
ylabel={$u_{y}$},
]
% begin plot
\addplot [color=black,solid,line width=0.5pt] table [x expr={(\thisrow{"Time"})}, y expr={(\thisrow{"velocity (1) (stats)"})}, col sep=comma] {data/fallingcyl_data_h0_1.csv};
\addplot [color=red,densely dotted,line width=0.5pt] table [x expr={(\thisrow{"Time"})}, y expr={(\thisrow{"velocity (1) (stats)"})}, col sep=comma] {data/fallingcyl_data_h0_2.csv};
\addplot [color=black,only marks,mark=+] table [x expr={(\thisrow{"time"})}, y expr={(\thisrow{"yvel"})}, col sep=comma] {data/fallingcyl_hashemi2012_data_vel.csv};
% end plot
\end{axis}
\end{tikzpicture}
}
% end subfigure
\caption{A rigid circular disk falling in a fluid column: vertical position~$r_{y}$ and vertical velocity~$u_{y}$ of the center of the circular disk in the fluid column computed with the proposed formulation and an initial particle spacing of $\Delta{}x = 2.0 \times 10^{-4}$ (red dashed line) and $\Delta{}x = 1.0 \times 10^{-4}$ (black solid line) compared to the reference solution~\cite{Hashemi2012} (crosses).}
\label{fig:example_fallingcylinder_pos_vel}
\end{figure}

The obtained vertical velocity and horizontal position of the center of the circular disk in the box over time~$t$ are displayed in Figure~\ref{fig:example_fallingcylinder_pos_vel} for two different values of the initial particle spacing~$\Delta{}x$ compared to the reference solution~\cite{Hashemi2012}. The results obtained with different initial particle spacing~$\Delta{}x$ show only minor differences. The terminal velocity of the rigid circular disk is slightly smaller than given in the reference solution~\cite{Hashemi2012}. It shall be noted that, in contrast to~\cite{Hashemi2012}, contact of the rigid circular disk and the wall of the box is explicitly considered. Consequently, the rigid circular disk comes at rest when approaching the bottom wall of the box.

% begin figure
% end figure

%%
\subsection{Melting and solidification of powder grains in a melt pool} \label{subsec:numex_melting}

In metal powder bed fusion additive manufacturing (PBFAM), structural components are created utilizing a laser or electron beam to melt and fuse metal powder, layer per layer, to form the final part. PBFAM has the potential to enable new paradigms of product design, manufacturing and supply chains. However, due to the complexity of PBFAM processes, the interplay of process parameters is not completely understood, creating the need for further research, amongst others in the field of computational melt pool modeling~\cite{Meier2017b,Furstenau2020}. For this purpose, an SPH formulation for thermo-capillary phase transition problems with a focus on metal PBFAM melt pool modeling has recently been proposed~\cite{Meier2020}. For simplicity, this and other state-of-the-art approaches in the field consider powder particles that are spatially fixed. In the real physical process, however, it is observed that, depending on the processing conditions, melt evaporation and thereby induced vapor and gas flows in the build chamber may result in powder particle entrainment and ejection, i.e., a considerable degree of material re-distribution during the melting process. On the one hand, this effect considerably affects process stability and mechanisms of defect creation, on the other hand, it can not be represented by state-of-the-art approaches restricted to immobile powder particles~\cite{Meier2020}.

The purpose of this example is not to study PBFAM in detail but to showcase the general applicability of the proposed formulation to capture the dynamics of mobile powder particles undergoing temperature-induced phase transitions, i.e., melting and solidification, while being exposed to a gas flow. To this end, surface tension and wetting effects as well as the influence of evaporation-induced recoil pressure, as discussed in~\cite{Meier2020}, are neglected. The focus is set on the investigation of highly dynamic motion and interaction of powder grains with each other, the liquid melt phase and a surrounding gas phase, undergoing reversible phase transitions, i.e., melting and solidification. This example is solely intended to demonstrate the model capabilities for this type of application, while keeping the overall example simple in this method-focused contribution. For this reason, non-physical parameter values and boundary conditions are chosen in the following.

\begin{figure}[htbp]
\centering
% begin subfigure
\subfigure [time $t = 0.0$]
{
\includegraphics[width=0.31\textwidth]{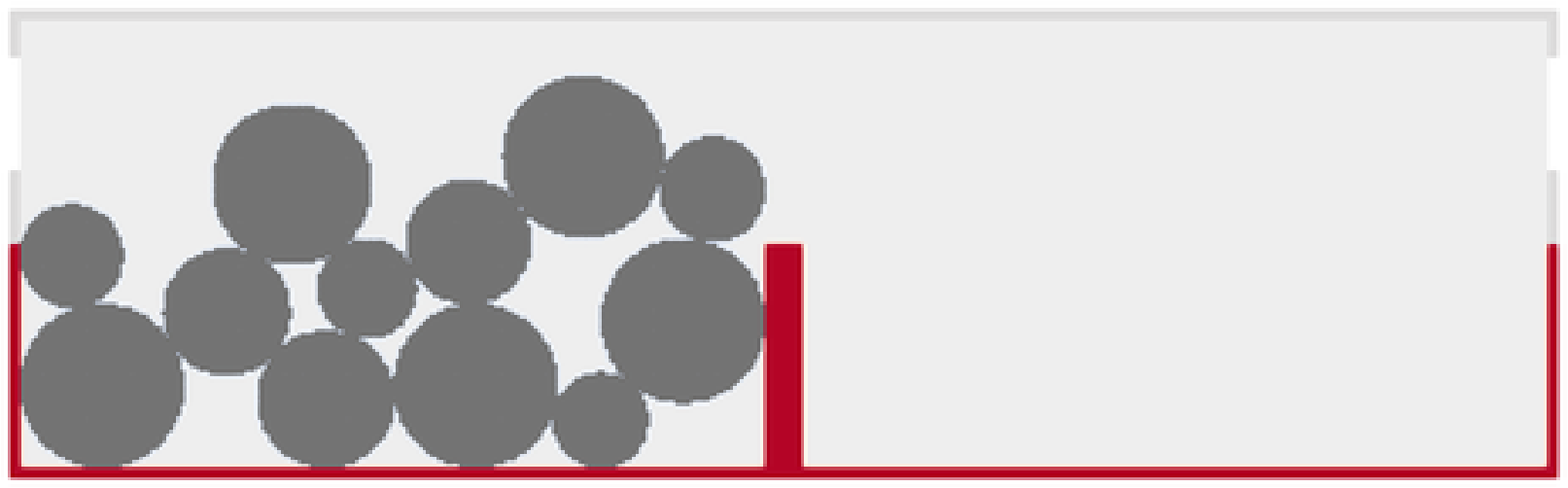}
\label{fig:example_melting_temp_init}
}
% end subfigure
% begin subfigure
\subfigure [time $t = 0.125$]
{
\includegraphics[width=0.31\textwidth]{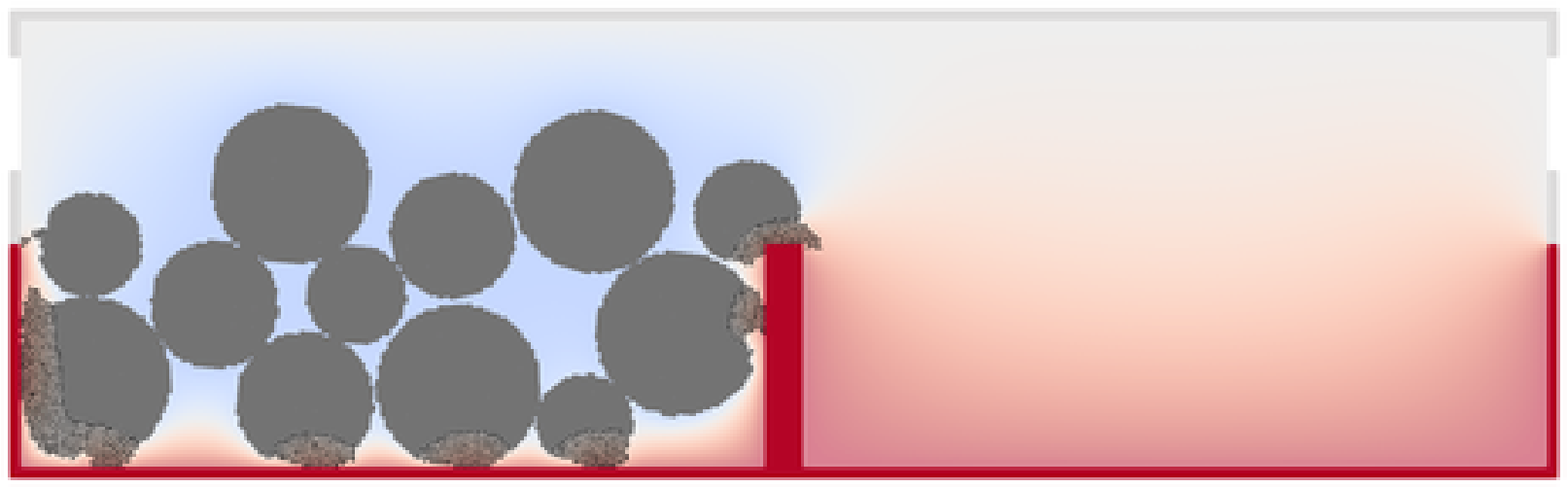}
}
% end subfigure
% begin subfigure
\subfigure [time $t = 0.25$]
{
\includegraphics[width=0.31\textwidth]{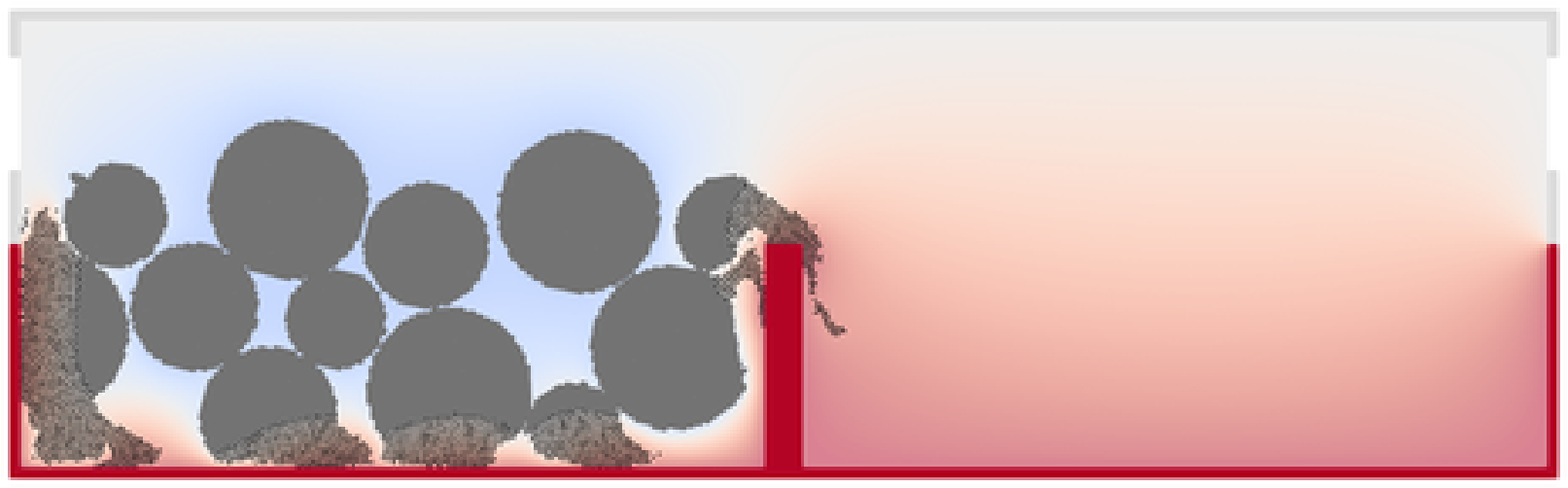}
}
% end subfigure
\\
% begin subfigure
\subfigure [time $t = 0.28125$]
{
\includegraphics[width=0.31\textwidth]{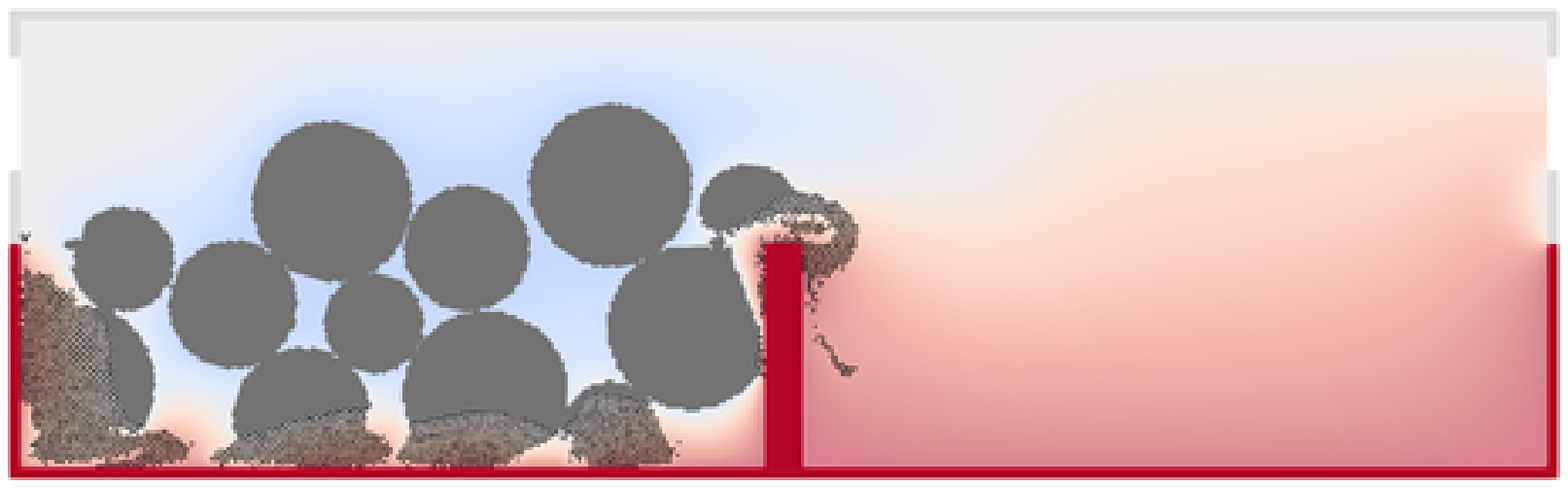}
}
% end subfigure
% begin subfigure
\subfigure [time $t = 0.3125$]
{
\includegraphics[width=0.31\textwidth]{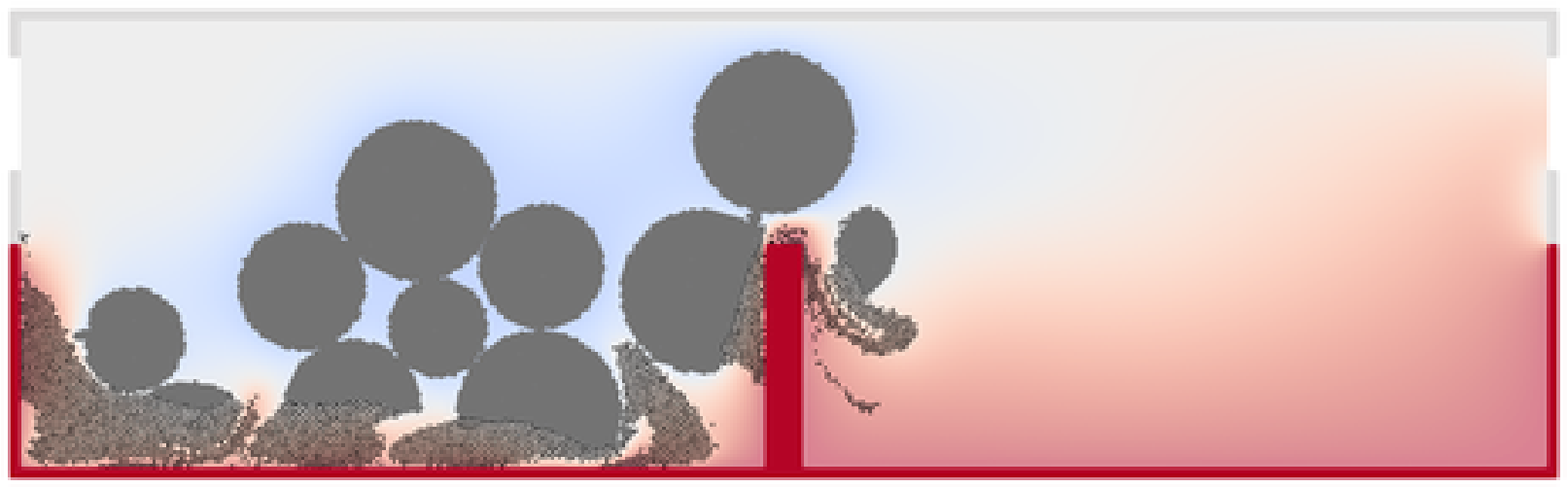}
}
% end subfigure
% begin subfigure
\subfigure [time $t = 0.34375$]
{
\includegraphics[width=0.31\textwidth]{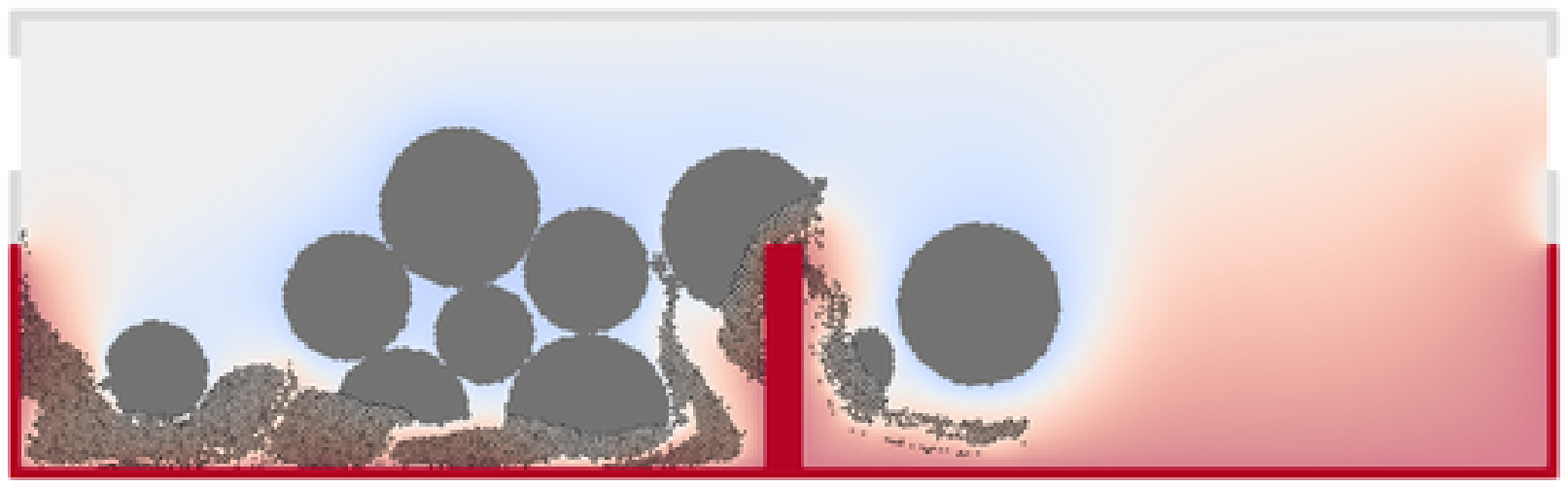}
}
% end subfigure
\\
% begin subfigure
\subfigure [time $t = 0.375$]
{
\includegraphics[width=0.31\textwidth]{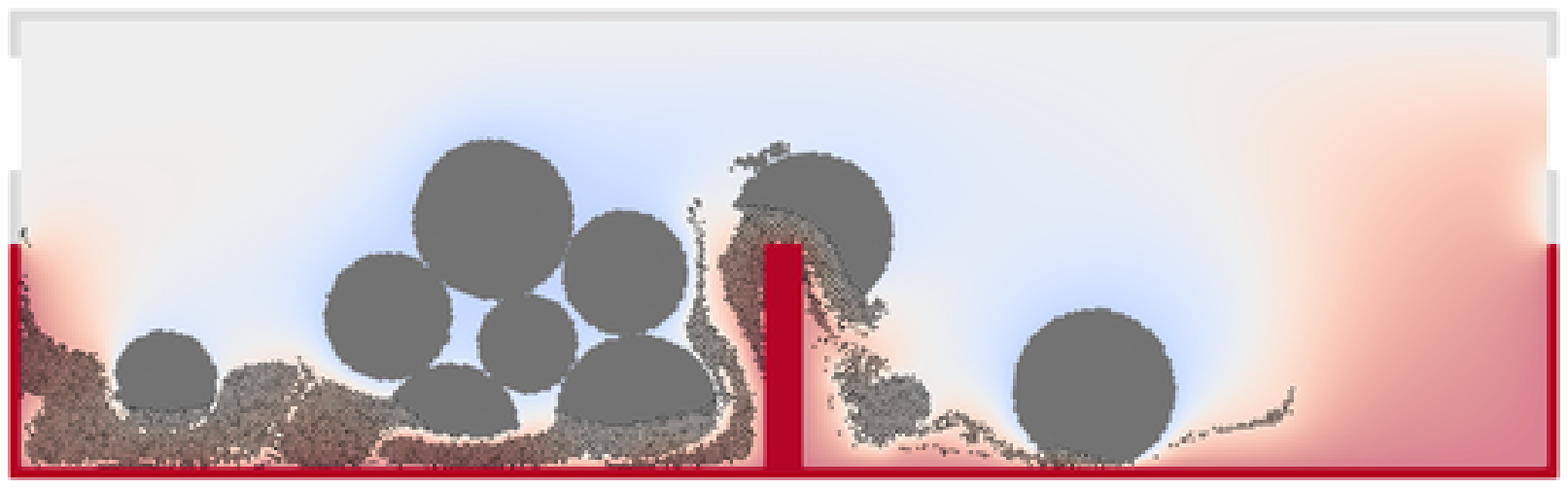}
}
% end subfigure
% begin subfigure
\subfigure [time $t = 0.4375$]
{
\includegraphics[width=0.31\textwidth]{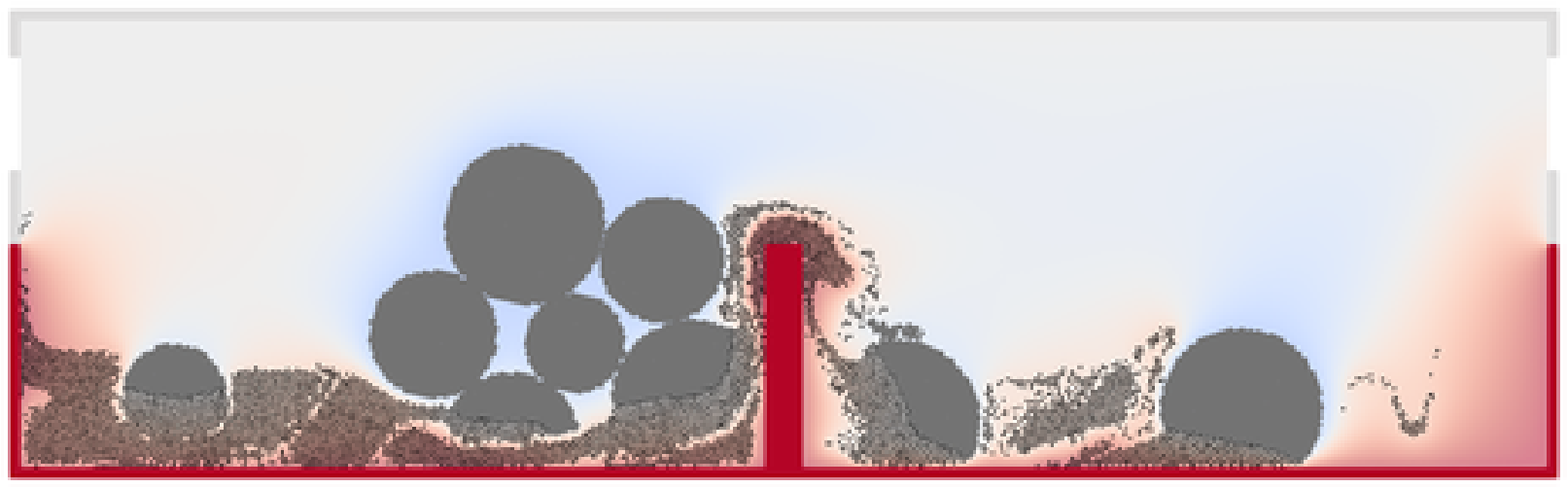}
}
% end subfigure
% begin subfigure
\subfigure [time $t = 0.5$]
{
\includegraphics[width=0.31\textwidth]{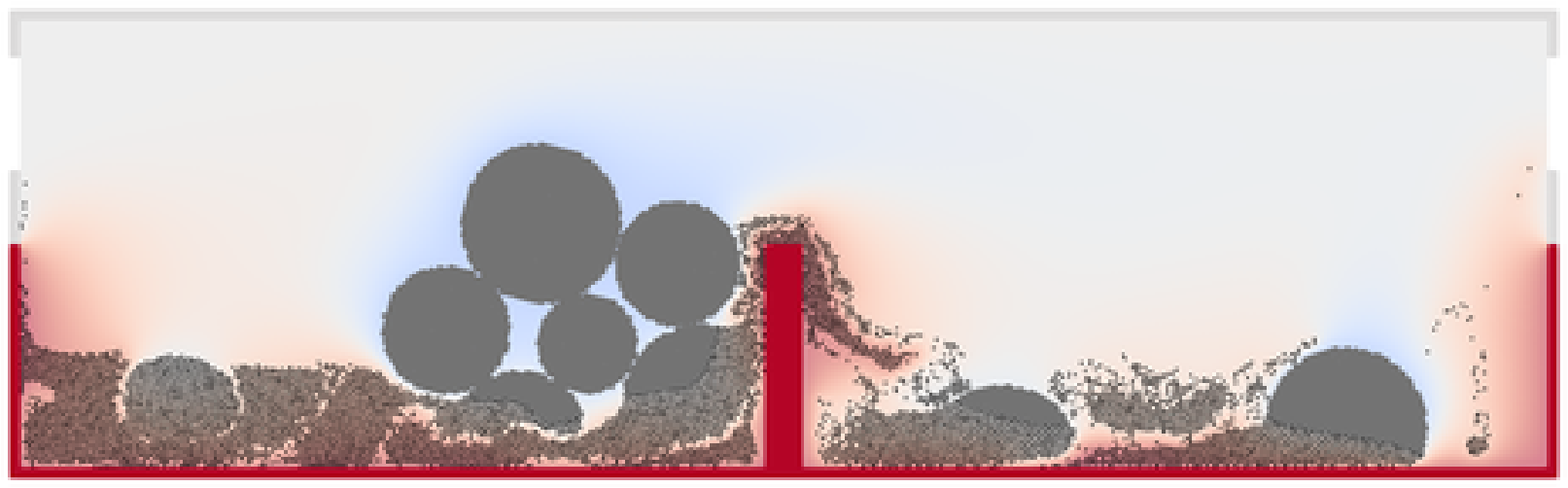}
}
% end subfigure
\\
% begin subfigure
\subfigure [time $t = 0.53125$]
{
\includegraphics[width=0.31\textwidth]{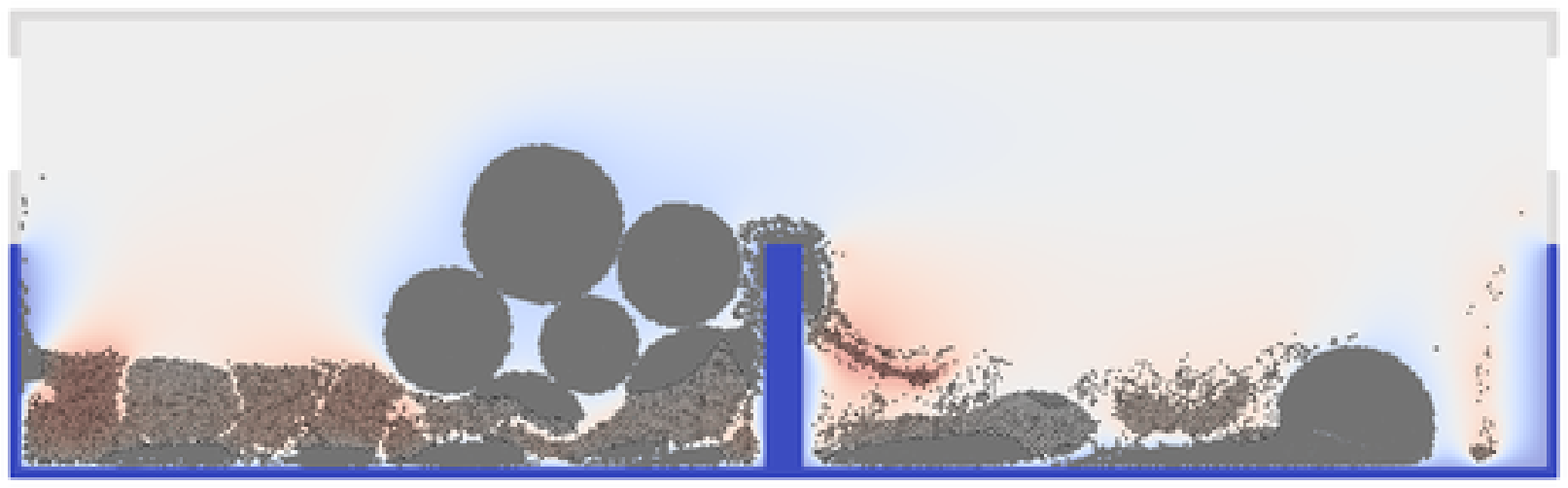}
}
% end subfigure
% begin subfigure
\subfigure [time $t = 0.5625$]
{
\includegraphics[width=0.31\textwidth]{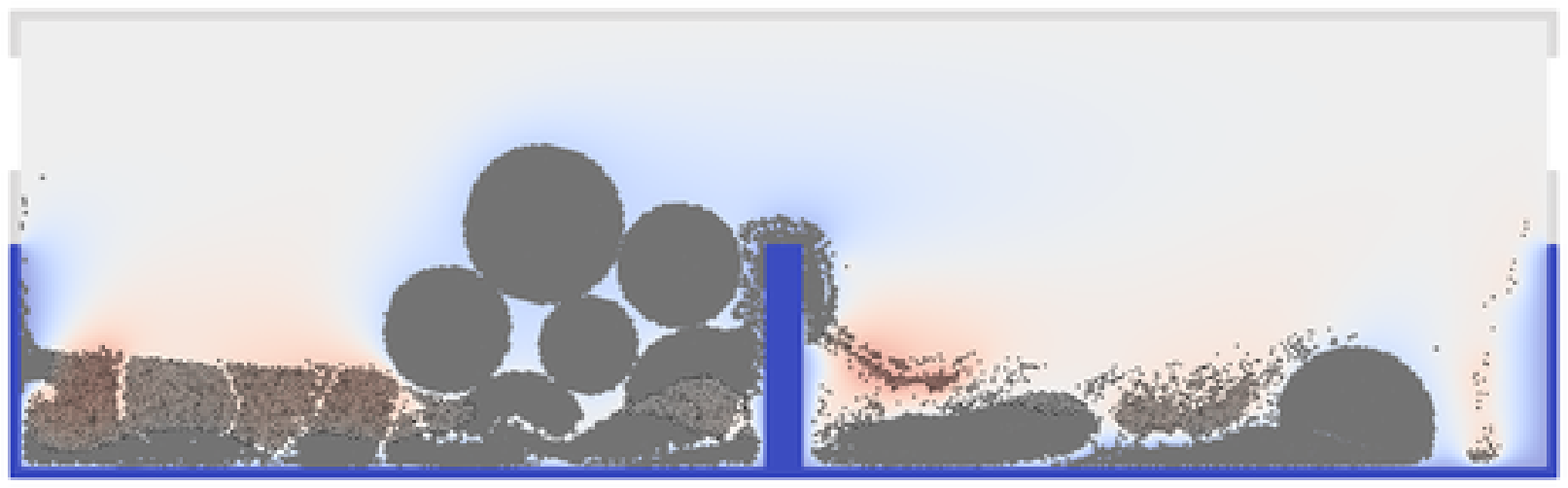}
}
% end subfigure
% begin subfigure
\subfigure [time $t = 0.59375$]
{
\includegraphics[width=0.31\textwidth]{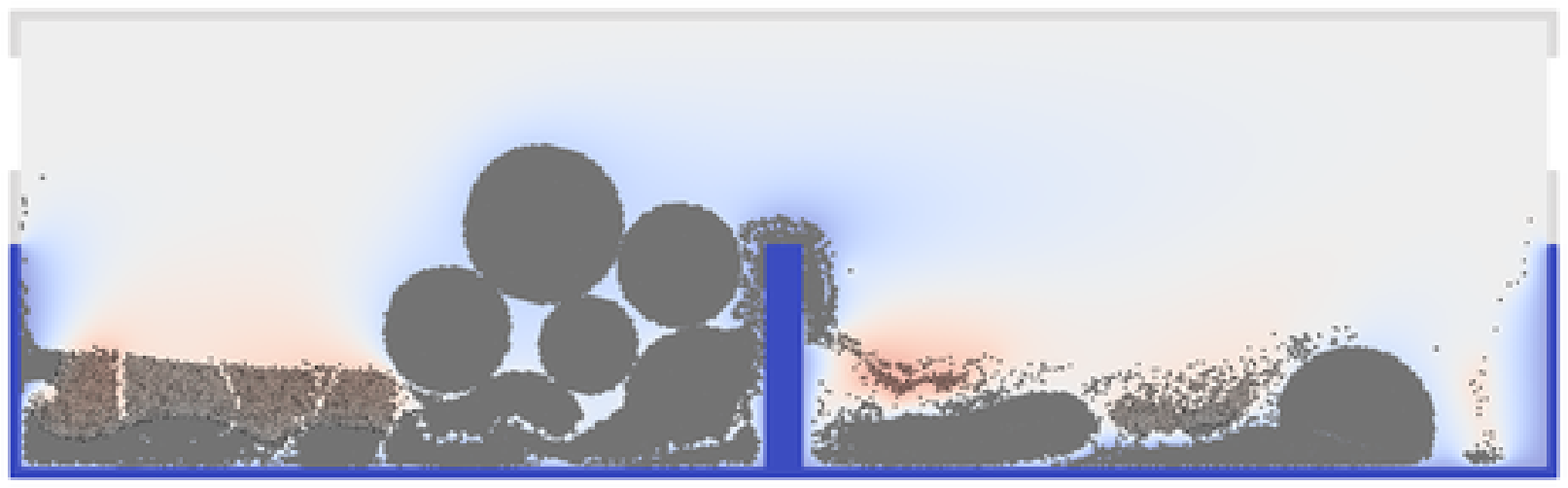}
}
% end subfigure
\caption{Melting and solidification of powder grains in a melt pool: time series of the obtained results with temperature field ranging from $0.0$ (blue) to $100.0$ (red).}
\label{fig:example_melting_temp}
\end{figure}

\begin{figure}[htbp]
\centering
% begin subfigure
\subfigure [time $t = 0.0$]
{
\includegraphics[width=0.31\textwidth]{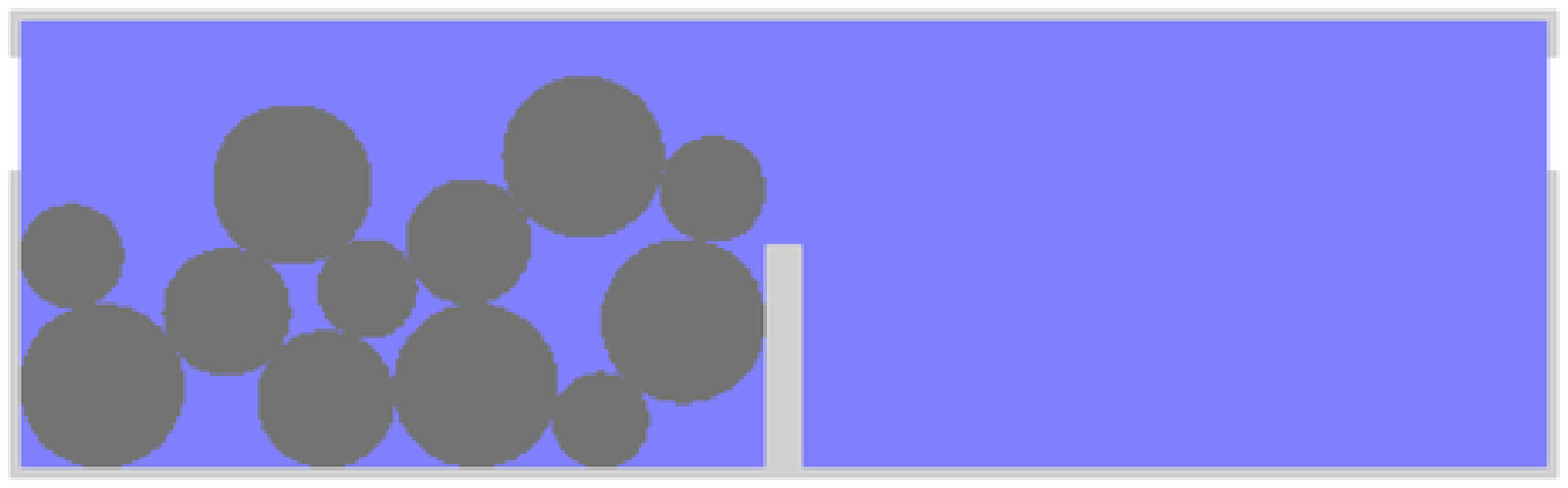}
\label{fig:example_melting_vel_init}
}
% end subfigure
% begin subfigure
\subfigure [time $t = 0.125$]
{
\includegraphics[width=0.31\textwidth]{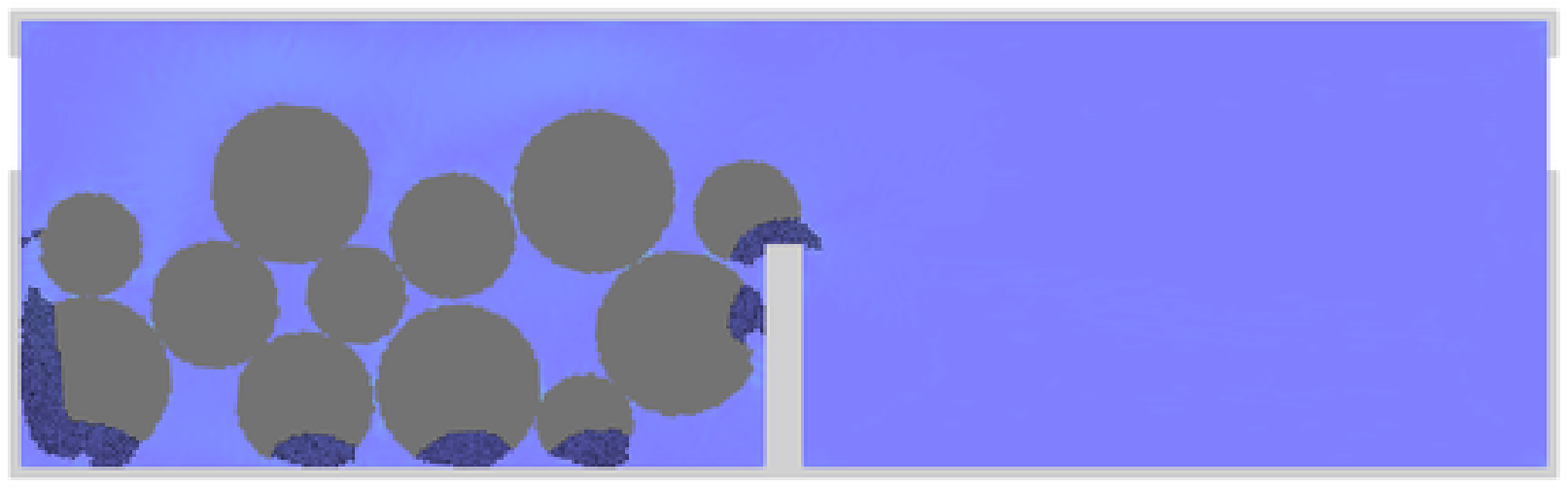}
}
% end subfigure
% begin subfigure
\subfigure [time $t = 0.25$]
{
\includegraphics[width=0.31\textwidth]{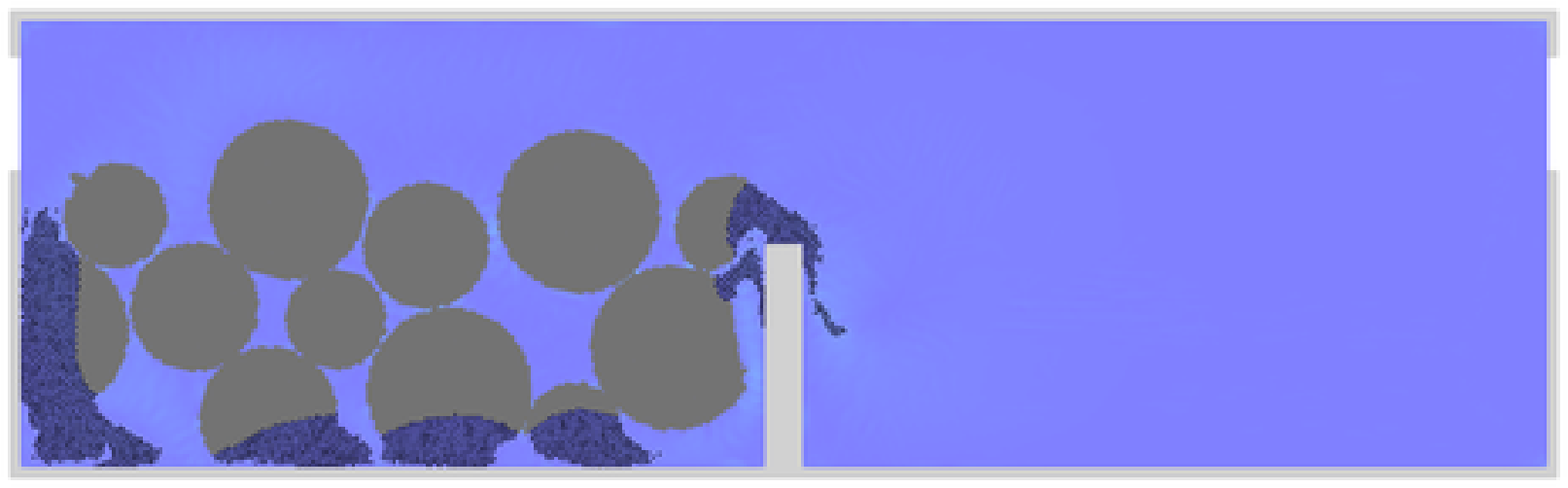}
}
% end subfigure
\\
% begin subfigure
\subfigure [time $t = 0.28125$]
{
\includegraphics[width=0.31\textwidth]{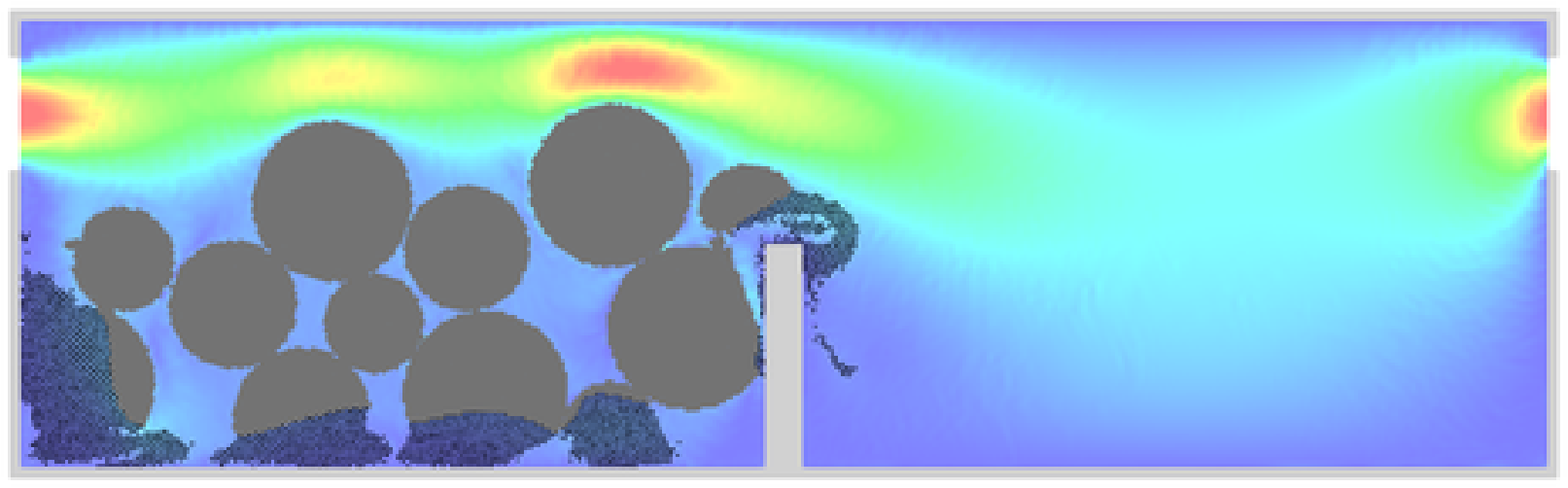}
}
% end subfigure
% begin subfigure
\subfigure [time $t = 0.3125$]
{
\includegraphics[width=0.31\textwidth]{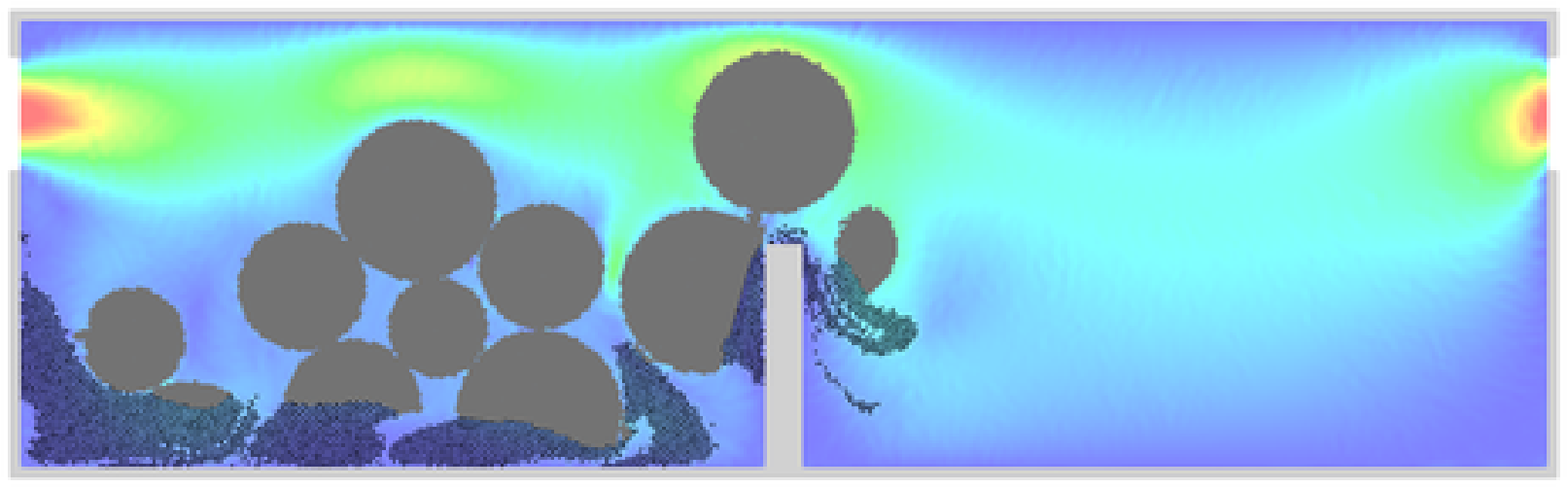}
}
% end subfigure
% begin subfigure
\subfigure [time $t = 0.34375$]
{
\includegraphics[width=0.31\textwidth]{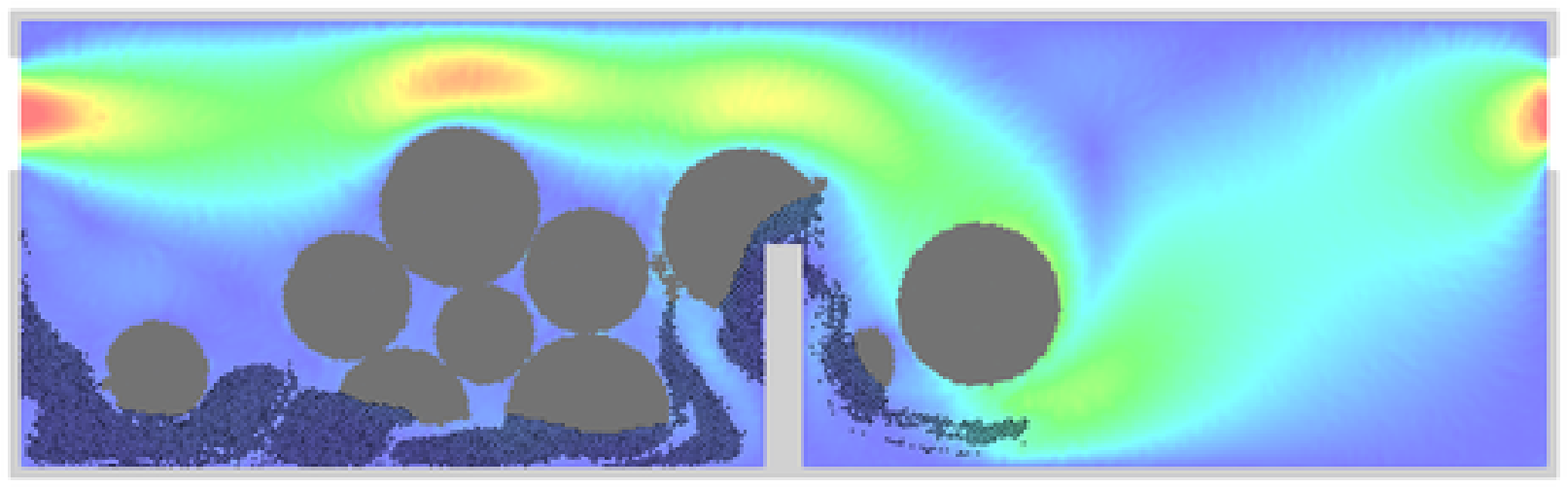}
}
% end subfigure
\\
% begin subfigure
\subfigure [time $t = 0.375$]
{
\includegraphics[width=0.31\textwidth]{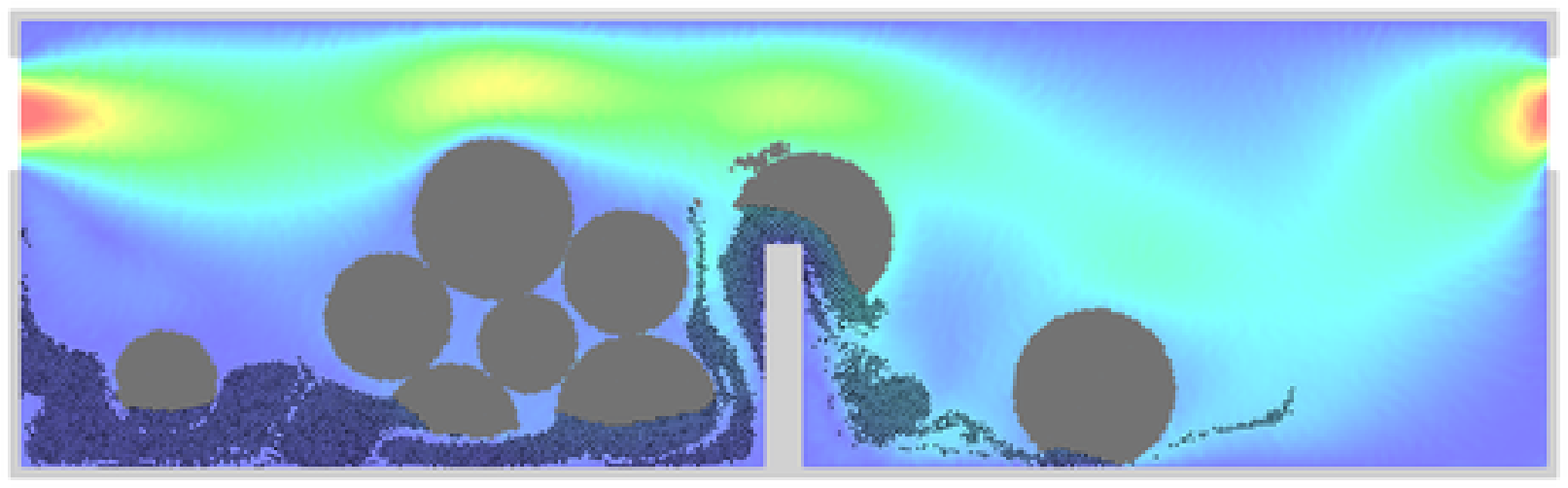}
}
% end subfigure
% begin subfigure
\subfigure [time $t = 0.4375$]
{
\includegraphics[width=0.31\textwidth]{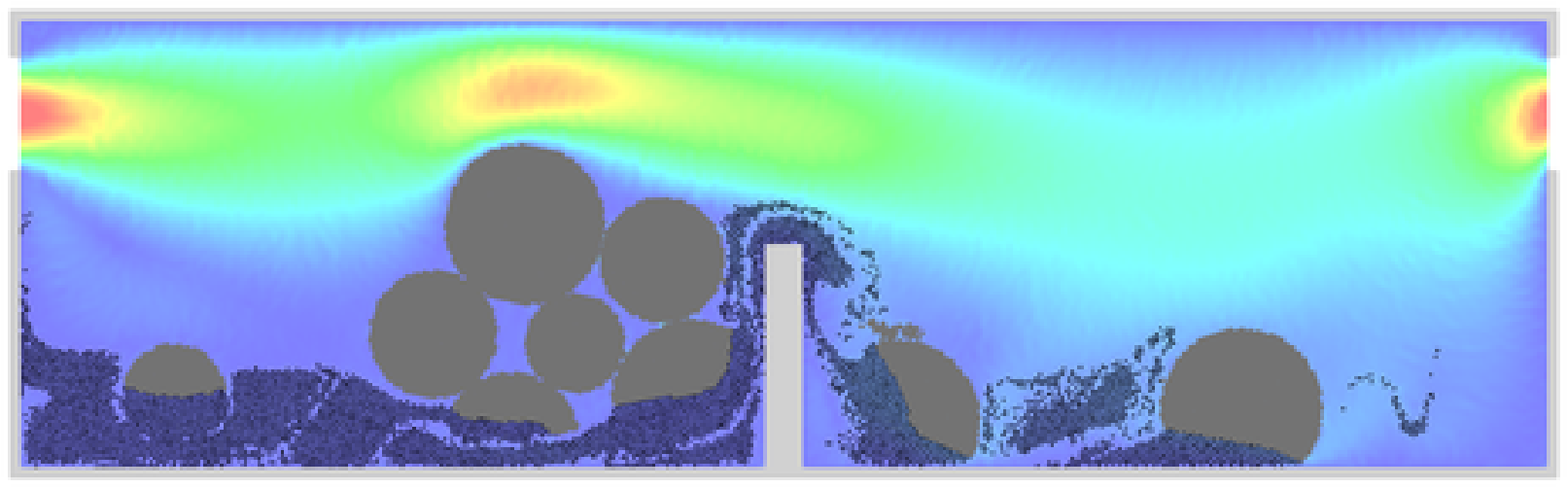}
}
% end subfigure
% begin subfigure
\subfigure [time $t = 0.5$]
{
\includegraphics[width=0.31\textwidth]{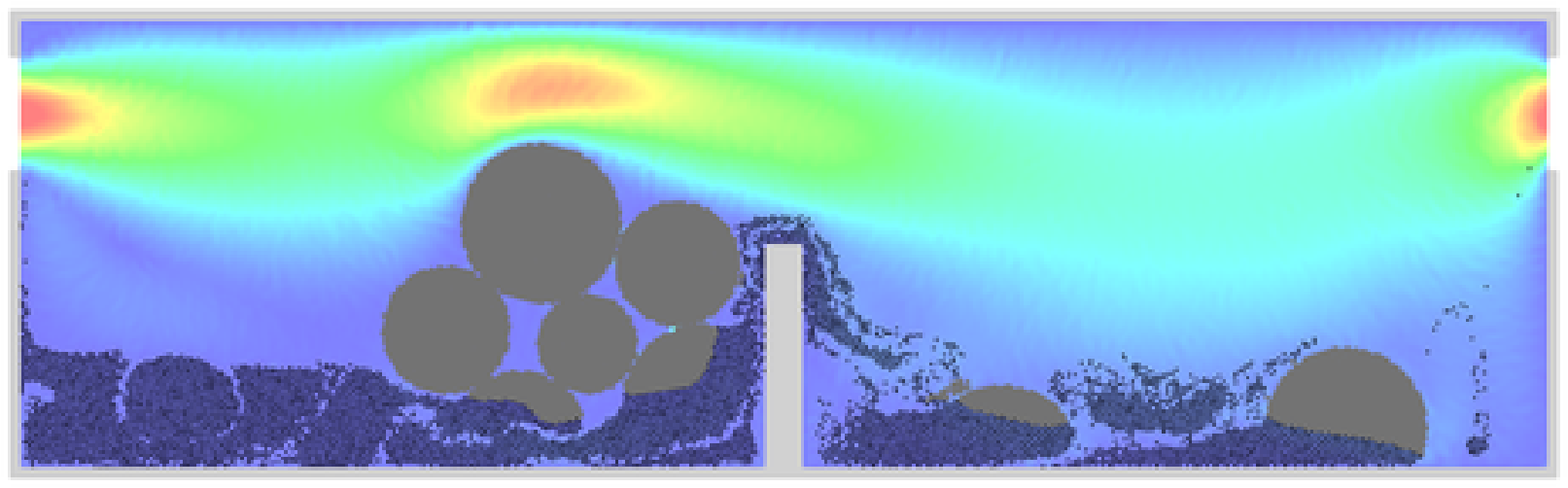}
}
% end subfigure
\\
% begin subfigure
\subfigure [time $t = 0.53125$]
{
\includegraphics[width=0.31\textwidth]{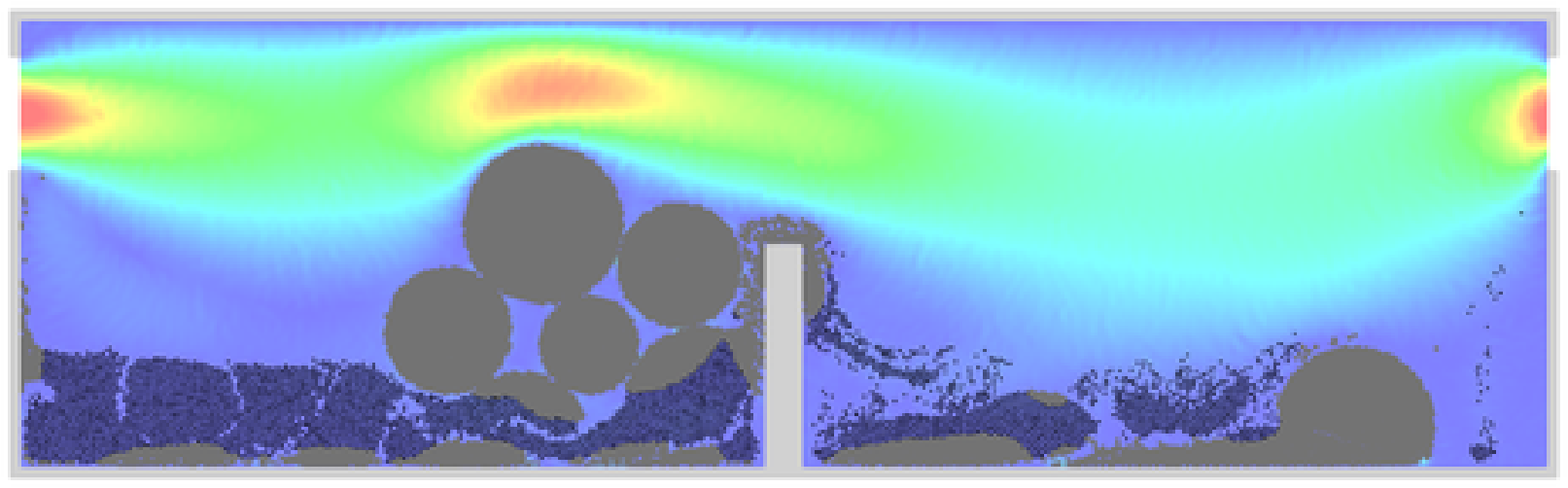}
}
% end subfigure
% begin subfigure
\subfigure [time $t = 0.5625$]
{
\includegraphics[width=0.31\textwidth]{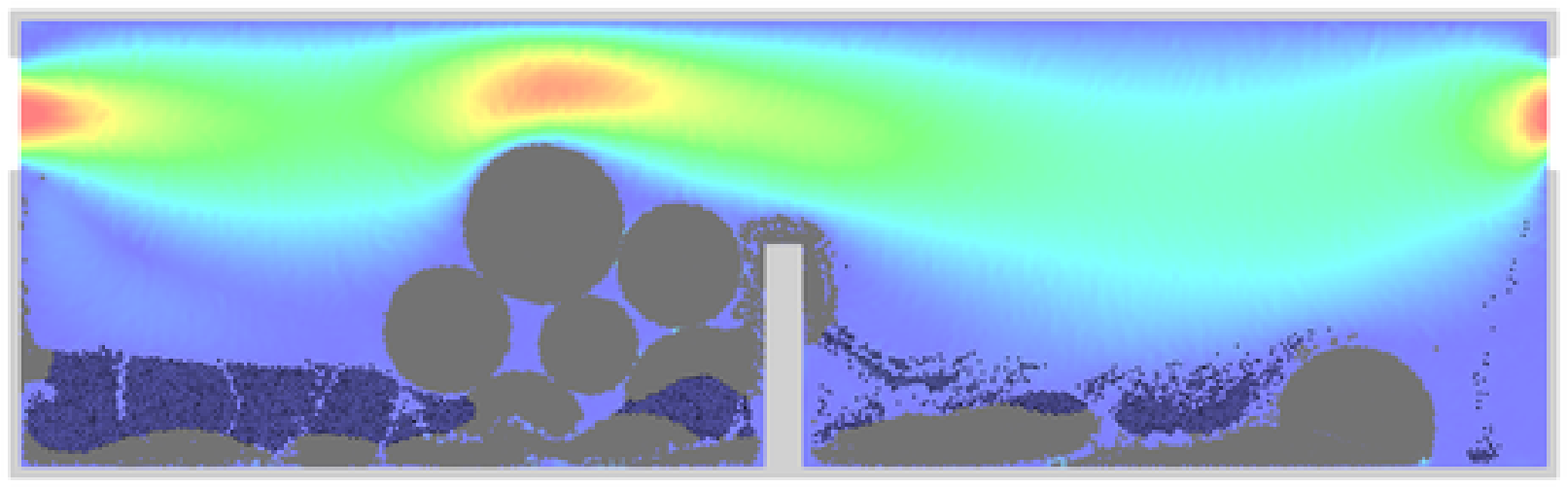}
}
% end subfigure
% begin subfigure
\subfigure [time $t = 0.59375$]
{
\includegraphics[width=0.31\textwidth]{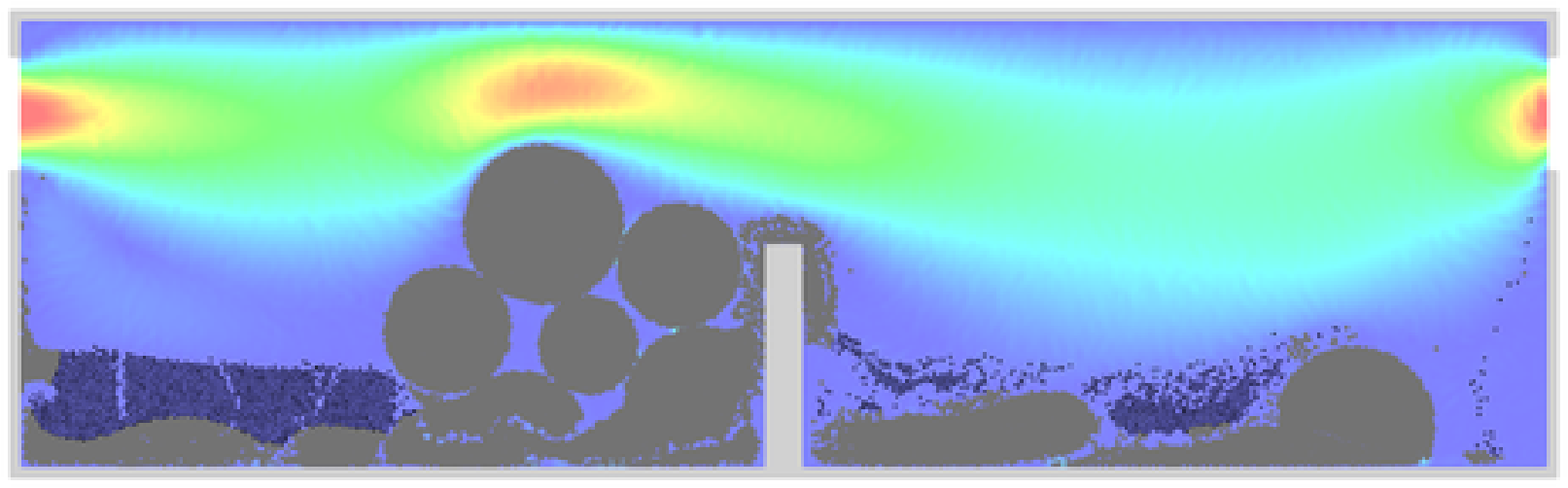}
}
% end subfigure
\caption{Melting and solidification of powder grains in a melt pool: time series of the obtained results with magnitude of the velocity field ranging from $0.0$ (blue) to $600.0$ (red).}
\label{fig:example_melting_vel}
\end{figure}

A rectangular box is composed of two chambers, each with width $20.0$ and height $12.0$, that are connected by an opening spanning the upper half of the box. An inlet and outlet of width $3.0$ are located at the top left and top right end of the box. Powder grains of a solid metal phase (density~$\rho^{s} = 1.0$, heat capacity~$c_{p}^{s} = 1.0$, thermal conductivity~$\kappa^{s} = 10.0$) with diameters between $2.5$ and $4.4$ are placed initially at rest inside the left chamber of the box. The initial positions of the powder grains can, e.g., be obtained in a pre-processing step based on the dicrete element method (DEM) and a cohesive powder model~\cite{Meier2019a,Meier2019b}. The remainder of the box is initially filled with a gas phase (Newtonian fluid, density~$\rho^{g} = 0.1$, kinematic viscosity~$\nu^{g} = 100.0$, heat capacity~$c_{p}^{g} = 0.01$, thermal conductivity~$\kappa^{g} = 0.1$). The temperature is initialized to $T^{s}_{0} = 25.0$ within the solid metal phase and to $T^{g}_{0} = 50.0$ within the gas phase. In the upper half of the box walls the temperature is fixed to $\hat{T} = 50.0$ at all times. In the lower half of the box walls the temperature is set to $\hat{T} = 100.0$ until time $t \leq 0.5$, and to $\hat{T} = 0.0$ for time $t > 0.5$. Refer to Figures~\ref{fig:example_melting_temp_init} and~\ref{fig:example_melting_vel_init} for an illustration of the initial configuration. Reversible phase transitions between solid metal phase and liquid metal phase (Newtonian fluid, density~$\rho^{l} = 1.0$, kinematic viscosity~$\nu^{l} = 100.0$, heat capacity~$c_{p}^{l} = 1.0$, thermal conductivity~$\kappa^{l} = 10.0$) is assumed to occur at a transition temperature of $T_{t} = 50.0$. With the goal to evoke drag forces acting on the powder grains, for times $t > 0.25$ a parabolic inflow respectively outflow of the gas phase with mean velocity $420.0$ is prescribed at the inlet and outlet of the box. A gravitational acceleration of magnitude~$\qty|\vectorbold{g}| = 1.0 \times 10^{4}$ is acting downwards, set as body force (per unit mass) of all involved phases.

For both fluid phases (liquid metal and gas), the reference pressure of the weakly compressible model is set to $p_{0} = 16.0 \times 10^{6}$, and the background pressure~$p_{b}$ of the transport velocity formulation is set equal to the reference pressure~$p_{0}$. The wall of the box is modeled using boundary particles. The inflow and outflow conditions are modeled similar as described in~\cite{Fuchs2020}. The problem is solved with initial particle spacing~$\Delta{}x = 0.1$ for times~$t \in \qty[0, 0.75]$ with a time step size of~$\Delta{}t = 0.625 \times 10^{-5}$ based on conditions~\eqref{eq:nummeth_timint_timestepcond}.

A time series of illustrations of the obtained results is given in Figures~\ref{fig:example_melting_temp} and~\ref{fig:example_melting_vel}. The solid metal phase is visualized in grey color. The particles discretizing the liquid metal phase are displayed in black color. In the background, the temperature respectively velocity field of the combined liquid metal and gas phase are displayed. Thereto, both fields were post-processed applying SPH approximation~\eqref{eq:nummeth_sph_postprocessing} and visualized by a color code. In Figure~\ref{fig:example_melting_temp} additionally the temperature of the walls is shown. First, the powder grains are heated and gradually start melting into liquid metal where in close contact to the hot wall. Eventually, after time $t=0.25$, powder grains are subjected to the gas flow through the box. Some (partially melted) powder grains are swept into the right chamber of the box, where melting after contact with the hot wall continues. A non-smooth and strongly distorted interface topology between liquid metal and gas phase develops, especially in the right chamber of the box, because surface tension and wetting effects are neglected. Finally, with the temperature in the lower half of the box set to $\hat{T} = 0.0$ after time $t = 0.5$, the liquid metal phase is cooled down drastically and eventually resolidifies.

% begin figure
% end figure

% begin figure
% end figure

This example demonstrates that highly dynamic motion of arbitrarily-shaped powder grains as relevant, e.g., for PBFAM melt pool modeling, can be captured along with melting and solidification by the proposed formulation in a robust manner. Consequently, the proposed formulation can be recommended as a useful extension of the SPH formulation for mesoscale melt pool modeling as proposed by~\cite{Meier2017b} allowing for more detailed studies.

\subsection{Gastric disintegration of food boluses} \label{subsec:numex_gastric}

Examination of gastric fluid mechanics plays an important role for modeling digestion of food in the human stomach. The digesta are characterized by a multiphasic nature consisting of fluid (gastric juice and chyme) and solid (food boluses) phases~\cite{Brandstaeter2019}. Intragastric fluid motion is driven by the propagation of so-called antral contraction waves (ACWs), i.e., circular constrictions of the gastric wall due to smooth muscle contractions~\cite{Brandstaeter2018}. The ACWs are initiated at the pacemaker region of the stomach and travel along the greater curvature towards the pylorus both mixing and grinding the digesta. Concurrently, absorption of gastric juice fosters chemical and mechanical breakdown of food boluses into chyme~\cite{Ferrua2011}. At low viscosity, i.e., following intragastric dilution of the digesta with gastric juice, retropulsive jet-like fluid motion between the ACWs can be observed~\cite{Pal2004,Kong2008,Ferrua2014}.

\begin{figure}[htbp]
\centering
% begin subfigure
\subfigure [time $t = 0.0$]
{
\includegraphics[width=0.31\textwidth]{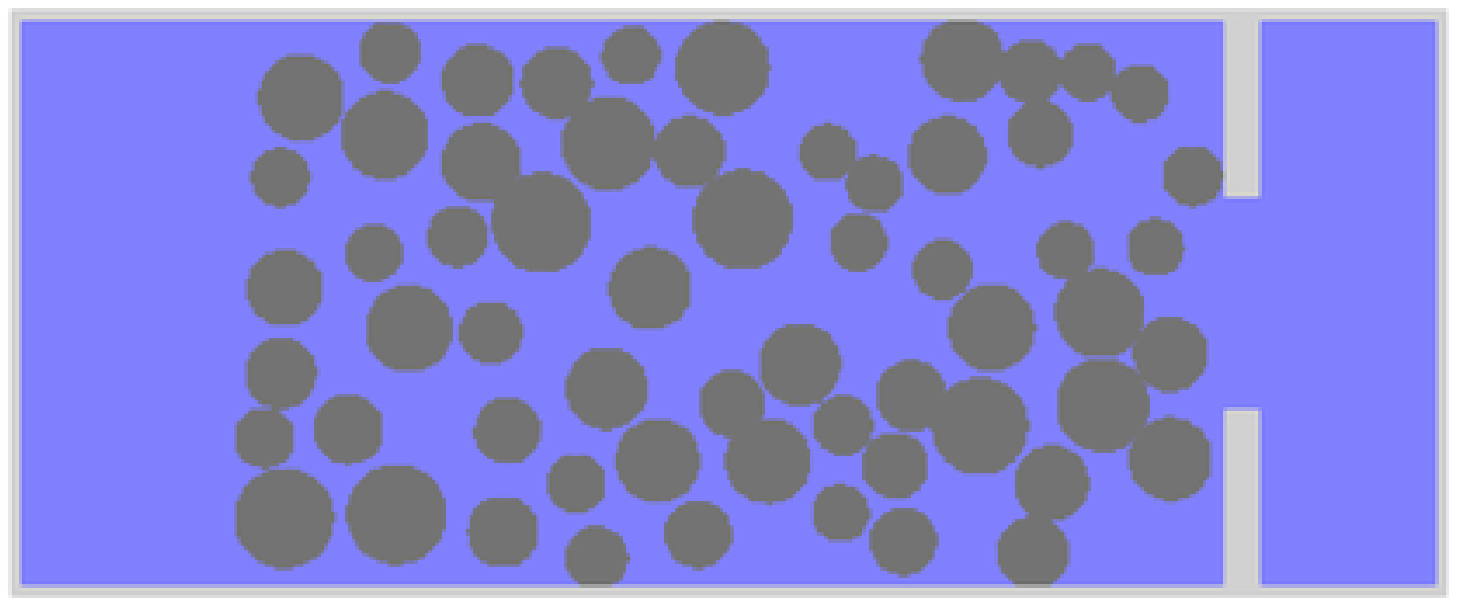}
\label{fig:example_gastric_init}
}
% end subfigure
% begin subfigure
\subfigure [time $t = 0.125$]
{
\includegraphics[width=0.31\textwidth]{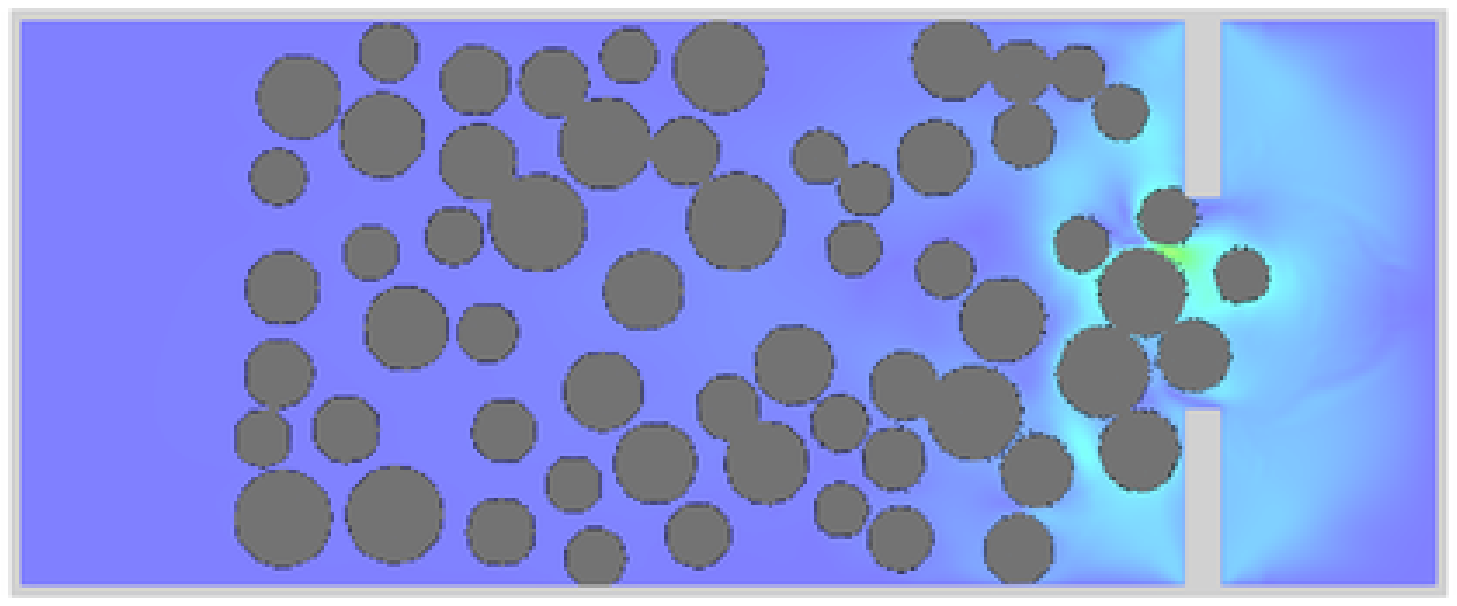}
}
% end subfigure
% begin subfigure
\subfigure [time $t = 0.25$]
{
\includegraphics[width=0.31\textwidth]{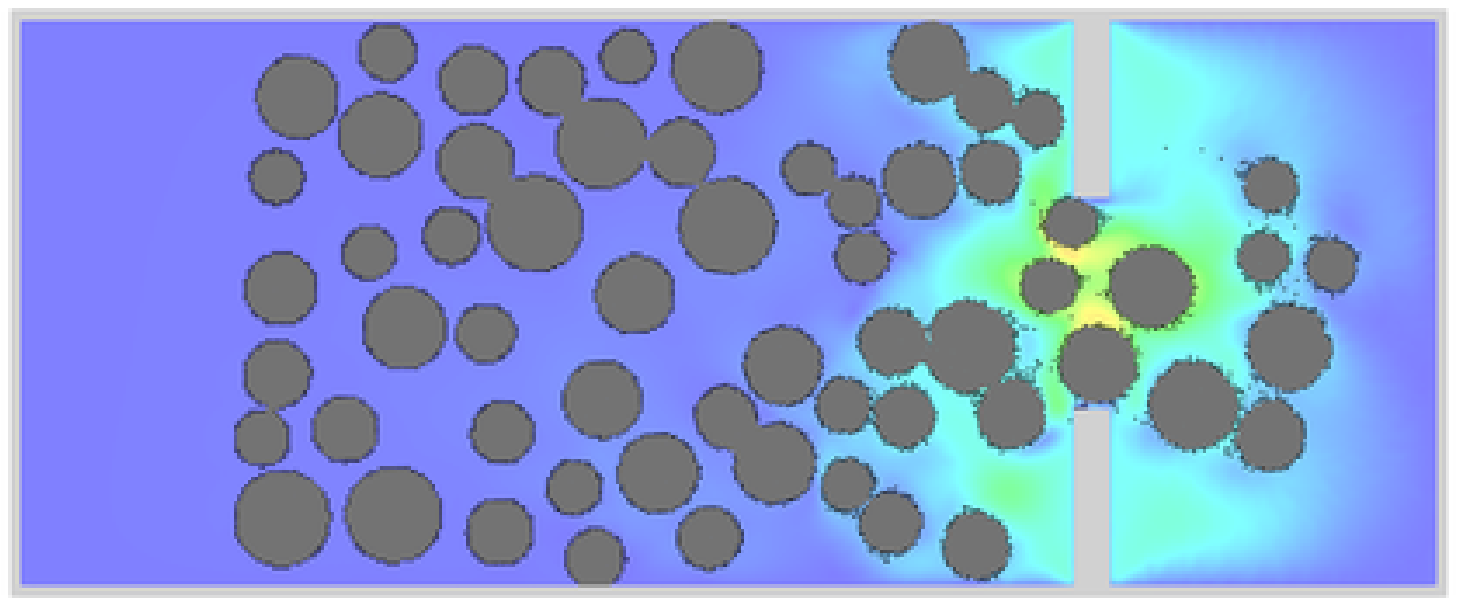}
}
% end subfigure
\\
% begin subfigure
\subfigure [time $t = 0.375$]
{
\includegraphics[width=0.31\textwidth]{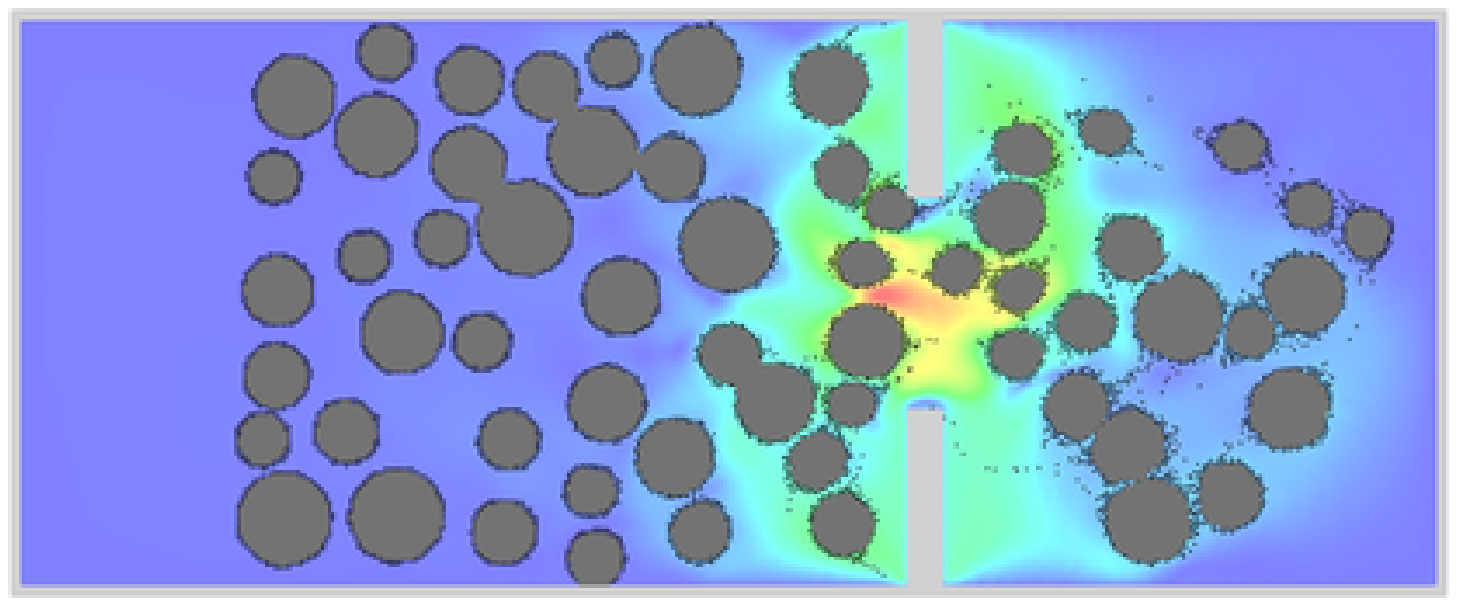}
}
% end subfigure
% begin subfigure
\subfigure [time $t = 0.5$]
{
\includegraphics[width=0.31\textwidth]{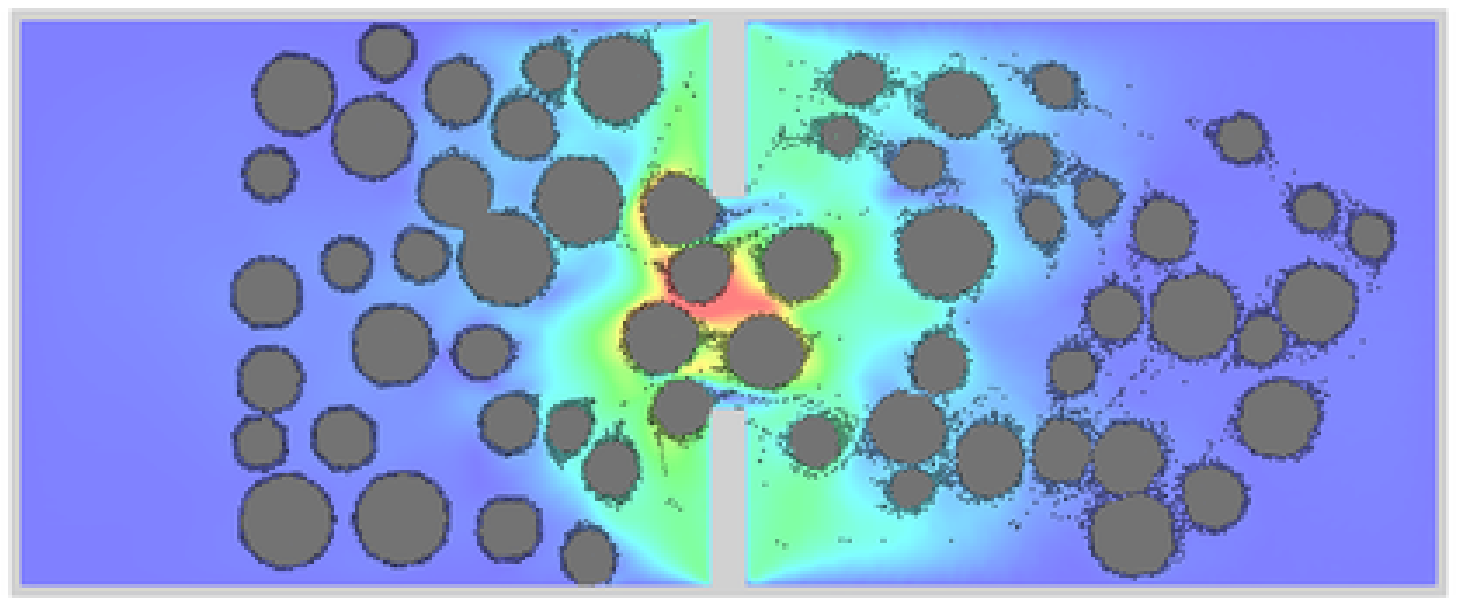}
}
% end subfigure
% begin subfigure
\subfigure [time $t = 0.625$]
{
\includegraphics[width=0.31\textwidth]{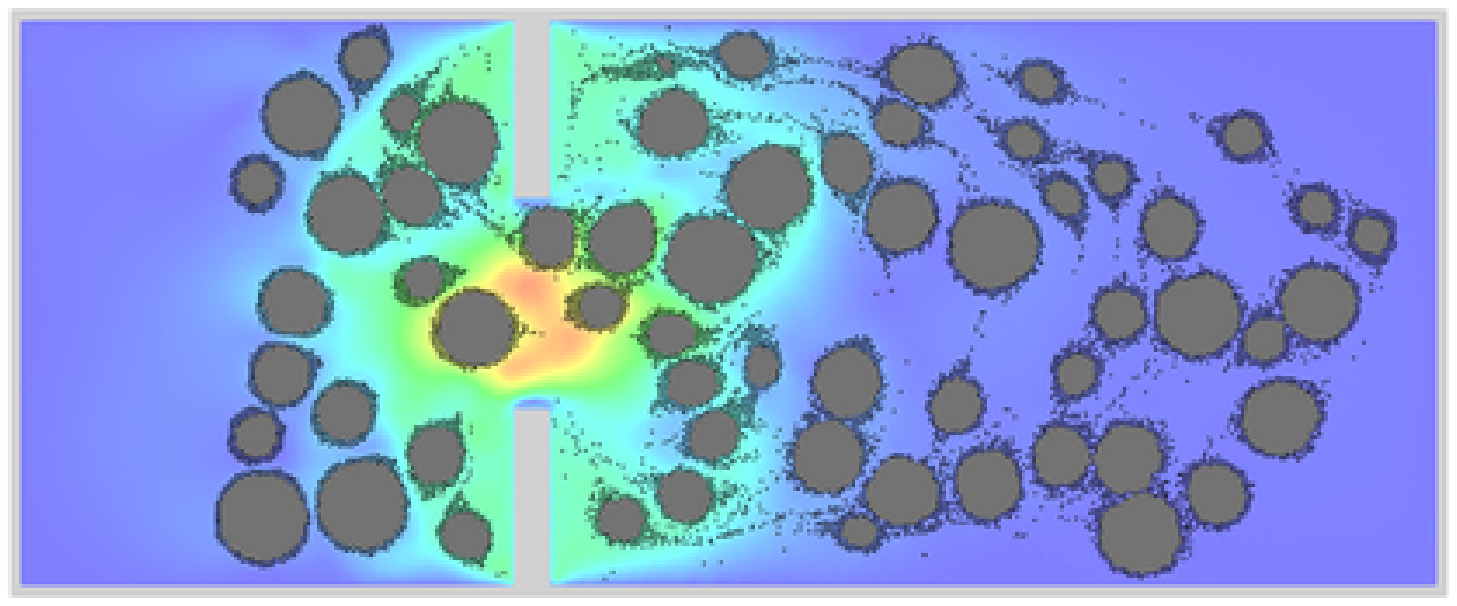}
}
% end subfigure
\\
% begin subfigure
\subfigure [time $t = 0.75$]
{
\includegraphics[width=0.31\textwidth]{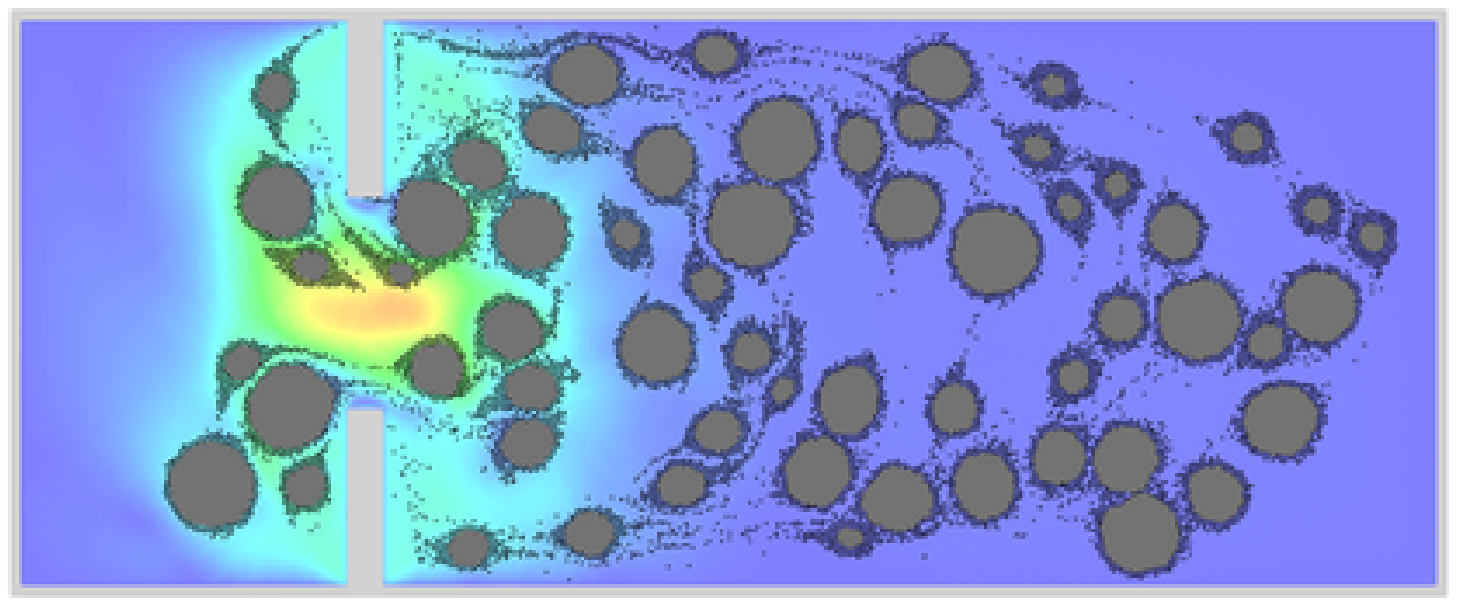}
}
% end subfigure
% begin subfigure
\subfigure [time $t = 0.875$]
{
\includegraphics[width=0.31\textwidth]{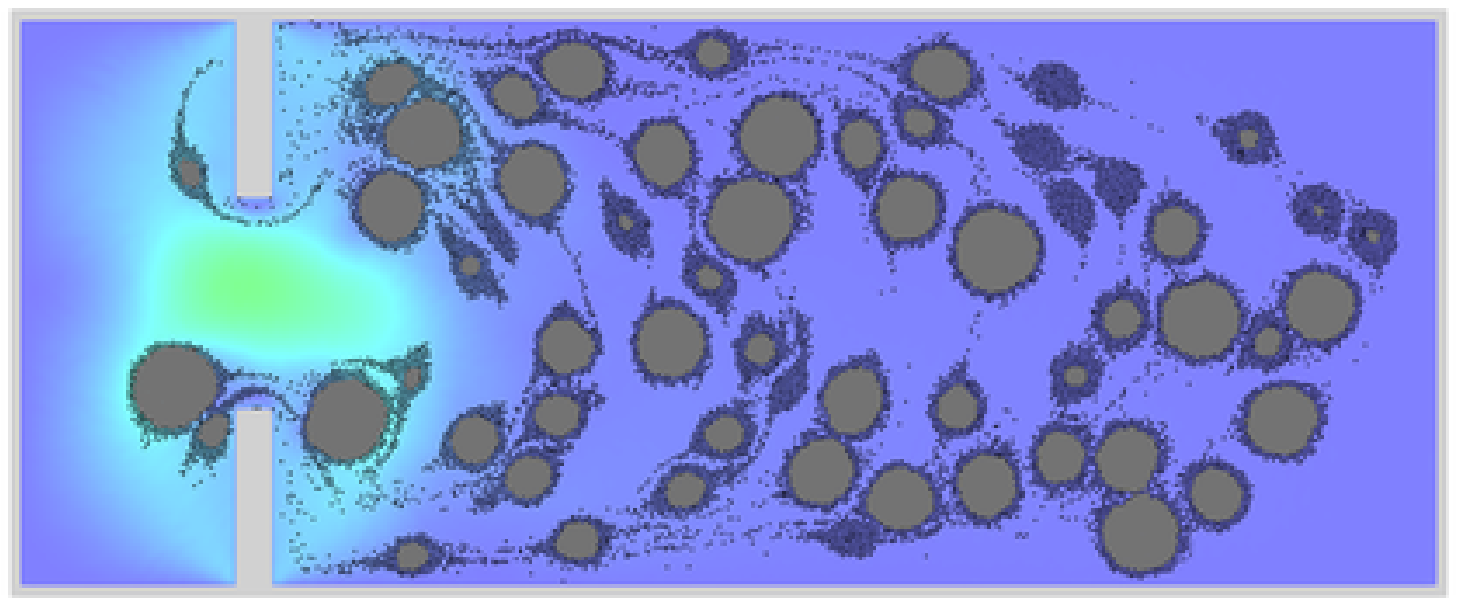}
}
% end subfigure
% begin subfigure
\subfigure [time $t = 1.0$]
{
\includegraphics[width=0.31\textwidth]{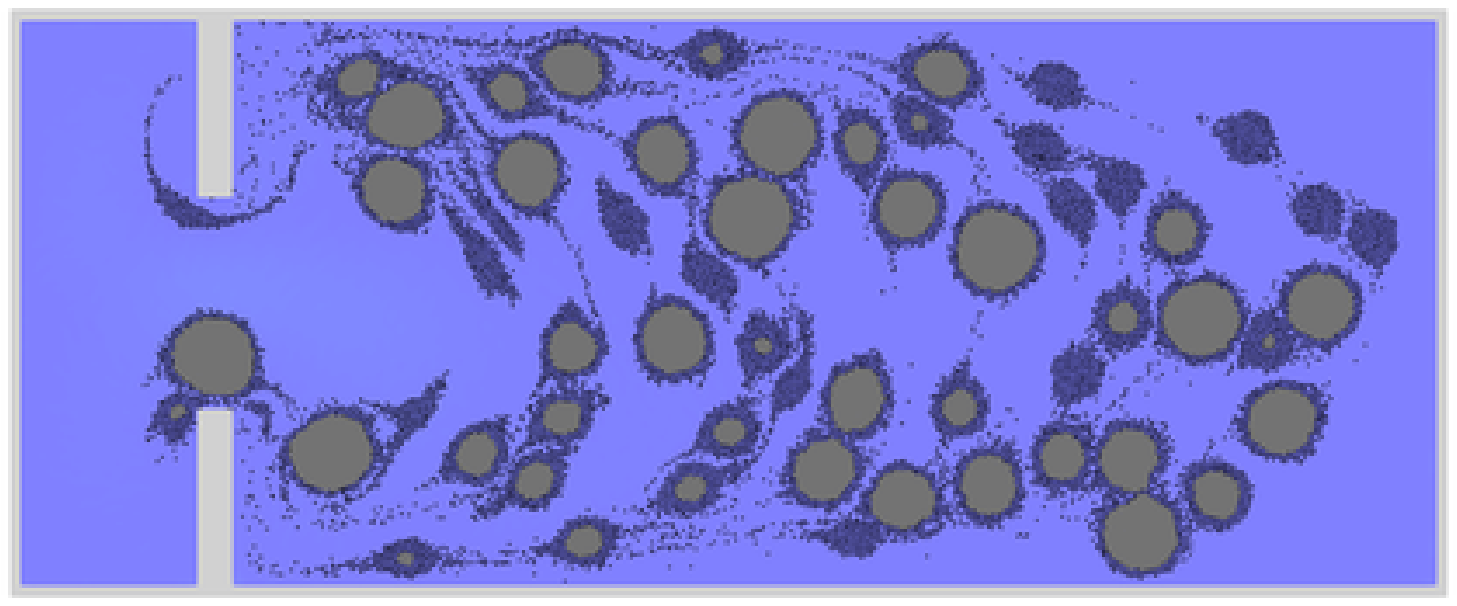}
}
% end subfigure
\caption{Gastric disintegration of food boluses: time series of the obtained results with magnitude of the velocity field ranging from $0.0$ (blue) to $120.0$ (red).}
\label{fig:example_gastric_overview}
\end{figure}

This example aims to demonstrate the capability of the proposed formulation to replicate typical gastric flow patterns including phase transitions. As compared to this complex application scenario, the configuration of the example is kept simple to focus on the principal effects. Consequently, non-physiological parameter values and boundary conditions are applied. Consider a rectangular box of width~$40.0$ and height~$16.0$ (coordinate system in the center) with a mobile constriction. A total of 60 food boluses (density~$\rho^{s} = 1.0$, diffusivity~$D^{s} = 0.25$), represented by mobile rigid bodies with diameters between $1.6$ and $2.8$, are placed at random positions inside the box. The remainder of the box is initially filled with gastric juice (Newtonian fluid, density~$\rho^{g} = 1.0$, kinematic viscosity~$\nu^{g} = 100.0$, diffusivity~$D^{g} = 1.0$). Both the food boluses and the gastric juice are initially at rest. The initial configuration of the example is depicted in Figure~\ref{fig:example_gastric_init}. Over time, food boluses disintegrate into chyme (Newtonian fluid, density~$\rho^{c} = 1.0$, kinematic viscosity~$\nu^{c} = 200.0$, diffusivity~$D^{c} = 0.25$). Herein, this is modeled considering the transport of a concentration~$C$ within the food boluses and chyme, resembling some kind of moisture penetration, by solving a diffusion equation, cf. Remark~\ref{rmk:goveq_diffusion}. Accordingly, the concentration within the food boluses is initialized with $C_{0} = 0.0$, while the concentration within the gastric juice is fixed to $\hat{C} = 1.0$ at all times. Phase transitions from food boluses to chyme is assumed to occur at a transition concentration of $C_{t} = 0.8$. The propagation of an ACW is modeled by the movement of the mobile constriction in the box with a time dependent horizontal velocity of $-14.5 \pi \sin\qty(\pi t)$ from horizontal position $14.5$ to $-14.5$, cf. Figure~\ref{fig:example_gastric_overview}.

For both fluid phases (gastric juice and chyme), an artificial speed of sound~$c = 1.0 \times 10^{3}$ is chosen, resulting in a reference pressure~$p_{0} = 1.0 \times 10^{6}$ of the weakly compressible model, with background pressure of the transport velocity formulation set to $p_{b} = 5 p_{0}$. The wall of the box and the mobile constriction are modeled using (moving) boundary particles. The problem is solved with initial particle spacing~$\Delta{}x = 0.1$ for times~$t \in \qty[0, 1.0]$ with a time step size of~$\Delta{}t = 1.25 \times 10^{-5}$ based on conditions~\eqref{eq:nummeth_timint_timestepcond}.

\begin{figure}[htbp]
\centering
% begin subfigure
%\subfigure [time $t = 0.375$]
%{
%\includegraphics[width=0.23\textwidth]{data/retrojetdisint_detail_t0_375}
%}
% end subfigure
% begin subfigure
\subfigure [time $t = 0.5$]
{
\includegraphics[width=0.31\textwidth]{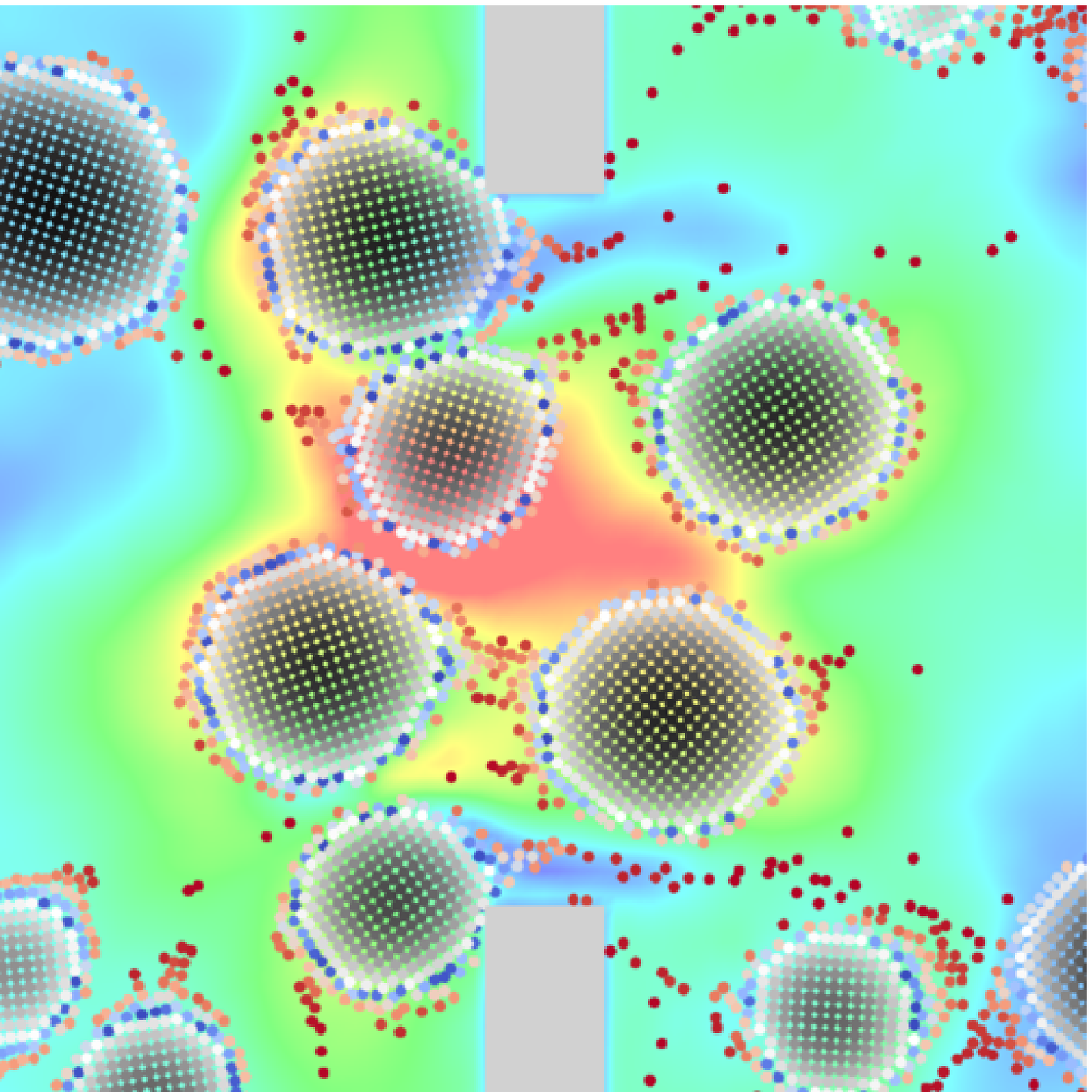}
}
% end subfigure
% begin subfigure
\subfigure [time $t = 0.625$]
{
\includegraphics[width=0.31\textwidth]{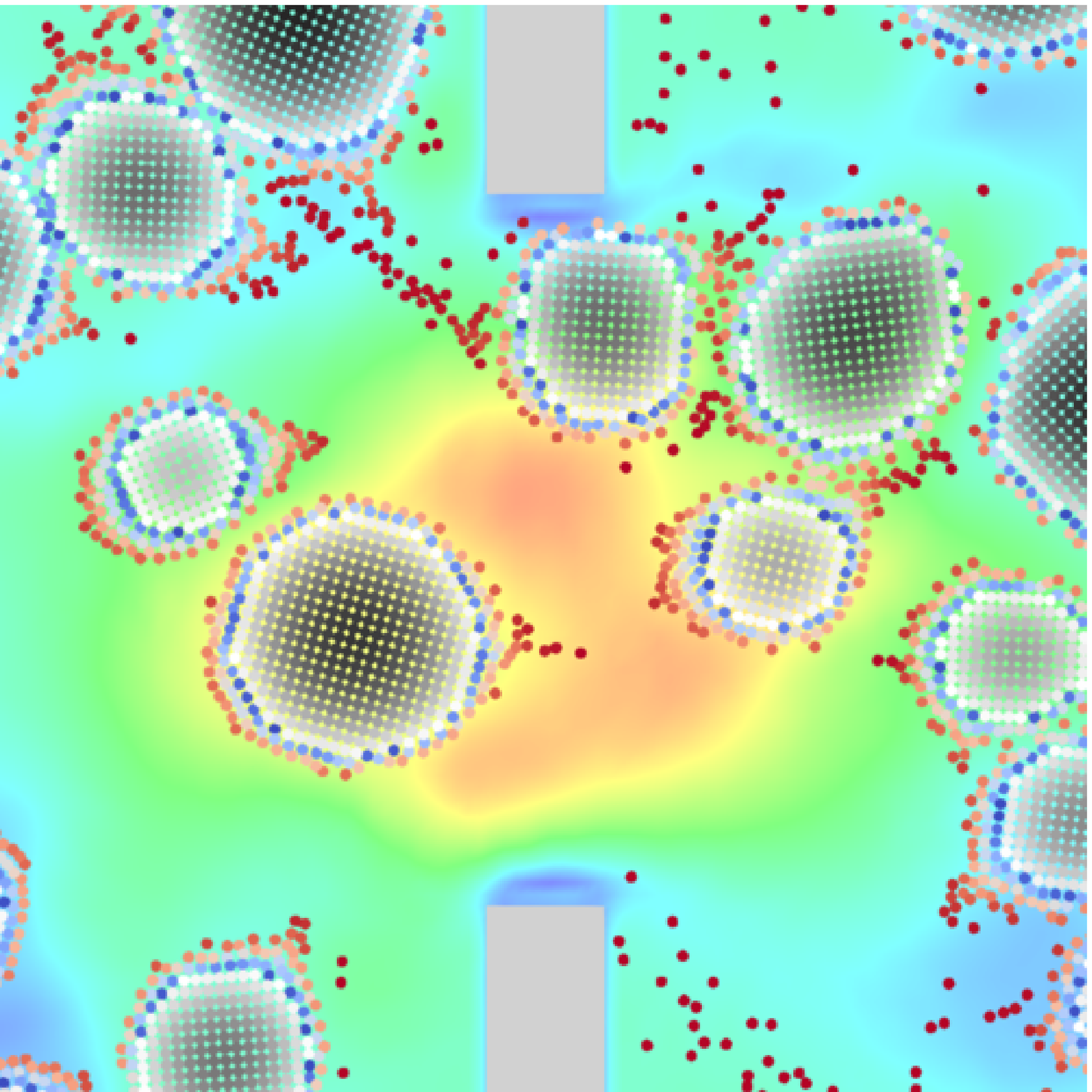}
}
% end subfigure
% begin subfigure
\subfigure [time $t = 0.75$]
{
\includegraphics[width=0.31\textwidth]{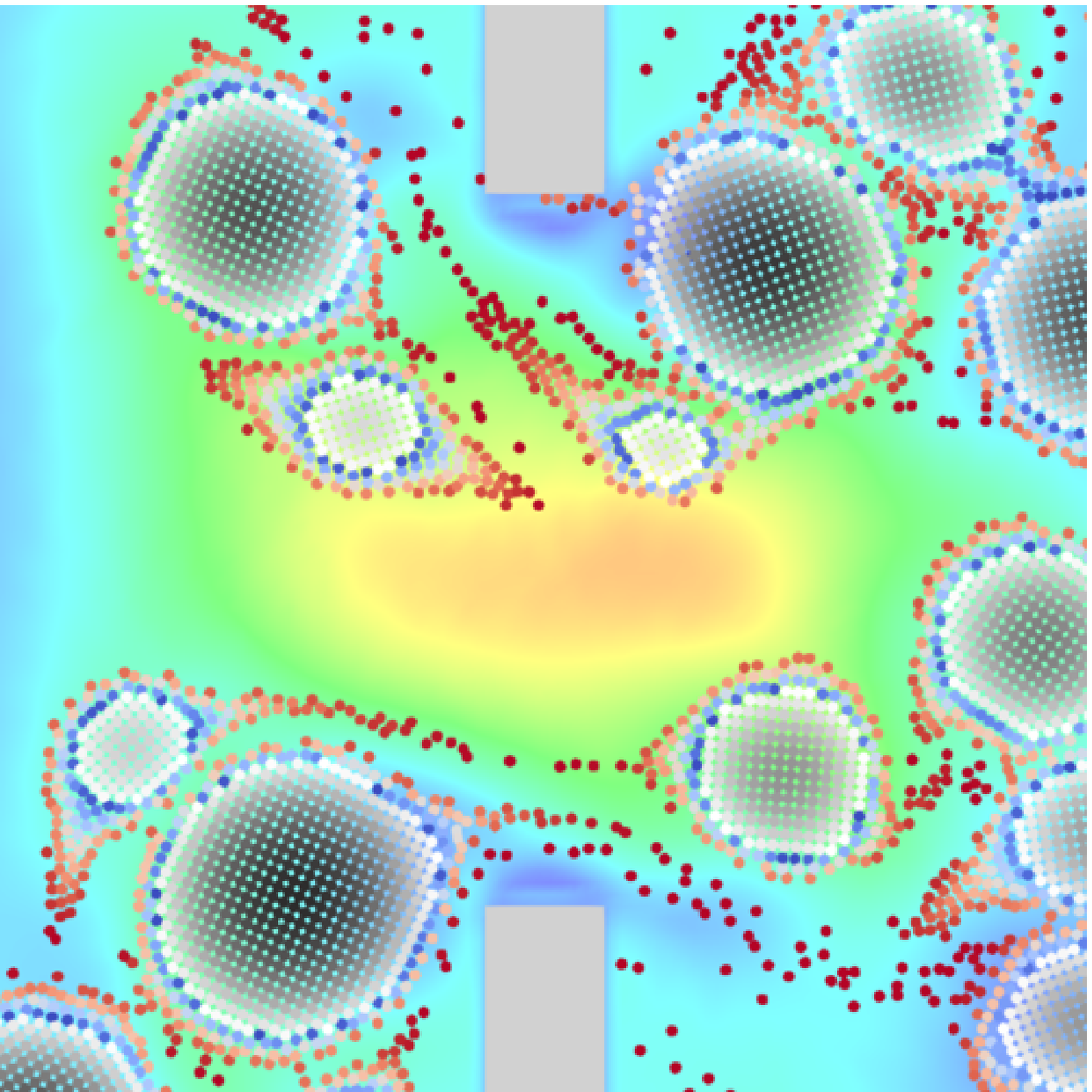}
}
% end subfigure
\caption{Gastric disintegration of food boluses: detailed view of region at the mobile constriction for selected points in time. The concentration~$C$ is ranging from 0.0 (black) to $C_{t} = 0.8$ (white) within the food boluses and from $C_{t} = 0.8$ (blue) to 1.0 (red) within the chyme. In the background, the magnitude of the velocity field is ranging from $0.0$ (blue) to $120.0$ (red).}
\label{fig:example_gastric_detail}
\end{figure}

Figure~\ref{fig:example_gastric_overview} shows a time series of illustrations of the obtained results. The food boluses are visualized in grey color. The particles discretizing the chyme are displayed in black color. In the background, the velocity field of both gastric juice and chyme is post-processed applying SPH approximation~\eqref{eq:nummeth_sph_postprocessing} and visualized by a color code. Clearly, the typical retropulsive jet-like fluid motion induced by the moving constriction can be observed. As a consequence, the food boluses are entrained with the fluid flow through the opening while coming into contact with each other. At the same time, disintegration of food boluses into chyme gradually takes place. After time $t = 0.875$ some food boluses are completely dissolved. A detailed view of the region at the mobile constriction is given in Figure~\ref{fig:example_gastric_detail} for selected points in time. Here, the particles discretizing the food boluses and the chyme are colored based on the concentration field, for distinction, utilizing two different color maps with transition concentration~$C_{t}$ as upper respectively lower value. A progressive mixing of gastric juice and chyme can be observed primarily driven by the fluid motion.

% begin figure
% end figure

% begin figure
% end figure

Note that the main purpose of this example is to show the robustness of the proposed formulation in the context of highly dynamic fluid flow and phase transitions, e.g., as occurring in the form of retropulsive jet-like fluid motion during digestion of food in the human stomach. For the sake of simplicity, non-physiological parameter values are applied. Amongst others, the time scales of ACW propagation and disintegration of food boluses are in a mismatch. In addition, the employed phenomenological digestion model does not explicitly resolve the influence of chemical and mechanical breakdown taking place in reality. In conclusion, this example demonstrates that typical gastric flow patterns including phase transitions are fully captured in a stable and robust manner.

\subsection{Strong scaling analysis of parallel computational framework} \label{subsec:numex_strongscaling}

The purpose of this example is to demonstrate the capability and efficiency of the proposed parallel computational framework in handling systems constituted of a large number of particles. To this end, a three-dimensional example consisting of a total of approximately $3.79 \times 10^{6}$ particles is examined on two different parallel systems. Conclusions are drawn concerning the parallel behavior of the parallel computational framework.

A total of 216 spherical-shaped mobile rigid bodies with diameter~$D = 2.5$ and density~$\rho^{s} = 10.0$ are placed on a regular grid in a cubic box of edge length~$L = 30.0$. The rigid bodies are initially at rest and not in contact with each other or the walls of the box. The remainder of the box is occupied by a Newtonian fluid initially at rest with density~$\rho^{f} = 1.0$ and kinematic viscosity~$\nu^{f} = 1.0$. A gravitational acceleration of magnitude~$\qty|\vectorbold{g}| = 1.0$ is acting in downward direction, i.e., the body forces (per unit mass) of fluid and solid field are given to $\vectorbold{b}^{f} = \vectorbold{g}$ and $\vectorbold{b}^{s} = \vectorbold{g}$.

% problem size:
% 3796416 total particles
% 3154464 fluid particles, 220536 rigid particles, 421416 boundary particles

For the fluid phase, an artificial speed of sound~$c = 50.0$ is chosen, resulting in a reference pressure~$p_{0} = 2.5 \times 10^{3}$ of the weakly compressible model, with background pressure~$p_{b}$ of the transport velocity formulation set equal to the reference pressure~$p_{0}$. The wall of the box is modeled using boundary particles. The complete domain is discretized by particles with initial particle spacing~$\Delta{}x = 0.2$ resulting in a total of approximately $3.79 \times 10^{6}$ particles, thereof $3.15 \times 10^{6}$ fluid particles, $2.20 \times 10^{5}$ rigid particles, and $4.21 \times 10^{5}$ boundary particles. Following a spatial decomposition approach, cf. Section~\ref{subsec:nummeth_parallelization}, the computational domain is divided into $48 \times 48 \times 48$ cubic cells of edge length $0.65$ resulting in approximately 34 particles per cell. The problem is solved for times~$t \in \qty[0, 30.0]$ with a time step size of~$\Delta{}t = 1.0 \times 10^{-3}$.

\begin{figure}[htbp]
\centering
% begin subfigure
\subfigure [time $t = 0.0$]
{
\includegraphics[width=0.2\textwidth]{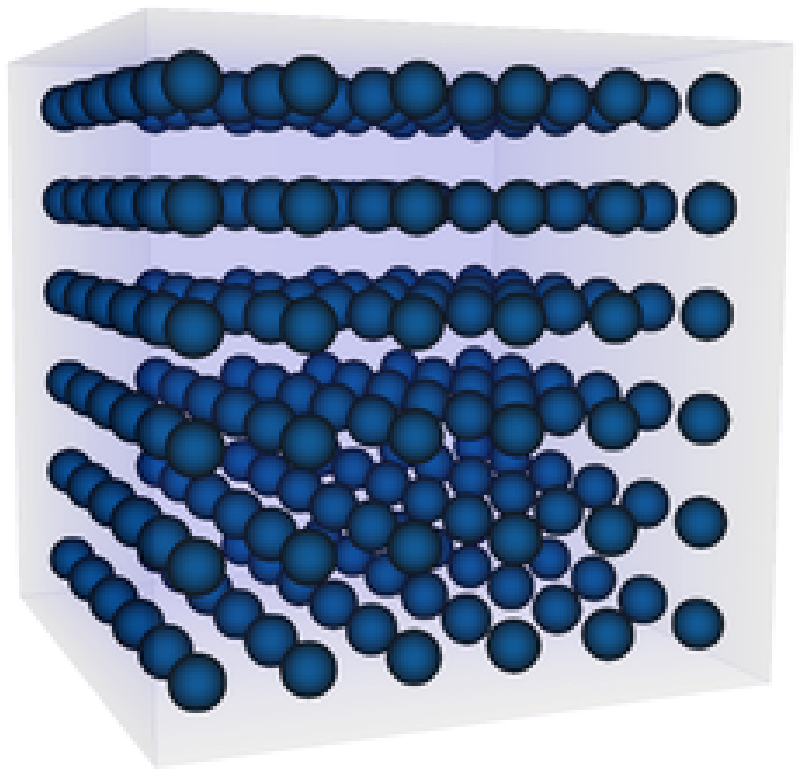}
}
% end subfigure
% begin subfigure
\subfigure [time $t = 10.0$]
{
\includegraphics[width=0.2\textwidth]{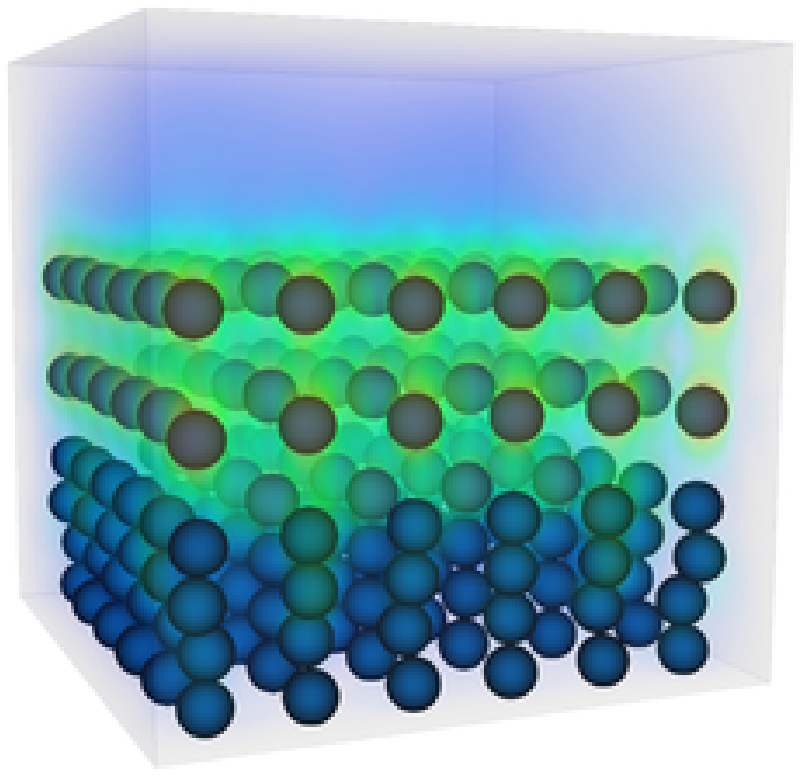}
}
% end subfigure
% begin subfigure
\subfigure [time $t = 20.0$]
{
\includegraphics[width=0.2\textwidth]{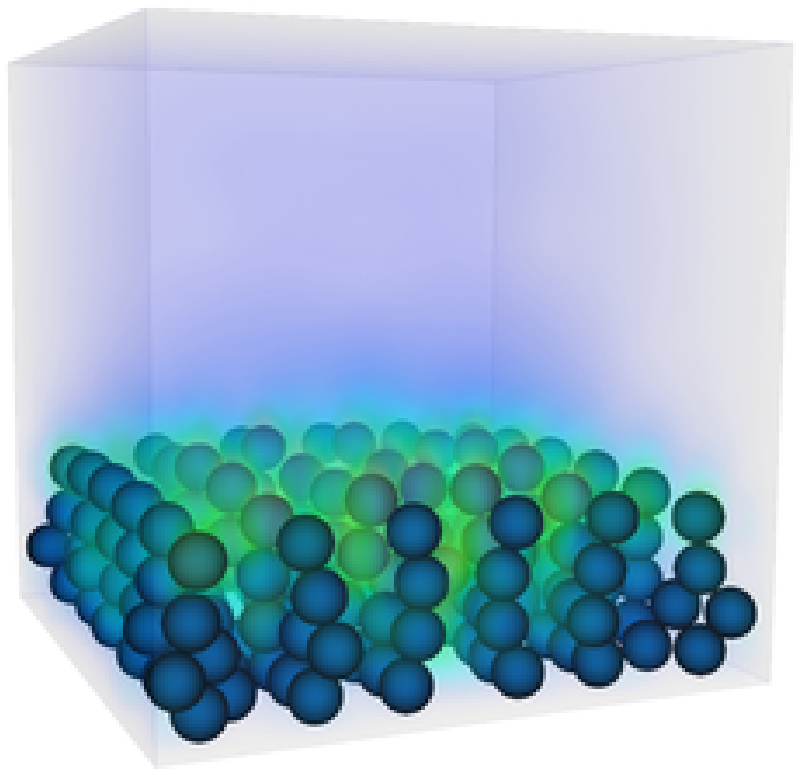}
}
% end subfigure
% begin subfigure
\subfigure [time $t = 30.0$]
{
\includegraphics[width=0.2\textwidth]{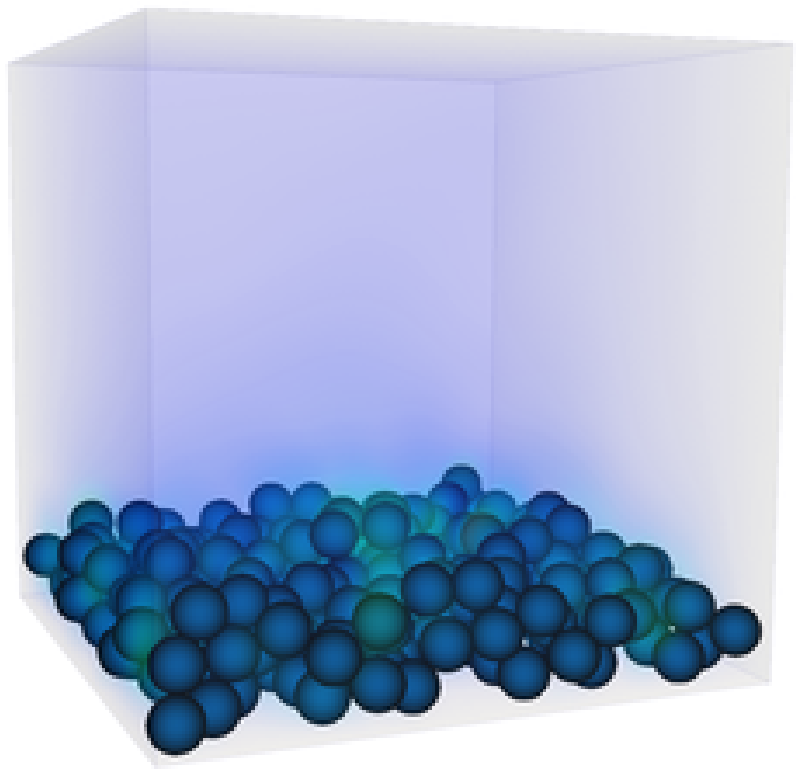}
}
% end subfigure
\caption{Strong scaling analysis of parallel computational framework: position of spherical-shaped rigid bodies within the box for different points in time with magnitude of the velocity field ranging from 0.0 (blue) to 1.25 (red).}
\label{fig:example_strongscaling_visualization}
\end{figure}

For the purposes of illustration, the spherical-shaped rigid bodies within the box are shown in Figure~\ref{fig:example_strongscaling_visualization} for the initial setup at $t = 0.0$ and later points in time. In addition, the velocity field of the surrounding fluid is post-processed applying SPH approximation~\eqref{eq:nummeth_sph_postprocessing} and visualized by a color code with opacity. The rigid bodies are falling freely in the viscous fluid under gravity due to density ratio~$\flatfrac{\rho^{s}}{\rho^{f}} = 10.0$ until contact with the bottom wall of the box occurs, as first observed after $t \approx 2.5$, or with neighboring rigid bodies, as first observed after $t \approx 5.0$. The rigid bodies begin piling up at the bottom wall of the box and are nearly at rest at $t = 30.0$.

% begin figure
% end figure

% hardware
%
% "bruteforce" consisting of up to 52 nodes with 2 x 12 cores (Intel Xeon E5-2680 v3 Haswell, 2.5 GHz, 64 GB RAM)
%
% "deep batch" consisting of up to 26 nodes with 2 x 8 cores (Intel Xeon E5-2630 v3 Haswell, 2.4 GHz, 64 GB RAM)
%
% "deep gold" consisting of up to 8 nodes with 2 x 20 cores (Intel Xeon Gold 6230 Cascade Lake, 2.1 GHz, 96 GB RAM)

\begin{figure}[htbp]%
\centering
% begin subfigure
\subfigure %[caption of subfigure]
{
\begin{tikzpicture}[trim axis left,trim axis right]
\begin{axis}
[
width=0.35\textwidth,
height=0.18\textwidth,
grid=major,
xmode=log,
ymode=log,
log basis x={2},
log basis y={10},
log ticks with fixed point,
xmin=16, xmax=1024,
ymin=0.1, ymax=10,
xtick={16, 32, 64, 128, 256, 512, 1024},
xticklabels={16, 32, 64, 128, 256, 512, 1024},
ytick={0.1, 1, 10},
xlabel={number of cores},
ylabel={solver time $\qty[\texttt{s}]$},
% define legend
legend columns=0,
legend entries={
Intel Xeon E5-2680 v3 Haswell,
Intel Xeon E5-2630 v3 Haswell,
linear scaling
},
legend to name=legend_strongscaling_analysis,
]
% begin plot
%\addplot [color=red,solid,line width=0.5pt,mark=square] table [x=cores, y=walltimeperstep, col sep=comma] {data/strongscaling_bruteforce_4e6_p16.csv};
\addplot [color=black,solid,line width=0.5pt,mark=square] table [x=cores, y=walltimeperstep, col sep=comma] {data/strongscaling_bruteforce_4e6_p24.csv};
\addplot [color=blue,solid,line width=0.5pt,mark=triangle] table [x=cores, y=walltimeperstep, col sep=comma] {data/strongscaling_deep_4e6_p16_batch.csv};
\addplot [color=red,dashed,line width=0.5pt] coordinates{(16,8) (1024,0.125)};
% end plot
\end{axis}
\end{tikzpicture}
\label{fig:example_strongscaling_analysis_time}
}
% end subfigure
\hspace{0.12\textwidth}
% begin subfigure
\subfigure %[caption of subfigure]
{
\begin{tikzpicture}[trim axis left,trim axis right]
\begin{axis}
[
width=0.35\textwidth,
height=0.18\textwidth,
grid=major,
xmode=log,
log basis x={2},
log ticks with fixed point,
xmin=16, xmax=1024,
ymin=0, ymax=100,
xtick={16, 32, 64, 128, 256, 512, 1024},
xticklabels={16, 32, 64, 128, 256, 512, 1024},
ytick={0, 25, 50, 75, 100},
xlabel={number of cores},
ylabel={parallel efficiency $\qty[\%]$},
]
% begin plot
%\addplot [color=red,solid,line width=0.5pt,mark=square] table [x=cores, y=strongscalingefficiency, col sep=comma] {data/strongscaling_bruteforce_4e6_p16.csv};
\addplot [color=black,solid,line width=0.5pt,mark=square] table [x=cores, y=strongscalingefficiency, col sep=comma] {data/strongscaling_bruteforce_4e6_p24.csv};
\addplot [color=blue,solid,line width=0.5pt,mark=triangle] table [x=cores, y=strongscalingefficiency, col sep=comma] {data/strongscaling_deep_4e6_p16_batch.csv};
\addplot [color=red,dashed,line width=0.5pt] coordinates{(16,100) (1024,100)};
% end plot
\end{axis}
\end{tikzpicture}
\label{fig:example_strongscaling_analysis_efficiency}
}
% end subfigure
% begin legend
\ref{legend_strongscaling_analysis}
% end legend
\caption{Strong scaling analysis of parallel computational framework: solver time per time step (left) and parallel efficiency given in percent of linear scaling (right) for a problem consisting of approximately $3.79 \times 10^{6}$ particles on up to 768 cores.}
\label{fig:example_strongscaling_analysis}
\end{figure}
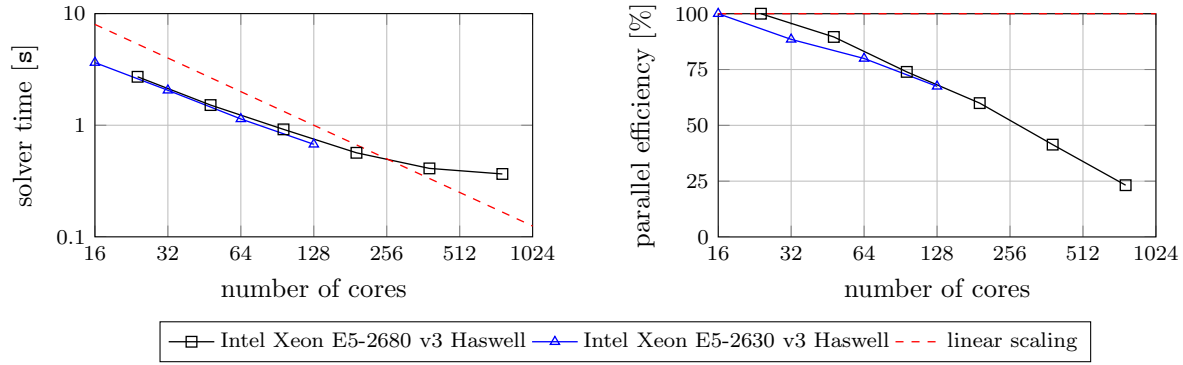

To showcase the capability and efficiency of the parallel computational framework, a strong scaling analysis is performed utilizing two different parallel systems: The first one consisting of 32 nodes with $2 \times 12$ cores (Intel Xeon E5-2680 v3 Haswell, 2.5 GHz) and the second one consisting of 8 nodes with $2 \times 8$ cores (Intel Xeon E5-2630 v3 Haswell, 2.4 GHz). The parallel behavior of the proposed computational framework is given in Figure~\ref{fig:example_strongscaling_analysis}, illustrating the obtained solver time per time step, cf. Figure~\ref{fig:example_strongscaling_analysis_time}, and the parallel efficiency given in percent of linear scaling, cf. Figure~\ref{fig:example_strongscaling_analysis_efficiency}. The parallel efficiency is computed as $\flatfrac{t_{1}}{ \qty( n \cdot t_{n} ) } \cdot 100 \, \%$, where $t_{1}$ and $t_{n}$ are the times to solve the problem on one node respectively $n$ nodes.

% begin figure
% end figure

The parallel computational framework scales almost linearly on both parallel systems for up to 128 cores respectively 192 cores. In this regime a parallel efficiency of more than $60 \, \%$ can be observed. For larger numbers of cores, the scalability deteriorates and the parallel efficiency drops to under $50 \, \%$. This can be explained with an increasing communication overhead, cf. Remark~\ref{rmk:nummeth_parallelization_overhead}. Comparable results of a strong scaling analysis for an SPH implementation are given, e.g., in \cite{Oger2016} ($\flatfrac{r_{c}}{\Delta{}x} = 2.5$) and \cite{Yang2020} ($\flatfrac{r_{c}}{\Delta{}x} = 2.4$), however, in contrast to this example ($\flatfrac{r_{c}}{\Delta{}x} = 3.0$) with a smaller ratio of the support radius~$r_{c}$ and the initial particle spacing~$\Delta{}x$, resulting in a lower influence on the communication overhead, cf. Remark~\ref{rmk:nummeth_parallelization_overhead}. As a conclusion one can state that the parallel computational framework is capable of efficiently solving systems constituted of a large number of particles on multiple cores. Looking at the parallel behavior, the obtained results confirm that the proposed framework meets all requirements necessary for detailed and accordingly computationally expensive studies.

\section{Conclusion} \label{sec:concl}

In this work, an approach for fluid-solid and contact interaction problems including thermo-mechanical coupling and reversible phase transitions is presented. All fields are spatially discretized using smoothed particle hydrodynamics (SPH). Being a mesh-free discretization scheme, SPH is, compared to mesh-based methods, especially suitable in the context of continually changing interface topologies and dynamic phase transitions by avoiding additional methodological and computational effort to capture such phenomena. A detailed concept for the parallelization of the computational framework, especially for an efficient evaluation of rigid body motion, is an essential part of this work.

The accuracy and robustness of the proposed formulation are demonstrated by several numerical examples studying a single rigid body in fluid flow, cf. Sections~\ref{subsec:numex_cylindershearflow} and~\ref{subsec:numex_fallingcylinder}. The obtained numerical results are in very good agreement with the literature. Also two complex examples close to potential applications scenarios in the fields of engineering and biomechanics were studied. First, motivated by metal PBFAM melt pool modeling, melting and solidification of powder grains subject to highly dynamic fluid motion was simulated, cf. Section~\ref{subsec:numex_melting}. Second, inspired by multiphysics modeling of the human stomach, gastric disintegration of food boluses is considered, cf. Section~\ref{subsec:numex_gastric}. Both examples confirm that highly dynamic motion of arbitrarily-shaped rigid bodies embedded in a complex fluid flow and including reversible phase transitions can be captured by the proposed framework in a stable and robust manner. Finally, the parallel computing abilities of the proposed computational framework were demonstrated by a strong scaling analysis of a three-dimensional example with $3.79 \times 10^{6}$ particles revealing a parallel efficiency of more than $60 \, \%$ on up to 192 cores, cf. Section~\ref{subsec:numex_strongscaling}.

To the best of the authors’ knowledge, the proposed parallel computational framework is the first of its kind modeling rigid body motion while simultaneously considering thermal conduction, reversible phase transitions, and multiple (liquid and gas) phases. In summary, it has the ability to accurately model a host of complex multiphysics problems, and it can thus be expected to become a valuable tool for detailed studies in engineering, e.g., metal additive manufacturing, and biomechanics, e.g., digestion of food in the human stomach.

\section*{Acknowledgments}

Funded by the Deutsche Forschungsgemeinschaft (DFG, German Research Foundation) - 350481011, 437616465, and 414180263. In addition, the authors thank Bugrahan Z. Tem\"ur for preliminary work on a serial rigid body implementation.

%% \section{} \label{}
%%
\bibliographystyle{elsarticle-num} 
\bibliography{collection.bib}
\end{document}